\documentstyle [12pt,mytesis,epsfig]{informe}

\begin{document}
\hyphenation {es-tre-llas es-tre-lla ga-la-xia ga-la-xias bri-llan-tes
}
\parindent=2em
%
%
\def\PLANCK{\frac{2\pi h \nu ^{3}}{c^{2}} \frac{1}{e^{
\frac{h\nu}{KT_{B}}}-1}}
\def\BOLTZMANN{\frac{n_{j}}{n_{i}}= \frac{g_{j}}{g_{i}}
e^{-\frac{h\nu_{ij}}{KT_{ex}} } }
\def\MAXWELL{\frac{4}{\sqrt {\pi}}(\frac{m}{2kT_{k}})^{\frac{3}{2}}
v^{2}e^{-\frac{mv^{2}}{2kT_{k}}} }
%
%
\def\lr#1{\hbox{\rm#1}}
\def\kms{\relax \ifmmode {\,\rm km\,s}^{-1}\else \,km\,s$^{-1}$\fi}
\def\ha{\relax \ifmmode {\rm H}\alpha\else H$\alpha$\fi}
\def\hb{\relax \ifmmode {\rm H}\beta\else H$\beta$\fi}
\def\hi{\relax \ifmmode {\rm H\,{\sc i}}\else H\,{\sc i}\fi}
\def\hii{\relax \ifmmode {\rm H\,{\sc ii}}\else H\,{\sc ii}\fi}
\def\h2{\relax \ifmmode {\rm H}_2\else H$_2$\fi}
\def\lha{\relax \ifmmode L_{{\rm H}\alpha}\else $L_{{\rm H}\alpha}$\fi}
\def\shi{\relax \ifmmode \sigma_{{\rm HI}}\else $\sigma_{\rm HI}$\fi}
\def\sh2{\relax \ifmmode \sigma_{{\rm H}_2}\else $\sigma_{{\rm H}_2}$\fi}
\def\degr{\hbox{$^\circ$}}
\def\arcmin{\hbox{$^\prime$}}
\def\arcsec{\hbox{$^{\prime\prime}$}}
\def\deg{\hbox{$^\circ$}}
\def\min{\hbox{$^\prime$}}
\def\sec{\hbox{$^{\prime\prime}$}}
\def\fdg{\hbox{$.\!\!^\circ$}}
\def\fs{\hbox{$.\!\!^{\rm s}$}}
\def\farcm{\hbox{$.\mkern-4mu^\prime$}}
\def\farcs{\hbox{$.\!\!^{\prime\prime}$}}
\def\degd#1.#2{ #1\fdg#2 }               
\def\mind#1.#2{ #1\farcm#2 }               
\def\secd#1.#2{ #1\farcs#2 }               
\def\hhh{\ifmmode {\rm ^h}              
         \else {${\rm ^h}$}
         \fi}
\def\sss{\ifmmode {\rm ^s}              
         \else {${\rm ^s}$}
         \fi}
\def\hms#1h#2m#3s{                      
                  \relax
                  \ifmmode #1^{\rm h}\,#2^{\rm m}\,#3^{\rm s}
                  \else \hbox{$#1^{\rm h}\,#2^{\rm m}\,#3^{\rm s}$}
                  \fi
                 }
\def\dms#1d#2m#3s{                      
                  \relax
                  #1\degr\,#2\arcmin\,#3\arcsec 
                 }
\def\hmsd#1h#2m#3.#4s{                  
                      \relax
                      \ifmmode #1^{\rm h}\,#2^{\rm m}\,#3\fs#4
                      \else \hbox{$#1^{\rm h}\,#2^{\rm m}\,#3\fs#4$}
                      \fi
                     }
\def\dmsd#1d#2m#3.#4s{                  
                      \relax
                      #1\degr\,#2\arcmin\,#3\farcs#4
                     }
\def\mag{\relax                          
        \ifmmode ^{\rm m}
        \else $^{\rm m}$
        \fi
       }
\def\magd#1.#2{                          
              \relax
              \ifmmode #1^{\rm m}
                       \hskip-0.55em.\hskip0.22em#2
              \else \hbox{#1$^{\rm m}
                    \hskip-0.55em.\hskip0.22em$#2}
              \fi
             }

\newcommand{\pixel}{{\it pixel \/}}
\newcommand{\pixels}{{\it pixels \/}}
\newcommand{\bias}{{\it bias \/}}
\newcommand{\ff}{{\it flat-field \/}}
\newcommand{\aap}{{\em Astron. Astrophys., \/}}
\newcommand{\apj}{{\em Astrophys. J., \/}}
\newcommand{\aass}{{\em Astron. Astrophys. Suppl. Ser., \/}}
\newcommand{\apjss}{{\em Astrophys. J. Suppl. Ser., \/}}
\newcommand{\ass}{{\em Astrophys. Space Sci., \/}}
\newcommand{\aj}{{\em Astron. J. \/}}
\newcommand{\mnras}{{\em Mon. Not. Roy Astr. Soc., \/}}
\newcommand{\araa}{{\em Ann. Rev. Astron. Astrophys., \/}}
\newcommand{\pasp}{{\em Pub. Astr. Soc. Pac., \/}}

\input epsf

\copyrightfalse
\signaturesfalse
\proyectofalse
\figurespagetrue
\tablespagetrue

\title{Dust and gas in Active Galaxies}

\author{M. Montserrat Villar Mart\'\i n}

\beforepreface

\newpage

~

\newpage

{\large ~

\vspace{8cm}

~~~~~~~~~~~~~~~~~~~~~~~~~~~~~~~~~~~~~~~~~~~~~~~ A mis queridos padres}

\newpage
~
\newpage

\prefacesection{Resumen}

Existen numerosas evidencias de la presencia de polvo en galaxias activas. Al estudiar la naturaleza del n\'ucleo activo gal\'actico  y las condiciones f\'\i sicas del medio interestelar de estas galaxias es necesario entender c\'omo el polvo interacciona con la radiaci\'on y las part\'\i culas del gas. Si no tenemos en cuenta en nuestro an\'alisis los efectos derivados, posiblemente llegaremos a interpretaciones err\'oneas. Existen muchas preguntas 
relacionadas con la existencia de polvo en galaxias activas como: ?`En qu\'e condiciones permite el continuo extremadamente duro del n\'ucleo activo la supervivencia de los granos de polvo? ?`Es la naturaleza del polvo en galaxias
activas la misma que en nuestro medio interestelar? ?`C\'omo se distribuye el polvo en relaci\'on con el gas altamente ionizado por el AGN central? ?`Existe polvo en radio galaxias con desplazamientos al rojo muy altos? El trabajo desarrollado en esta tesis trata de encontrar respuesta a algunas de estas preguntas, mediante un estudio detallado, tanto observacional como te\'orico, de los mecanismos que controlan la interacci\'on entre el polvo, la radiaci\'on y las part\'\i culas del gas. Los efectos observables del polvo sobre el espectro de l\'\i neas de emisi\'on son tambi\'en analizados en profundidad. El objetivo es entender cuestiones m\'as generales como el origen del gas emisor, el/los mecanismos de ionizaci\'on, la geometr\'\i a, la conexi\'on entre galaxias activas a alto y bajo redshift o la validez del modelo de unificaci\'on. En pocas palabras, este trabajo intenta alcanzar una visi\'on m\'as clara del mundo de las galaxias activas.

\newpage
~
\vspace{2.5cm}

{\Huge \bf Summary}

\vspace{1.5cm}

There are strong evidences which favour the existence of dust in active galaxies.
Understanding the way in which dust interacts with the radiation and influences the physical
conditions of the gas is crucial if we want to learn about the nature
of the central active nucleus and about the physical conditions of the ISM
in such galaxies. Not taking into account such effects may lead us
towards misleading interpretations.  Many intriguing questions concerns to the nature and the existence of
dust in active galaxies:  for instance, under which
conditions does the very hard ionizing continuum of an AGN allows
the survival of dust grains? Is the composition and size distribution of the dust the same 
as in our local interstellar medium? How is dust distributed compared to
the gas which is at least in part highly ionized by the central AGN?  Does dust also exist in radio
galaxies at very high redshifts? The work developed in this thesis tries to find answers to some of these
questions, through a detailed theoretical and observational research
of the mechanisms which control the interaction of dust with the radiation and 
with the ions. The observable effects of the dust on the emission line
spectrum are also analyzed in detail. The final
goal has been  to give clues about more general questions: origin of the emitting gas, ionization mechanisms, geometry, connection between low and high redshift active galaxies or the validity of the
unification scenario. This thesis tries, in summary, to provide a clearer understanding of active galaxies in general.


\afterpreface

\newpage
~
\newpage

\chapter{Introduction}

\section{The importance of studying dust in active galaxies.}
	
	Dust is an important constituent of active galaxies
and must be included in any physical description of them.  It is
necessary to reach a clear understanding of its influence, not only on
the radiation, but also on the physical conditions of dusty
nebulae. Ignoring dust in the interpretation of the observations is an
oversimplification which may lead us to false conclusions. An example
will describe clearly the kind of misunderstandings which can be reached
when skipping dust in our interpretations.

	 {\it High z} radio galaxies (z$>$0.5) show very blue
colors and emit a very strong Ly$\alpha$ line. The initial
interpretation has been
that these objects were galaxies in the process of formation and that the
blue young stars could explain the observed properties. In this
scenario, the continuum is
produced by the stars while Ly$\alpha$ results from
recombination of gas from HII nebulosities. Further investigations revealed
that the continuum was extended and closely aligned with the radio
axis. This indicated a common mechanism for both the radio
and the UV conti\-nuum emission. Polarization studies (e.g. Tadhunter,
Fosbury \& di Serego Alighieri 1988; Cimatti et al. 1993) revealed
that the continuum was highly polarized and demonstrated that the
extended radiation could be scattered nuclear radiation by dust within the
interstellar medium. This implies that the star formation rate is
lower than previously believed, also lower than we thought if these
objects were protogalaxies. There is no clear evidence that these
objects are galaxies being born, not even at redshift $\sim$ 4.  Some
evidences rather demonstrate the existence of an old stellar
population (this will be treated in detail in Chapter 6).  The
redshift corresponding to galaxy formation seems to be much further
back than the currently 
highest redshift quasar known at $z= 4.9$ (Shneider, Schmidt \& Gunn 1991).

The regions responsible for the emission of the narrow
lines in active galaxies might well be mixed with dust resulting in
selective absorption of the impinging ionizing
photons and, therefore, affecting
the ionization structure of the gas. The physical conditions of the
gas are also directly
influenced: the temperature and the (gas phase) abundances of the heavy elements are sensitive
to the presence dust. All these effects will manifest themselves in the emission line
spectrum.

  In this work we have developed a complete and detailed study
(observational and theoretical) of the effects of internal (i.e. mixed
with the ionized gas) as well as external (i.e. intervening) dust, on the emission line spectra
of narrow line active galaxies. We recall that the emission lines
observed in the spectra of
this type of objects are not contaminated by a broad component emitted
within the so-called Broad Line Region, which  makes our study much simpler.

	 This thesis has consisted of three main projects and attempts
to address and
solve some of the important problems encountered nowadays in the field of
active galaxies. These often involve the presence of dust:

1) Effects of dust on the Ly$\alpha$ and Balmer decrements in the
Narrow Line Region of Seyfert 2 galaxies. {\it What is the proper geometry
describing the Narrow Line Region (NLR)?}

2) Does dust
exist as a component which is mixed with the ionized gas observed in  Extended Emission Line Regions of Radio
galaxies? {\it What  is the origin and fate of this extended gas?}

3) Effects of dust and of resonant scattering on the UV spectrum of high
$z$ radio galaxies. {\it Which factor plays a major role on the UV
line ratios at high $z$: geometry or internal dust?}

 These specific objectives are quite different in terms of the type of
object considered. In effect, we move from low
redshift ($z$) to very high redshift objects (which correspond to much
earlier epochs) and consider regions which strongly differ by their
physical conditions. However, the goal remains the same: {\it to
understand the nature of active galaxies}.  Many researchers are
working on this thrilling field, trying to answer many questions which
remain open, using different theoretical and observational techniques.
In our work, we have modeled the emission line spectra of the regions
under study. Starting from a given set of input parameters (some of
them well established by the observations), we create theoretical
spectra which are then compared with the observations. The discrepancies or
agreements resulting from the comparison provide important
information about the nature of the regions and objects under study.

\newpage

\section{Evidences of dust in active galaxies.}

\subsection{Absorption}

 Extinction occurs whenever electromagnetic radiation is propagated
through a medium containing small dust particles.  In general, the
transmitted beam is reduced in intensity by two physical processes,
absorption and scattering. The energy of an absorbed photon is
converted into internal energy of the particle, which is thus heated,
while a scattered photon is deflected from the line of sight. There is
direct evidence for dust in our Milky Way in the form of dark lanes
and rifts that we can see with the unaided eye. Also in some active
galaxies, we can detect the presence of dust from the existence of
obscured regions, like Cen A (NGC5128), the nearest giant double radio
galaxy. In 1847, Sir John Herschel observed for the first time the
nowadays familiar dark lane contrasted against the
diffuse background of light. Currently, it is known that the galaxy
consists of an ellipsoidal stellar component bisected by a dense dusty
layer across the apparent minor axis (see Fig.~1) (e.g. Bland et
al. 1987; Morganti et al. 1991). Recent HST WFPC2 images of 3CR radio
galaxies in the redshift range 0.0$<z<$0.5, show lanes, patches and
wisps that the authors attribute to dust out to a redshift of
$\sim$0.48 in $\sim$ 30-50\% of the sample (Baum et al.~1995).  At
$z>$0.1, dust is more difficult to detect and one cannot easily separate
dust obscuration from emission clumpiness.

\begin{figure}[htb]
\vspace{3.5in}
\caption[The dusty lane in Cen A]{}{\small The dusty lane in the active galaxy Centaurus A. A deep (1 hr) image taken at the prime focus of the Anglo-Australian Telescope. (Reproduced from Morganti et al.~1991)}

\end{figure}

 Selective extinction (or reddening) is also evident in the spectra
of Narrow Line Regions (NLR) in Seyfert
2 galaxies. In effect,   the Ly$\alpha$/H$\beta$
and H$\alpha$/H$\beta$ ratios indicate a mixture of reddening due to dust {\it mixed}
with the emitting clouds and of resonant scattering of Ly$\alpha$ by
neutral hydrogen (Binette et al. 1993, see also Chapter 2) (some
authors propose reddening by {\it pure} dust (i.e. gas-free dust!), like Ferland and 
Osterbrock 1986, and Wills et al. 1993).
 
The narrow emission line profiles provide another indication of dust
in the NLRs. Many lines have a noticeable blue asymmetry which is
interpreted as a combination of radial motion and dust obscuration
(e.g. de Robertis \& Osterbrock 1984).  The asymmetry parameter
(defined by Heckman, Miley \& Breugel 1981) measures the asymmetry of
the line profiles. Dahari \& de Robertis (1988) searched for
correlations between this parameter and features attributed to dust such as
reddenning (meaasured from the Balmer decrement, H$\alpha$/H$\beta$) far IR luminosity (60-100 $\mu$m) and the flux ratio
F(60$\mu$m)/F($\lambda$5007). The results showed a good correlation in
Seyfert 2 galaxies between the asymmetry parameter and these three
variables. The authors propose that such correlations suggest a direct
relation between the asymmetry parameter of the lines and the presence
of dust in the NLR.

\subsection{Scattering and Polarization}

 The existence of scattered light from the interstellar dust grains
was discovered by means of long exposure photographs of systems such
as the Pleiades cluster of stars. The  associated nebulosity is
referred to as reflection nebulosity.  The light of the stars in the
cluster illuminates the dust grains which then scatter the light, most
efficiently at
blue wavelengths, explaining the blue colors associated with the
nebulosity.  The scattering of radiation has a useful observational
consequence: the scattered light is polarized. This was first
demonstrated by Hall (1949) and Hiltner (1949) who showed that the
light of reddened stars is partially plane polarized, typically at the
1-5\% level.

	Polarization has often been observed in active galaxies and
 the interpretation generally proposed has been scattering of the radiation by an agent
 which nature does not reveal clearly: either electrons or dust
grains. Antonucci and
 Miller (1985) discovered that in the Seyfert 2 galaxy NGC1068, both
 the nonstellar nuclear continuum and the weak broad wings to the Balmer
 lines and the FeII emission are linearly polarized with the
 polarization plane perpendicular to the symmetry axis of the nuclear
 radio morphology. The most plausible interpretation is the existence
 of a Seyfert 1 type nucleus and of a Broad Line Region hidden from our
 line of sight by an obscuring torus. The continuum and broad lines
 which are detected in polarized light, are emitted originally within these hidden
 regions and later scattered towards the observer. The most plausible
 scattering agent is, in this case, electrons, although dust could
 also play a role.

 Since Antonucci and Miller's discovery, the number of Seyfert 2
galaxies that harbor hidden BLRs visible only in polarized flux
spectra has noticeably increased: Mrk 3, Mrk 48, NGC 7212, etc
(e.g. Miller \& Goodrich 1990; Tran 1995). The nature of the
scattering material is not clear, dust and/or electrons? But, at least
for some objects, scattering by dust seems to be the dominant
mechanism (e.g. NGC 7674, Tran 1995). Recently, Inglis et al.(1995)
have mapped the polarization of NGC1068 also in regions out of the
nucleus. In off-nuclear regions, dust is also responsible for the
scattering.

	The low redshift radio galaxy PKS2152-69 (see Fig.~2) shows
extra nuclear scattering light too. It contains a very blue cloud at 8 kpc
from the nucleus which is  aligned with the radio axis and emits
highly polarized flux. Dust (and electrons) in this cloud could be the
scattering agent of the central UV radiation emitted by the nuclear source.

\begin{figure}[htb]
\vspace{2.8in}
\caption[PKS2152-69.]{}{\small A WFPC2 UV-continuum (F300W) image of
PKS2152-69.  The angular separation between the nucleus (lower right)
and the jet-cloud interaction front is about 8 arcsec. The interaction site
emits X-rays (ROSAT HRI), optical/UV emission lines, a blue/UV
continuum as well as radio waves. (Fosbury et al. in prep)}

\end{figure}

Scattering and polarization (maybe by dust) are also detected at very
high redshifts. High redshift radio galaxies (z$>$0.5) present
elongated structures in the UV rest frame, which are aligned with the radio
axis. This extended UV radiation is polarized, with the electric vector
perpendicular to the axis of the optical structures. The observations indicate that the
radiation emitted by a hidden quasar is being scattered by dust and/or
cool electrons in the extended ISM. The discovery in several radio
galaxies (e.g. di Serego et al. 1996; Dey \& Spinrad 1995) of broad polarized
MgII$\lambda$2798 emission line (originally emitted in the Broad Line
Region which is also hidden from our line of sight), supports this
scenario.
 
\subsection{Thermal emission of warm dust}

	  Diffuse emission from interstellar dust was predicted by van
de Hulst (1946) as a consequence of absorption: the  energy
absorbed by dust grains must re-emerge in the infrared. In a typical
interstellar environment, a dust particle gains energy mainly from the
absorption of ultraviolet photons from the ambient interstellar
radiation field. The dust grains will emit in the IR a power equal to
that absorbed. These effects are also clearly seen in active and
starburst galaxies.

	Within the framework of unification theories of AGNs (see Antonucci 
1993 for a
review), it is believed that a dusty torus surrounds the central
accretion disk and the broad line region of active galactic nuclei (AGNs).
The main continuum source and the broad lines would be obscured from
direct view by a torus in narrow line
active galaxies. The dust in the torus should absorb much of the
nuclear radiation and re-emit it in the infrared.  Measurements by
Rieke \& Low (1975) and Stein \& Weedman (1976) showed already in the
70's that many Seyfert galaxies have large infrared excesses.  In
fact, the overall characteristics of the spectra of AGN show a certain
gross uniformity with much of the energy emitted in the IR. This
emission is attributed to warm dust, heated either by the central
AGN or by recently formed stars, or both.  At very high $z$, the
dust thermal emission is redshifted to the far IR and the submm
waveband, depending on $z$. Several high redshift radio galaxies have
been detected with excesses in these spectral ranges, which indicates
the existence of warm dust also at much earlier epochs of the Universe
(e.g. Downes et al. 1992; Isaak et al. 1994).

	The unified models of AGNs propose that the diversity of AGNs
is smaller than believed and that many of the observed differences are
due to orientation effects: the obscuring torus determines a radically
different appearance of the AGN at different viewing angles, AGNs of
different orientation will be assigned to different classes. This
scenario has been tested by studying dust emission from the torus
(Hess 1995): for instance, if radio galaxies and quasars are
intrinsically the same objects, the torus should intercept the same
amount of ultra-violet photons in both types of objects and we would
expect that they emit similar output of reprocessed infrared
emission.  IRAS data indicate that quasars are brighter at 60
$\mu$m than narrow line radio galaxies, which contradicts the unified
models.  This could be explained if the obscuring circumnuclear torus
is still optically thick at 60$\mu$m. Other mechanisms dependent on
orientation could explain the differences if for instance the IR
emission  was not
only due to thermal dust emission but also due to  a non thermal
component. In this case, relativistic beaming of the  core component 
might influence the 60 $\mu$ emission.

\subsection{Spectral absorption features}

 Interstellar absorption features in the spectra of reddened stars and
infrared sources are attributed to solid particles on the basis of
position, width, shape and continuity of profile. The absorption lines
due to gas phase atomic or molecular species in the interstellar
medium are, in general, extremely sharp (FWHM of 1 km s$^{-1}$ or
less), reflecting the conditions of low temperature and pressure in
which they are formed.  Solid state spectral features are
intrinsically broad and continuous and cannot be resolved into
discrete lines, in contrast to the vibration-rotation bands of many
gas phase molecules.  Of the observed absorption features attributed
to interstellar dust, the $\lambda$1275 feature in the mid-ultraviolet
is the strongest and, in frequency units, the broadest, forming a
prominent peak in the interstellar extinction curve.  In principle,
 dust-related spectral features provide a direct means of
identifying the chemical composition of interstellar grains.

Active galaxies also show absorption features due to dust. Mrk 231 is
a Seyfert 1 galaxy with very strong IR emission. Its optical spectrum
is very heavily reddened. Relatively high resolution infrared
measurements by Rieke (1976) show the redshifted interstellar
absorption feature due to silicates at 10 $\mu$m, which means
absorption certainly arises in Mrk 231.

However, the absence of spectral features which are customarily attributed
to dust in our
galaxy should not be interpreted necessarily as indicating  
the absence of dust.  The dust in
external galaxies might not have exactly the same extinction
properties as in our interstellar medium. Even in our own Galaxy, there
are regional variations in the optical properties of the interstellar
dust. Already in 1937, Baade and Minkowsky found that the extinction
curves for stars in the Orion Nebula differ from the ``mean'' galactic
extinction curve. Selective removal of the small particles is the
reason. Environmental influences (grain growth by coagulation, size
dependent destruction, etc) can strongly influence the composition and
size distribution of the dust grains and, therefore, the extinction
properties.  Dust in active galaxies may exist under very different
conditions than in the ISM of our galaxy.

\subsection{Spectral emission features}

 The principal emission signatures attributed to cosmic dust arise in
silicates and hydrocarbons.  The PAH spectrum is a series of narrow
emission features with principal wavelengths at 3.3, 6.2, 7.7, 8.6 and
11.3 $\mu$m. They are now widely attributed to polycyclic aromatic
hydrocarbons, a class of organic molecules composed of benzene
rings. Their infrared spectra are characterized by a number of
resonances, excited by absorption of ultraviolet photons (Duley and
Williams 1981; L\'eger and Puget 1984). The PAH features are commonly
observed in the lines of sight to planetary nebulae, HII regions and
reflection nebulae.

	PAH features are observed in the nuclei of many external
galaxies. They are particularly prominent in systems with nuclear HII
regions and are therefore indicative of recent OB star formation in
which an intense
ultraviolet radiation field very soft beyond the Lyman limit can excite the
different transitions. Indeed, the occurrence of PAH features appears
to be a useful discriminator between starburst gala\-xies (dominated by
rapid star formation) and active galaxies (dominated by compact,
energetic nuclei). PAH emission occurs in the spectra of starburst
galaxies because many luminous HII region are included in a resolution
element.  In contrast, PAH features are generally absent from the
spectra of active galaxies, suggesting that the carriers are destroyed
by hard UV radiation in the vicinity of the nuclear source (Aitken \&
Roche 1985; D\'esert \& Dennnefeld 1988).

\subsection{Depletion of heavy elements}

 The term depletion refers to the
underabundance of gas phase elements with res\-pect to the solar standard as
a result of their presence in the dust grains. Ultraviolet spectroscopy of interstellar gas
indicates a depletion in the abundances of many of the heavy elements with respect to
solar values, which is most readily explained if the missing atoms are tied up in
solid particles. 

	 Ferland (1993) demonstrated the presence of dust grains in
the narrow line region gas. He showed that in the absence of dust the
[CaII] near infrared forbidden lines ($\lambda\lambda$7291,7324) would
be among the strongest lines in the optical to infrared spectrum of
the NLR, assuming cosmic Ca abundance. He interpreted the non
detection of these lines as due to the depletion of calcium in the
NLR, showing that grains must be present, mixed with the ionized gas.

\newpage

\section{The projects developed in this thesis}

\subsection{Effects of dust on the Ly$\alpha$ and Balmer decrements in the 
Narrow Line Region of Seyfert 2 galaxies. {\it (Chapter 2)}} 

	Despite its simple atomic configuration, the hydrogen line
spectrum of AGNs is not well understood.  In 1984, Gaskell and Ferland
showed that the theoretical spectra of many NLR models of Seyfert 2
define a narrow band in the Ly$\alpha$/H$\beta$ {\it vs.}
H$\alpha$/H$\beta$ diagnostic diagram named the intrinsic band. The
displacement inside this band is explained by variations in
metallicity, hardness of the ionizing continuum and electronic
density. However, the observed data locate many Seyfert 2 galaxies
outside the theoretical band, sometimes quite far away from it, being
Ly$\alpha$ much fainter than expected with respect H$\alpha$ and
H$\beta$.  Previous studies (e.g. Ferland \& Osterbrock 1986; Wills et
al. 1993) propose a combination of atomic processes and dust
absorption to explain the observations.

	In this work we propose a new scenario that solve the
discrepancies between the HI observed and predicted ratios of the NLR,
a scenario where dust plays an important role, but also geometry, a
factor which has not been considered before and which strongly influences the
line emission, specially the resonant line Ly$\alpha$.

\subsection{Does dust exist within the ionized gas of the Extended Emission 
Line Regions of Radio galaxies? {\it (Chapters 3 and 4)}}

	The main goal of this project has been to investigate the
origin of the gas which forms the extended emission line regions
(EELRs) and which is observed to extend  up to several hundreds of kpc in powerful
radiogalaxies. In some cases, morphological similarities with
filamentary systems
surrounding galaxies near the centers of rich clusters
suggest that the origin of the EELR consist of gas cooling from the hotter
phase, similar to the X-ray emitting corona which surround many cD
galaxies in clusters.  In other cases, morphological and kinematical
studies of the gas often suggest the existence of a collision or
merger in the recent past.

	The existence or not of dust in such regions has important
consequences for the origin of the gas: if the gas had cooled from a
hotter X-ray phase, any dust introduced in the intracluster medium
would have been sputtered and rapidly destroyed. In the cooling process there
would have been not enough time for the dust to form. On the other
hand, if the gas consists of galactic debris, the gas/dust ratio is expected to
have a value appropriate to the chemical composition of a normal
galaxy.

	We have investigated the existence of dust in the EELRs of a
sample of low $z$ radio galaxies in order to discriminate between
these two main theories relative to the origin of the gas.
	
\subsection {Effects of dust and resonant scattering on the UV spectrum of high
$z$ radio galaxies. {\it (Chapter 5)}}

	Radio galaxies at very high $z$ ($z>$2) have been until very recently  the youngest
galaxies we know of and, therefore, good laboratories to study 
galaxies at early epochs.  In order to determine correctly the epoch
when they were formed, we need
to establish the age of the oldest stars in such systems. However, we don't have yet a
clear picture of the processes which are going on in these special
objects. The presence of an active (very powerful) nucleus influences
strongly the ISM of the galaxy and it is difficult to separate the
different components (including stars) which contribute to the observed
radiation.  We need, therefore, a clear understanding of the physical
processes involved in the formation of the various lines and of the continuum
in order to be able to disentangle the stellar from the AGN-related sources.

	In our work we interpret the UV (rest-frame) emission line
spectrum of a sample of high redshift radio galaxies. Due to the short
wavelengths and the presence of several resonant lines, this spectral
range is very sensitive to geometry and dust. Comparing our
photoionization models with the observed line ratios, we study in
detail the effects that both, dust and geometry have on the UV line
ratios. Our study provides results about the presence of dust and the
3--dimensional structure of these galaxies, which can be very useful for
understanding the nature of such extreme objects.  

\begin{figure}[htb]
\vspace{3in} 
\caption[Composite High $z$ Radio Galaxy
spectrum]{}{\small A composite radio galaxy spectrum in the UV
rest-frame constructed from observations of 0.1$<z<$3 galaxies (from
McCarthy 1993).}
\end{figure} 
\vspace{0.5cm} 
\newpage
\addcontentsline{toc}{part}{Bibliography} 
\parskip=1ex
{\Large \bf References}

\vspace{0.5cm}

\noindent Aitken D.K., Roche P.F., 1985, \mnras 213, 777

\noindent Antonucci R.R.J., Miller J.S., 1985, \apj 297, 621

\noindent  Antonucci R.R.J., 1993, {\it ARA\&A} 31, 473

\noindent Baade W., Minkowsky R., 1937, \apj 86, 123

\noindent Baum S., de Koff S., Sparks W., Miley G., Biretta J., Golombek D., Macchetto D., McCarthy P., poster contribution for the meeting ``Cold Gas at high z'', Hoogeveen, August 1995 (in press)

\noindent Binette L., Wang J., Villar-Mart\'\i n M., Martin P.G., Magris C.M., 1993, \apj 414, 535 

\noindent Bland J., Taylor K. \& Atherton P.D., 1987, \mnras 228, 595

\noindent Cimatti A., di Serego Alighieri S., Fosbury
R.A.E.,
Salvati M., Taylor D., 1993, \mnras 264, 421

\noindent Cimatti A., Freudling W., 1995, {\it {\it A\&A}}, 300, 366

\noindent Dahari O., De Robertis M.M., 1988, \apj 331, 727

\noindent de Robertis M.M., Osterbrock D.E., 1984, \apj 286, 171

\noindent D\'esert F.X., Dennefeld M., 1988, {\it A\&A}, 206, 227

\noindent Dey A., Spinrad H., 1995, \apj in press

\noindent di Serego Alighieri S., Cimatti A., Fosbury
R.A.E., 1994, \apj 431, 123

\noindent di Serego Alighieri S., Cimatti A., Fosbury
R.A.E., P\'erez-Fournon, 1996, \mnras submitted

\noindent Downes D., Radford S.J.E., Greve A., Thum C., Solomon P.M., Wink J.E., 1992, \apj 398, L25

\noindent Duley W.W., Williams D.A., 1981, \mnras 196, 269

\noindent Ferland G.J., Osterbrock D.E., 1986, \apj 300, 658

\noindent Ferland G.J., 1993, in Proc. Madrid Meeting on The Nearest Active Galaxies,
ed. J.E. Beckman, H. Netzer \& L. Colina, p. 75

\noindent Hall J.A., 1949, Science, 109, 166

\noindent Heckman T.M., Miley G.K., Breugel W.J.M., 1981, \apj 247, 403

\noindent Hess R., 1995, Ph.D. Thesis, University of Groningen

\noindent Hiltner W.A., 1949, Science, 109, 165

\noindent Inglis M.D., Young S., Hough J.H., Gledhill T., Axon D., Bailey J.A., Ward M.J., 1995, \mnras 275, 398

\noindent  Isaak K.G., McMahon R.G., Hills R.E., Withington S., 1994, \mnras 269, L28

\noindent L\'eger A., Puget J.L., 1984, {\it A\&A}, 137, L5

\noindent McCarthy P.J., 1993, {\it ARA\&A} 31, 639

\noindent Miller J.S., Goodrich B.F., 1990, \apj 335, 456

\noindent Morganti R., Robinson A., Fosbury R.A.E., di Serego Alighieri S., Tadhunter C.N., Mailn D.F., 1991, \mnras 249, 91

\noindent Rieke G.H., Low F.J., 1975, \apj 200, L67

\noindent Rieke G.H., 1976, \apj 210, 5

\noindent Schneider D.P., Schmidt M., Gunn J.E., 1991, \aj 102, 837

\noindent Stein W.A., Weedman D.W., 1976, \apj 205

\noindent Tadhunter C.N., Fosbury R.A.E., di Serego
Alighieri S.,
1988, in Maraschi L., Maccacaro T. \& Ulrich M.H., eds., ``BL Lac Objects'',
Springer-Verlag, Berlin, p.79

\noindent Tran H.D., 1995, \apj 440, 565 (I, II \& III)

\noindent van de Hulst H.C., 1946, {\it Rech. Astron. Obs. Utrecht}, 11, 1

\noindent Wills B.J.,  Netzer H., Brotherson M.S., Han M., Wills D., Baldwin J.A., Ferland G.J., Browne I.W.A., 1993, \apj 410, 534

\chapter{Effects of internal dust on the Narrow-Line region Lyman and Balmer
decrements.}
\centerline{Binette, Wang, Villar-Mart\'\i n, Martin \& Magris 1993, ApJ, 414, 535}
\vspace{0.3cm}
 \pagestyle{myheadings}
 
\markright{Internal dust in the NLR of Seyfert2's}

\vspace{0.4cm}

{\LARGE  \bf Abstract}
 
\vspace{0.4cm}

	We present detailed calculations on the effects of internal dust on the Balmer
and Lyman decrements for a spherically symmetric distribution of low covering factor
clouds represented as slabs and consider the effects of Ly$\alpha$ resonance and
absorption by dust. We consider the important effects of perspective on the emergent
fluxes, which in our simplified scheme present either the photoionized face to the
observer (``f'') or the back (``b'') face. We adopt canonical values for the gas excitation
($U_f$) and for the ionizing energy distribution ($\alpha=-$1.4, F$_{\nu} \sim
\nu^{+\alpha}$) and compute sequences of photoionization models in which the
relative internal dust content by mass ($\mu$) is progressively increased to values
comparable to the local ISM (i.e., $\mu$=1). 
 We find that for moderate amounts of internal dust $\mu$=0.2-0.3,
radiation-bounded clouds result in Lyman and Balmer decrements in the range 32-37 and
3.0-3.1, respectively.  Our main result,
valid for the calculations with standard NLR input parameters and for an open geometry in
which multiple cloud covering is negligible, is that even with internal dust, while
H$\alpha$/H$\beta$ is reddened, Ly$\alpha$/H$\beta$ turns out not very different from
recombination case B. The Seyfert 2 observations show much lower Lyman decrements than
predicted from the spherically symmetric model and a critical study  is made of
various explanations: dust -or line scattering screens, a semiopen geometry can span
the region covered by the Seyfert 2 observations. We also study the possibility that
dust reddening of the {\it conti\-nuum} may account for the apparent deficit of ionizing
photons seen in many Seyfert 2's. This reddening may be present in addition to, or in
place of, the anisotropic beaming/occultation of ionized radiation (cf.``occultation/reflection picture'') that is generally invoked to explain the deficit.

\section{Introduction}

	The calculation of the intrinsic Balmer decrement emitted by the narrow-line
region (NLR) of Seyferts has attracted much interest in the past (e.g., Gaskell 1984;
Malkan 1983; Halpern \& Steiner 1983; and Ferland \& Netzer 1983) because it is the
most direct means of determining how much (intervening) extinction the observed
emission-line spectrum has undergone. This information can be used to deredden the
whole line spectrum for the effect of external dust along the line of sight. It has
been recognized early on by Halpern \& Steiner (1983) and Ferland \& Netzer (1983)
that the NLR intrinsic Balmer decrement was likely to be higher than that of
recombination case B as a result of collisional excitation within the partially
ionized zone. The resolution of the problem of intrinsic HI line ratios was further
advanced by combining the information provided by the Lyman decrement with that of the
Balmer decrement (e.g., Ferland \& Osterbrock 1985). Gaskell \& Ferland (1984) studied
in details how the NLR metallicity, density, and hardness of the ionizing continuum
affected both decrements. The most probable range for the NLR decrements which these
authors favor is H$\alpha$/H$\beta$=2.8-3.1 and Ly$\alpha$/H$\beta$=30-50 (similar to
values determined earlier by Ferland \& Osterbrock 1985). The observational
determination of the intrinsic HI line ratios is facilitated in Seyfert 2 galaxies since
one need not decompose the line profiles into BLR and NLR components. However, Seyfert
2's are on average significantly more reddened than Seyfert 1's (Gaskell 1984). De
Zotti \& Gaskell (1985) concluded that the dust was probably associated with the NLR
clouds or filaments owing to the extremely weak correlation of the extinction with
disk inclination angle and also due to the significantly lower extinction found in
Seyfert 1.5.

	If there is so much dust associated with the nuclear emission region, we might
expect some dust to be associated with the photoionized gas itself. The effect of
internal dust mixed with ionized gas for physical conditions pertaining to the
broad-line region (BLR) has been investigated by Ferland \& Netzer (1979). More
recently, however, Netzer (1993) concluded that the BLR was devoid of dust (but see Crosas \& Weisheit 1993). From considering
the ratios H$\alpha$/H$\beta$ {\it vs.} H$\gamma$/H$\beta$ in Seyfert 2's, Binette et
al. (1990) suggested the possibility of an intrinsic (NLR) Balmer decrement as high as
H$\alpha$/H$\beta \approx$ 3.4, which they modeled using photoionization calculations
with dust internally mixed with the ionized gas. 

	Internal dust results not only in somewhat higher Balmer decrements but can
furthermore increase the equilibrium temperature of the ionized gas. This property was
used by Magris, Binette, \& Martin (1993) to solve the problem of the high
electronic temperature (based on [OIII] lines) observed in the extended
ionized nebulosities of powerful radio galaxies (cf. Tadhunter, Robinson \&
Morganti 1989). Furthermore dust can reflect towards the observer a nonnegligible
fraction of the impinging optical radiation, an interesting property which appears
to explain quite well the extended blue polarized continuum observed in distant
radio galaxies (cf. Fosbury 1993; Cimatti et al.
1993; Binette et al. 1993a). Dust may also explain the absence of any intermediate
region between the BLR and the NLR according to Netzer \& Laor (1993), not to mention
its relation with the infrared emission (Clavel, Wamsteker, \& Glass 1989; Barvainis
1992; Sanders et al. 1989; Pier \& Krolik 1992).

	In this work , we expand upon the calculations of Binette et al. (1990) and
study the effects of internal dust on the intrinsic H$\alpha$/H$\beta$ by taking into
account the observer's perspective of the emitting clouds. In addition, we consider
the effects of dust on the Lyman decrement. Kwan \& Krolik (1981) and Puetter \&
Hubbard (1987) have shown using dust-free BLR calculations how significant is the
effect of perspective on resonant Ly$\alpha$. With dust mixed in, it becomes essential
to take into account satisfactorily the cloud's perspective and the global NLR
geometry when deriving the intrinsic decrements, as described in
$\S$2.2. We first investigate in $\S$2.3 the magnitude of the Ly$\alpha$
destruction due to internal dust in photoionization calculations and compare in
$\S$2.4 our calculations with the Seyfert 2 data of Kinney et al. (1991) (hereafter KAW3).

\section{Computational method}

We describe the photoionization code and the line transfer of Ly$\alpha$, H$\alpha$,
and H$\beta$ across the dusty medium and review the input parameters that enter our
calculations.

\subsection{The photoionization code}

	To compute HI lines we have employed the multipurpose photoionization-shock
code MAPPINGS (cf. Binette, Dopita, \& Tuothy 1985). To compute the hydrogen lines, we
treat the hydrogen atom as a six-level system (1$s$, 2$s$, 2$p$, 3, 4 and 5) plus
continuum. The hydrogen levels are populated through direct recombination (followed by
cascade) as well as by collisional excitation either by thermal electrons (using
Johnson 1972 rate coefficients) or by suprathermal electrons (adopting the method
developed by Shull \& Van Steenberg 1985) as earlier described in Binette et al.
(1993b, hereafter BWZM).

	In the computation of the ionization structure, we adopt a
simple-plane-parallel geometry whereby our putative thick gas cloud is represented by
a radiation-bounded slab of gas containing interstellar dust. The ionized slab is
subdivided into many small layers in which the equations of photoionization and
thermal equilibrium are solved by standard methods.

	The escape probability formalism is used to solve for the transfer of the
resonance lines. The effects of dust on the ionization structure as well as on the
thermal balance of the plasma are considered in detail by implementing the physical
processes relevant to dust mixed with emission plasma as described in Appendix C of
Baldwin et al. (1991). For instance, heating of the plasma by dust photoelectric
emission is considered. The dust grain charge is calculated self-consistently and the
formula describing the photoelectron energy distribution and the yield are from Draine
(1978), but with a cap of 0.2 for the yield at high photon energies as in Baldwin et
al. (1991).

	One interesting aspect of the new code MAPPINGS is that the effect of dust
scattering on the line transfer is explicitly solved using the numerical solution of
Bruzual, Magris \& Calvet (1988). This aspect of the transfer is particularly relevant
to the ``open geometry'' adopted for most of the calculations presented in this thesis.
Destruction of the resonant line by dust absorption is taken into account using the
results of  Hummer \& Kunasz (1980). We describe in more details in Appendix A of BWZM
the method implemented in MAPPINGS for solving the line transfer of emission plasma
with internally mixed dust.

\subsubsection{Trace elements and dust content}

	Even though this work only addresses the intensities of hydrogen lines, the
abundance of trace elements remains an important factor since they affect directly
the thermal balance of the photoionized plasma. Abundance gradients observed in spiral
galaxies point towards a metallicity even higher than solar for the nuclear region.
Many models of the NLR on the other hand have favored more often than not the
exploration of abundances lower than or equal to solar (e.g., Gaskell \& Ferland 1984;
Ferland \& Netzer 1983). As this work emphasizes the role of internal dust,
for definiteness we will assume that the abundances of metals appropriate to the NLR
are close to solar and adopt the solar system abundances of Anders \& Grevesse (1989)
(see list in Table 4 of Appendix A). We define Z as the total (gas+dust phases)
metallicity relative to solar which implies Z$\equiv$1 for this work.

	The absorption and scattering cross sections of dust (from the infrared to the
soft X-rays) were computed by Peter Martin (the curves are displayed in Fig.6 of
Martin \& Rouleau 1991) and corresponds to a dust models of the extinction curve of
the solar neighborhood interstellar medium. In MAPPINGS, we scale the scattering and
absorption cross sections (and, therefore, the dust content) by a dimensionless factor
$\mu$. The grain model of Martin \& Rouleau (1991) results in a dust-to-gas {\it mass}
ratio of 0.0064 $\mu ~(\rho_{dust}/\rho_{gas} \cong 0.71 \rho_{dust}/\rho_H)$ and to an
extinction opacity at 5500 \AA\,
$\tau_V(=$0.921A$_V$) = 4.8 $\times$ 10$^{-22} \mu N_H$
cm$^2$, where $N_H$(cm$^{-2}$) is the {\it total} hydrogen column density. When $\mu
\equiv$ 1, these quantities are all consistent with the standard extinction curve of
the local ISM (interstellar medium).

	To be consistent with the depletion of metals unto dust grains, we deplete the
abundances of metals in the {\it gas} phase, Z$_{gas}$, according to the dust content
$\mu$. The adopted scheme described in Appendix A uses the depletion indices listed in
Whittet (1992) and if necessary scales them with $\mu$/Z. Renormalization of the
derived gas phase abundances is performed until the mass locked into dust grains
becomes consistent with the dust-to-gas mass ratio implied by $\mu$. For instance,
when $\mu$=1, Z$_{gas}=$ 0.6 as a result of depletion. In this work, the {\it total}
metal abundances are always solar and, therefore, $Z=Z_{gas} + Z_{dust} \equiv 1$

\subsubsection{The slab geometry and the input parameters}

	All our calculations assume the same standard values for the input parameters,
that is: an isobaric density behaviour, a power law ionizing continuum of index
$\alpha = -$1.4 (F$_\nu \propto \nu^{+\alpha}$) and solar {\it total} abundances (but
depleted gas abundances: $Z_{gas}=Z-Z_{dust}=1-Z_{dust}$). Our constant pressure models
are parameterized in terms of the so-called ionization parameter, that is, the ratio of
the density of ionizing photons impinging on the slab to the density of the outermost
gas layer of the slab:

	$$ U_f = \frac{1}{cn_H^f} \int {\frac{\phi^f_{\nu}}{h\nu} d\nu} = 
\frac{\phi^f_{\nu}}{cn_H^f} ~~~ [1]$$

where $c$ is the speed of light and $\nu_0$ is the Lyman limit frequency,
$\phi^f_{\nu}$ is the ionizing photon flux (in photons cm$^{-2}$ s$^{-1}$) impinging
on the slab (quantities with superscripts $f$ or $b$ refer to their values at the
front or at the back of the slab, respectively). For the reference dust-free model
({\it filled star} in the figures) which is referred below, we adopted $n_H^f$=5000
cm$^{-3}$ and $U_f$=0.0015 as justified in $\S\S$ 2.3.1 and 2.3.3.

	Using the code MAPPINGS, the calculation of the ionization and emissivity
structure is carried inward in the slab either up to a column depth (i.e., $\int{N_H
dx}$) $N_H^{slab} < N_{H^*}$ in the case of matter-bounded calculations or, in the
case of radiation-bounded calculations, up to $N_{H^*}$. The column density $N_{H^*}$
gives the depth of the complete ``photoexcited'' region, that is, the depth at which the
incoming ionizing flux is exhausted. The operative definition of the boundary of the
photoexcited region (at depth $N_{H^*}$) is defined as the depth where the following
two conditions are simultaneously satisfied: (1) the inabsorbed ionizing flux 
$\phi^f_{\nu} < 1$ \%, of the impinging flux $\phi^f_{\nu}$ , and (2) the ionized fraction $n^b_{H^+}/n^b_H \leq 1$\%. The code
computes successively the emergent flux seen directly from the phtoionized face
(perspective ``f''; see Fig.4) as well as seen from the back (perspective ``b''). The
transfer solution used across the dusty medium not only considers absorption but dust
scattering as well (see BWZM).

\begin{figure}
\vspace{2.5in}
\caption[]{}{\small Adopted slab geometry for the constant pressure photoionization
calculations. The slab comprises a fully ionized zone  and a partially ionized zone
(PIZ) which adds up to the column density of the photoexcited zone $N_{H*}$.  Beyond the photoexcited regions a zone of neutral gas may exist of
column density $N_{H^0}^0$. Perspective ``f'' considers the emergent line fluxes seen
from the side which is photoionized (front), while perspective ``b'' corresponds to the
opposite back side. The dust extinction opacity of any zone is proportional to its
relative dust content $\mu$ and total hydrogen column density $N_H$.}

\end{figure}

	Because the harder photons of the power-law ionizing continuum (with
$\alpha=-$1.4) creates quite a large partially ionized region (hereafter PIZ) in which
the ionized gas very gradually becomes neutral, $N_{H^*}$ is a factor $\sim$ 10 higher
than the Str\"omgren depth (in cm$^{-2}$) which is technically defined as

\vspace{0.6cm}

\centerline {	$ N^S_{H^+} = c U_f/\alpha_B \approx 10^{23} U_f $ cm$^{-2}$ (T$_e$ = 10$^4$ K) ~~~ [2]}

\vspace{0.6cm}

where $\alpha_B$ is the recombination coefficient to excited states of hydrogen. In an
HII region, because of the softness of the ionizing continuum, the PIZ is negligible
and $N_{H^*} \simeq N^S_{H^+}$. For radiation-bounded clouds, equation (2) entails that
the fully ionized zone [$N_{H^+}(n_{H^+}/n_{H^0} \geq 10) \approx 0.8 N^S_{H^+}$]
which contributes most of the line luminosity nevertheless contributes only a small
fraction of the dust opacity inside $N_{H^*}$ and, therefore, the spectrum seen from
perspective ``b'' is more reddened than the one emerging from perspective ``f''.

By portraying the emitting NLR cloud as a slab, a determinant parameter with which the
dust opacity scales is the total column density $N_{H^+}^{slab} (=N_{H^0}^{slab} 
+ N_H^{slab})$ of the slab. In this work, in most instances, we assume the slab to be
not only radiation-bounded but in some cases to exceed $N_{H^*}$ due to a neutral 
$H^0$ zone of column density $N^0_{H^0}$ (beyond the PIZ) as depicted in Figure 4. The
dust inside this region can play a determinant role in reducing the emergent line
fluxes seen from perspective ``b''. Apart from absorption of the lines by dust, the
neutral zone also is assumed isothermal with a temperature taken to be equal to the
temperature at the last computed ionized layer. Changing this temperature even
arbitrarily has practically no effect on the zone's reflectivity to Ly$\alpha$ since
even without such a zone the already huge Ly$\alpha$ line opacity within the PIZ
would cause most of the Ly$\alpha$ to emerge from perspective ``f'' rather than from ``b'' (cf.2.3.2).

	In our nomenclature, the total column density of the slab is given by 
$N^{slab}_H = N_{H^*} + N_{H^0}^0$ (see Fig.4) if the slab is radiation-bounded
(otherwise, $N^{slab}_H < N_{H^*}$ if the slab is matter-bounded). We assume that
dust-to-gas ratio $\mu$ is uniform within each zone but may take different values in
the photoexcited zone from that in the neutral zone. Assuming that the same dust size
distribution and composition applies to the different zones (that is, the shape of the
extinction curve remains constant), we obtain that the total slab opacity in the V
band (5500 \AA\ ) is given by

$$ \tau^{slab}_V = \tau_V(N_{H^*}) + \tau_V(N_{H^0}^0) =$$

$$= 4.8 \times 10^{-22} [\mu(H^*)N_{H^*}) + \mu(H^0)N^0_{H^0}]~~~ [3]$$.

Due to the high reflectivity of Ly$\alpha$ by the H$^0$ zone as well as by substantial
amount of H$^0$ inside the transition zone, the emergent Ly$\alpha$ originating from
perspective ``b'' is generally negligible. The Balmer lines are generated isotropically
but are absorbed differently when emerging from pers\-pective ``b'' or ``f'' because of the
asymmetry in dust opacity introduced by the $N_{H^0}^0$ zone and the PIZ. If we
picture the NLR as an ensemble of clouds with very low covering factor symmetrically
distributed around a nuclear ionizing source, the {\it number} of cloud seen from the
back can be postulated to be equal statistically to that of clouds seen from the
front. To a first order the resulting line ratios for such an ``open'' geometry is
obtained by adding first with equal weight the line fluxed from the ``f'' and ``b''
perspectives and then taking the ratios of the summed fluxes.

	In $\S$ 2.3, we will first explore the effect of varying the internal
dust-to-gas ratio $\mu$ on the Ly$\alpha$/H$\beta$ and H$\alpha$/H$\beta$ line ratios
emerging from a symmetric distribution of clouds (i.e., the case represented by
``f''+``b''). After comparing models ($\S$2.4.1) with Seyfert 2 data of Kinnety et al. (1991) and after discussing the pure dust hypothesis ($\S$2.4.2), we will look
into the possibility of having more clouds seen from perspective ``b'' as well as the
limiting case of a spherically closed geometry ($\S$2.4.3).

\section{Results}

	Our main purpose is to explore the effect of internal dust on the HI line ratios
in the context of the standard photoionization model of the NLR. For this reason, we
have limited the scope of the calculations to restricted plausible values of input
parameters; we consider, for instance only solar abundances (but with progressive
depletion of gas phase metallicity $Z_{gas}$ with increasing dust content $\mu$) and
canonical $\alpha=-$1.4 ionizing power law \footnote{A much harder distribution (flatter power law) would give a rather poor fit to the low excitation lines assuming the photoionized clouds are radiation bounded.} (cf. Ferland \& Osterbrock 1986;
Kinney et al. 1991). The choice and role of the gas density is first discussed
($\S$2.3.1) as well as that of incomplete opacity to ionizing photons ($\S$
2.3.2) and the selection of the ionization parameter ($\S$2.3.3). We then discuss the
results of nebular models with internal dust $\S$2.3.4.

\subsection{Selection of a characteristic cloud density}

	The choice of a representative cloud density in the calculations is much more
problematic than the selection of other parameters. There is evidence of a wide
distribution of cloud densities in the NLR following the work of DeRobertis \&
Osterbrock (1984,1986) who have shown in many Seyferts the tendency of the forbidden
line profile widths to correlate with the respective deexcitation critical density of
the transition involved. This lends support to the picture of a wide gradient in cloud
densities (10$^3 \leq n_e \leq 10^6$ cm$^{-3}$) as well as in velocity dispersion as a
function of distance from the nucleus. In this picture, the bulk of a line luminosity
would be contributed by clouds of density approaching the line's critical density.
The few available direct measurement of the NLR electron densities have been mostly
provided by the ratio of the [SII]$\lambda$$\lambda$6716,6731 density-sensitive
doublet (Osterbrock 1989) which show densities which are relatively low (200-2000
cm$^{-3}$). These low densities do not necessarily contradict a wide range of cloud
densities within the NLR because the critical density of [SII] red lines is relatively
low and moreover the S$^+$ zone coincides with the partially ionized zone of the
emitting cloud where the electron density can be expected to be lower than in the fully
ionized zone. This can be demonstrated with constant pressure photoionization
calculations with densities covering the range 80$ < n_H^f < 6000$ cm$^{-3}$ which show
that the electronic density inferred from the [SII] line ratio is a factor 4-5 smaller
than the mean $< n_e >$ of the fully ionized zone. Finally, there also exists a
significant fraction of Seyferts in which line widths correlate with line excitation
(cf. Pelat, Alloin, \& Fosbury 1981; Filippenko 1985; DeRobertis \& Osterbrock
1984,1986) which could indicate that the density range within these objects is
narrower.

	Following the indications given by the [SII] densities and given that we will
not integrate the line contribution of clouds over different densities, we opt for a
single density of $n_H^f$=5000 cm$^{-3}$ which we expect to characterize the NLR gas
that emits the bulk of the HI lines. Before discussing models with dust we first
summarize the dependence of HI lines on density.

	The direct dependence of H$\alpha$ and H$\beta$ intensities on the gas density
is very weak when the line is produced by recombination alone (Hummer \& Storey 1987).
However, in a power law photoionized plasma, both lines become indirectly coupled to
density when substantial collisional excitation takes place within the PIZ. The reason
is that increasing the density reduces the effectiveness of cooling of many important
forbidden lines as a result of collisional deexcitation, thereby increasing the
equilibrium plasma temperature which in turn increases the rate of collisional
excitation of HI in the PIZ. In the fully ionized zone, there is too little HI to
excite and the Balmer lines are produced only by recombination independent of the
plasma temperature. In the case of Ly$\alpha$, the dependence of the line intensity on
the electron density is much more direct. When the density is increased much above
10$^3$ cm$^{-3}$, there is an increasing conversion of the level 2s population
otherwise responsible for the 2$q$ continuum emission into Ly$\alpha$ emission by
excitation into the 2$p$ level (see Osterbrock 1989; Gaskell \& Ferland 1984). In the
low density regime case B in the absence of the Ly$\alpha$ sinks (e.g., dust
absorption), the fraction of energy emitted as Ly$\alpha$ versus 2$q$ emission from
recombination alone is $\alpha_{2p}^{eff}/\alpha_B \approx 0.67 (T_e/10^4K)^{-0.054}$
 and tends toward unity for densities above 10$^5$  cm$^{-3}$.

\begin{figure}
\vspace{5in}
\caption[]{}{\small Sequences of dust-free solar metallicity photoionization
calculations with varying gas density $n_H^f$. Results for two different energy
distributions  of the ionizing continuum are represented: the power law with $\alpha
=-$1.4 ({\it solid line}) adopted in this work and a black body of 40,000 K ({\it
dotted line}) for which recombination is the only process generating HI lines. The
ionization parameter $U_f$ is constant along both sequences and takes on the values of
0.0015 and 0.010 for the power law and for the blackbody sequences, respectively. Panel
$(a)$ shows the Ly$\alpha$/H$\beta$ ratio (``f''+``b'') as a function of the mean electron
density $<n_e>$ characterizing the isobaric photoexcited zone, while panel $(b)$ shows
the corresponding behavior of H$\alpha$/H$\beta$. The star represents the standard
model ($U_f$=0.0015, $n_H^f$= 5000 cm$^{-3}$) discussed in $\S$2.3.}

\end{figure}

	To illustrate quantitatively the effect of density, we have calculated a
sequence of constant pressure models covering the range 50$< n_H^f < 2 \times 10^6$ 
cm$^{-3}$. The results of integrated nebulae calculations as function of the mean
slab electron density are presented in Figure 5. The case of a pure
recombination-dominated HII region model ({\it dotted line}) is also shown for
comparison with the power-law model ({\it solid line}). In Figure 5$a$, it is evident
that both ionizing energy distributions result in a qualitatively similar functional
dependence of Ly$\alpha$/H$\beta$ on $< n_e >$, except that Ly$\alpha$/H$\beta$ is
consistently higher for the power law due to collisional excitation within the PIZ.
For the Balmer decrement (Fig. 5$b$), at low densities ($< n_e > \ll 5 \times 10^4$
cm$^{-3}$), however there is a sharp increase in H$\alpha$/H$\beta$ owing to the
enhancement of collisional excitation that occurs when the gas temperature increases due
to gradual suppression of the forbidden lines.

	Subsequent models with internal dust have all been calculated with the single
representative density of n$^f_H$= 5000 cm$^{-3}$. This reference density coupled with
the adopted $U_f$ introduced below provides us with a comparative zero point (i.e.,
dust-free case) in our Ly$\alpha$/H$\beta$ versus H$\alpha$/H$\beta$ diagrams 
(as represented by the filled star). If one is interested in estimating the line
ratios different zero point in density space, Figure 5 can be used to determine very
approximately the differential change in the ratios due to density alone and apply it
to the models which contain dust.

\subsection{Asymmetric emissivity and internal dust absorption}

	Because line scattering increases the path length of Ly$\alpha$ photons, 
the efficiency of internal dust to absorb line photons is drastically higher for the
optically thick resonant Ly$\alpha$ line than for the optically thin Balmer lines.
Hummer \& Kunasz (1980) have derived scaling laws for the energy loss, the mean path
length, and the mean number of scattering suffered by resonant line photons in the
presence of continuous absorption (by dust and/or photoelectric absorption). (See
Neufeld 1990 for a discussion of these scaling laws using analytical methods). The
fraction of energy from a resonant line generated within a midplane layer which
escapes unabsorbed from a slab of total line opacity $\Gamma$ depends essentially on
the parameter $\beta \equiv k_c/k_L$ which is the ratio of continuous opacity to line
opacity. The results of Hummer $\&$ Kunasz (1980) concerning the transfer as well as
the interplay between escape probability and dust absorption have been incorporated
in MAPPINGS as described in detail in Appendix of BWZM.

	One aspect of the transfer particularly relevant to thick nebulae is the
asymmetry which characterized the Ly$\alpha$ flux emerging from the front or from the
back of the ionized slab. This is taken into account in the code through the definition
of an emissivity whose asymmetry depends on the local two-sided escape probability.
The asymmetry factor $A^f(0\leq A^f\leq 2)$ in our transfer solution is defined as
follows

	$$ A^f= 2 \epsilon(\tau,\beta)/[\epsilon(\tau,\beta) +
\epsilon(\Gamma-\tau,\beta')] ~~~ [4]$$

where $\tau$ is line opacity of the layer under consideration, $\Gamma(=\int{d\tau})$ is
the line opacity of the whole slab, $\epsilon(\tau,\beta)$ is the escape probability
from the back. (See Appendix A in BWMZ for further details on the escape asymmetry).
The implication of the two-escape scheme is that the emissivity of the resonant line is
asymmetric with the forward emissivity given by $j_L^b = j_L(2-A^f)$. The observer, in
this simple two-stream transfer scheme, is placed either in front of the slab or
behind it and the emergent line flux from either perspective ``f'' or ``b'' is obtained by
integrating the appropriate emissivity ($j_L^f$ or $j_L^b$).

	An important consequence of using equation (4) in integrating the line
emissivity is that for a slab illuminated on only one side by the ionizing radiation,
the Ly$\alpha$ intensity emerging from perspective ``f'' is much higher than that from
``b'' whenever the line is reasonably thick ($\tau > 10$), a condition almost always
satisfied except for extremely low column density matter-bounded slabs. For a slab of
total line opacity $\Gamma$, the depth $N_H$ at which line photons have equal
probability of escaping from the front (``f'') as from the back (``b'') occurs at
the depth $N_H(\tau=\Gamma /2)$, i.e., $N_H=N_H(\tau=\Gamma /2)$. Since
$d\tau/dN_H \propto n_{H^0}/n_H$ is a monotonically increasing function of $N_H$, 
the depth $N_H(\tau = \Gamma/2$) is quite generally located much deeper than the depth that
divides equal Ly$\alpha$ production. As a result, Ly$\alpha$ photons preferentially
escape from the front (``f''). This is true even for soft impinging ionizing continua as
occurs, for instance, in HII regions. For a power-law ionizing continuum, to the
extent that recombination still dominates Ly$\alpha$ production, most of Ly$\alpha$ is produced within the fully
ionized zone, i.e., at $N_H \leq 0.8 N_{H^+}^S$ which is much less than $N_H(\tau=\Gamma /2) \sim N_H{H^*}$ owing to the extended PIZ (cf. $\S$ 2.2.3). In the opposite case of the extremely low column density nebulae, the neutral fraction $n_{H^o}/n_H \approx$ constant so that $N_H(\tau=\Gamma /2) \approx N^{slab}_H$/2, which coincides approximately with the
depth dividing equal Ly$\alpha$ production. In this case, the Ly$\alpha$ flux emerges
symmetrically from the front (``f'') and the back (``b'').

	To illustrate how one progressively goes from symmetrically emerging
Ly$\alpha$ in the very thin slab case to the opposite asymmetric situation in the
radiation-bounded slab case, we have computed a sequence of matter-bounded models
 in which $N^{slab}_H$ is gradually increased in lock steps. Figure 6 gives the
fraction of Ly$\alpha$ which escapes from perspective ``b'' as a function of the
matter bounded slab's
thickness (expressed in a different way in each panel). Note that internal dust has
been included in these models, but it does not affect significantly the degree of
asymmetry except when $N_H^{slab}$ approaches $N_{H^*}$, the radiation-bounded case.
Two of the sequences correspond to $U_f$=0.015 with $\mu$=1.0. The results indicate
that the fraction of Ly$\alpha$ which escapes from the back is close to half only when
the matter-bounded slab is indeed very thin (i.e., for $\tau_{H^0}^c(912\AA\ ) <$ 10
or  $N_H^{slab}/N_{H^+}^S < 0.5-0.8)$.

\begin{figure}
\vspace{4in}
\caption[]{}{\small Matter-bounded calculations in which the ordinate represents the
fraction of Ly$\alpha$ flux which escapes from the back (perspective ``b'') as a
function of different quantities which measures the ionized slab's thickness. The
abscissae represent, panel by panel, the following: panel $(a)$: the fraction
of unabsorbed ionizing photons which leaks out from the back $\phi^b_H/\phi_H^f$;
panel $(b)$: the slab's column density to that of the Str\"omgren depth $N_{H^+}^S =
cU_f/\bar{\alpha_B}$, where $\bar{\alpha_B}$ is evaluated at the slab {\it mean} electronic
temperature; panel $(c)$: the hydrogen ionized fraction within the last layer at the
back; panel $(d)$: the slab Lyman-limit continuum opacity. The solid and short dash
curves were both calculated using the reference value of $U_f$ = 0.0015 (and $n_H^f$ =
5000 cm$^{-3}$; see $\S \S$ 2.3.2 and 2.3.3) but with an internal dust content of
$\mu$ = 1.0 and 0.2 respectively. The dotted line corresponds to calculations with
$U_f$=0.015 and $\mu$=1.0}

\end{figure}

	The effect of perspective need not be considered in previous work on the
intrinsic NLR Ly$\alpha$/H$\beta$ and H$\alpha$/H$\beta$ line ratios since it was
implicitly assumed that perspective effects averaged out due to the geometry of the
ensemble of NLR clouds. This might be a valid assumption in the dust-free case where
no Ly$\alpha$ destruction takes place despite the high number of scatterings. When dust
is present, however, the effect of perspective cannot be ignored since the Ly$\alpha$
transfer with internal dust present is nonconservative with respect to perspective with
most of the surviving Ly$\alpha$ photons escaping from the front.

	Perspective can also affect significantly the integrated Balmer lines of a
system of clouds with internal dust because the dust within the extensive PIZ
signi\-ficantly absorbs the Balmer lines which emerges from the back (producing at the
same time the steeper Balmer decrement). This effect is particularly important in
understanding the results presented in $\S$ 2.3.4. To illustrate the dominant
effects of both perspective and dust, we picture a thin fully ionized region which
generates the bulk of Ly$\alpha$, and, depending on how opaque the region is to dust
[$\tau_V(H^+)$], a substantial fraction of the resonant Ly$\alpha$ escapes from the
front while half of the Balmer lines' flux is required to transfer across the larger
and dustier PIZ before escaping from the back (the other half emerges from the front
after suffering little extinction).

\begin{figure}
\vspace{5in}
\caption[]{}{\small In panel $(a)$, Ly$\alpha$/H$\beta$ as a function of the
Lyman-limit opacity of the matter-bounded photoionized slabs. As in Fig.6, the solid
and short dashed curves were calculated with $U_f$=0.0015 and an internal dust content
of $\mu$=1.0 and 0.2 respectively. The dotted line corresponds to calculations with
$U_f$=0.0015 and $\mu = 1.0$. Panel $(b)$ plots the Balmer decrement as a function of
the Lyman-limit opacity.}

\end{figure}

	To illustrate the role of internal dust on the total emerging Ly$\alpha$ and
Balmer lines, we present in Figure 7 the line ratios resulting from adding together
the fluxes from perspective ``f'' and ``b'' (before taking the ratio) as a function of the
Ly$\alpha$ limit opacity $\tau_{H^0}^c(912\AA\ )$ for the matter bounded sequences
presented in Figure 6. The larger destruction rate of Ly$\alpha$ for the
high-ionization parameter ({\it dotted line}) model is apparent and is the result of a
higher value of $\beta$ throughout the nebula (cf. $\S$ 2.3.3). We also find for all
three sequences that for $\tau_{H^0}^c(912\AA\ )>100$, because dust in the PIZ starts
absorbing much of the H$\beta$ flux escaping from the back, the "f+b"
Ly$\alpha$/H$\beta$ ratio rebounds up (the Ly$\alpha$ flux actually remains steady).
The somewhat higher Balmer decrement in the matter-bounded regime in the higher $U_f$
model is generally caused by reddening [larger $\tau_V(H^+)$]. When 
$\tau_{H^0}^c(912\AA\ )< 10$, then in addition to reddening, for $\beta$ sufficiently
high, the normal conversion of the higher Lyman lines into Balmer or higher series is
reduced because of dust absorption of these Lyman line (this effect simulates
something resembling case A).

\subsection{Selection of the ionization parameter $U_f$}

The most simple geometrical model of the narrow-line region consists of a symmetric
distribution of ionized clouds surrounding the ionizing source. Any reasonable assumed
electron densities leads us to infer a very small {\it volume} filling factor for the
emitting gas. The ionization parameter which is inferred from the line ratios is
surprisingly uniform within a given class of AGN.

	To determine $U_f$, we have simply required the radiation-bounded calculations
to approximately fit the "mean" optical Seyfert 2 line spectrum derived by Ferland \&
Osterbrock (1986, hereafter FO86) in which a ratio of  [OIII]$\lambda$5007/H$\beta$
of $\sim$ 10 is observed. With the other parameters specified above, this leads us to
adopt $U_f$=0.0015 for the reference ionization parameter.

The choice of the ionization parameter affects strongly the role played by dust. For
instance, internal dust can absorb a fraction of the ionizing radiation and therefore
shrink the Str\"ongrem depth due to competitive absorption with H$^0$. In the case
of our reference model with $U_f$=0.0015, only 3\% of  $\phi^f_H$ is lost to dust
absorption when $\mu=1$. However, with an ionization parameter 10 times larger [recall
that $\tau_V(H^+)$ scales with $N^S_{H^+}$, we find that as much as 24 \% of
$\phi^f_H$ is absorbed by dust when $\mu$=1. This effect can be seen in the  $U_f$=0.015
sequence shown in Figure {\it 6b} where the dotted line curve is shifted to the left
because scaling is relative to the dust-free value of $N_{H^+}^S$.

	Another aspect where $U_f$ is important is the amount of destruction of
resonant Ly$\alpha$ by dust. To show this, let us define using equation (2) the pure
absorption opacity by dust at depth half way within the fully ionized zone:

	$$ \tau_{dust}^{abs}(1216 \AA\, {\footnotesize \frac{1}{2}} N_{H^+}^S)\cong 2.2 
\tau_V^{ext} \cong 60 \mu U_f ~~~ [5]$$ 

	The justification for choosing this position is that most of the Ly$\alpha$ is
generated as well as destroyed within the fully ionized region. Assuming $U_f =
0.0015$ and $\mu = 1.0$, we obtain $\tau_{dust}^{abs}(1216 \AA\ ) \approx$ 0.1. 
As a crude
estimate to illustrate the effect of dust on Ly$\alpha$, at this depth, the line
(scattering) opacity is $\sim$ 10$^5$ which allows us to estimate $\beta \sim $
10$^{-6}$ and derive the fraction of Ly$\alpha$ which escapes unabsorbed from this
layer at $\sim 50\%$ by using the scaled results of Hummer \& Kunasz (1980). On the
other hand, for the higher value of $U_f$=0.015, the line opacity at a similar
relative position corresponding to $\tau_{dust}^{abs}(1216\AA\ )  \approx$ 1 ($\mu$=1) is
higher ($\sim 3 \times 10^6$), and we estimate that with $\beta \sim 3 \times 10^{-7}$,
only 2\% of Ly$\alpha$ escapes the layer. For calculations with the adopted ionization
parameter $U_f$=0.0015, we see that Ly$\alpha$ destruction due to dust should be
moderate.
 
\subsection{Results for a symmetric cloud distribution}

	We approximate the typical NLR emission cloud as a slab of gas of constant
internal pressure with an outer density of $n_H^f =$5000 cm$^{-3}$ (cf.$\S$ 2.3.1).
Using MAPPINGS, we compute the line spectrum from photoionization calculations using
the ionization parameter $U_f$=0.0015 (cf.$\S$ 2.3.3) and an ionizing power-law
of index $\alpha=-$1.4. In all the calculations, the gas phase abundances are  modified
consistently with the dust content $\mu$ in such a way that the total metallicity
(gas+dust) remains always solar: $Z=Z_{gas}+Z_{dust}=$1 (cf.$\S$ 2.2.2 and Appendix
A). The assumed symmetric distribution of the clouds is taken into account in the
calculated spectrum by adding together spectral ``f'' and ``b'' of MAPPINGS calculations
which is equivalent to supposing that there are as many clouds seen face-on as are
seen from the back.

	Using this procedure, we have computed three sequences of models with the
dust content $\mu$  covering the range $0 \leq \mu \leq 1.25$. These three sequences
are presented in Figure 8 ({\it solid lines}) and correspond to three different total
slab column densities $N_H^{slab}$. The intermediate sequence corresponds to the
radiation-bounded case $N_H^{slab} = N_{H^*}$ (see $\S$ 2.3.3 for the operative
definition of the depth of the photoexcited region $N_{H^*}$), a second sequence
represents the matter-bounded case in which $N_H^{slab}$ corresponds to the depth 
where hydrogen is 10\% ionized (i.e., $N_H^{slab}\simeq$ 2.4 $\times$ 10$^{20}$
cm$^{-2}$), a third one is the case in which the NLR clouds are more massive than the
radiation-bounded case and has $N_H^{slab} \equiv N_{H^*} + N_{H^0}^0$ = 5 $\times$
10$^{21}$ cm$^{-2}$, which is about 6 times the $N_{H^*}$ of the dust-free case
\footnote{Note that $N_{H*}$ progressively increases with $\mu$, from 8 $\times$ 10$^{20}$ cm$^{-2}$ when $\mu$=0 to 2.6 $\times$ 10$^{21}$ cm$^{-2}$ when $\mu$=1.25. The main reason has to do with the criterion requiring that hydrogen be only 1\% ionized at the inner boundary of the photoexcited region. Because the recombination coefficient depends on $T_e$, $N_{H^*}$ becomes larger when the temperature of the PIZ increases as a result of depletion of the gas phase with increasing $\mu$. We have found that $T_e$ increases from 1200 K in the dust-free case to 3300 K at $\mu$=1.25}. The
zero-point dust-free reference model with $U_f$=0.0015 is represented in all figures
by the filled star. We emphasize that our simple geometry assumes a very low volume
filling factor of the emitting clouds so that as many ``f'' clouds as ``b'' clouds are
seen by the observer without any intervening obscuration.

\begin{figure}
\vspace{4in}
\caption[]{}{\small Ly$\alpha$/H$\beta$ as a function of the Balmer decrement  for the
three sequences with $U_f$=0.0015 in which $\mu$ is monotonically increasing (0$\leq
\mu \leq$ 1.25). Tick marks from left to right along the solid lines indicate values
of $\mu$ of 0.1, 0.2, 0.4, 0.6, and 1.0, respectively. The three sequences ({\it solid
lines}) differ only by the total column density of the slab $N^{slab}_H$. The
lowermost sequence is matter-bounded using the criterion that the hydrogen ionized
fraction is 10\% at the back layer. The intermediate sequence corresponds to the
radiation-bounded case without any additional neutral gas zone at the back. The
uppermost sequence includes a neutral back zone such that the {\it total} column
density of the slab is constant at 5 $\times$ 10$^{21}$ cm$^{-2}$. The neutral and
photoexcited zones are assumed to have the same dust content [$\mu (H^0) = \mu
(H^*)$]. The filled star represents the reference dust-free case ($\mu$=0). Also shown
in the figure is an ionization parameter sequence ({\it dotted line}) for
radiation-bounded slabs (with $N^0_{H^0}$ = 0) of constant dust content $\mu$ = 1.0
and covering the range 10$^{-4} \leq U_f \leq $ 10$^{-2}$. In all above calculations,
the ionizing energy distribution is a power law with index $\alpha = $-1.4, the gas
density is $n_H^f$ = 5000 cm$^{-3}$ at the face of the cloud, the metals are depleted
from the gas phase in accordance with the dust content $\mu$, and the line intensities
are given by the sum of the ``f'' and ``b'' perspective. }

\end{figure}

	The most important result of the calculations shown in Figure 8 is to prove
that despite large amounts of internal dust, the Ly$\alpha$/H$\beta$ calculated using
canonical input parameters is not radically different from recombination case B, and,
if anything, it can even be larger when $\mu \rightarrow 1$. The expected range for
Ly$\alpha$/H$\beta$ in the dust-free recombination case B is between 23 and 34
depending on the gas density. If, in addition, collisional excitation is present due
to the hardness of the continuum or to low gas metallicity, both the Balmer and the
Lyman decrement become larger (see Gaskell \& Ferland 1984), a situation not unlike
that shown in Figure 8 except that here it is the result of internal dust.

	The two main factors explaining the relatively high Ly$\alpha$/H$\beta$ are
that, first, with $U_f$=0.0015, Ly$\alpha$ destruction is modest in the clouds since
the column density of ionized gas and therefore $\tau_V(H^+)$ is not large, and
second, the portion of the Balmer lines' flux which escape from the back is subject to
considerable absorption by the dust inside the PIZ or within the neutral core of the
cloud. It is therefore not surprising that the models which depart most from case B
Ly$\alpha$/H$\beta$ are those of the matter-bounded sequence, comparatively little
absorption occurs for the Balmer lines while destruction of Ly$\alpha$ photons (which
occurs mostly in the fully ionized zone) remains significant (cf.$\S$ 2.3.2). To
illustrate to what extent much higher values of $U_f$ would be catastrophic to the
escape of Ly$\alpha$ photons, the dotted line in Figure 8 represents an ionization parameter
sequence with models of constant dust content ($\mu$=0.1). Of course, these models are
of limited relevance since they result in a forbidden-line spectrum incompatible with
observed optical line ratios.

	Since in the dust-free case we used solar metallicity for the gas, little
collisional excitation takes place within the PIZ and the resultant Balmer decrement is
not significantly above that of a dust content of $\mu$=1, however, the Balmer
decrement in the radiation-bounded case is shown in Figure 8 to approach 3.3. On the
other hand, moderate amounts of internal dust $\mu \sim 0.2-0.4$ would favor an
intrinsic Balmer decrement in the range 3.0-3.1 assuming the clouds are
radiation-bounded (the corresponding range the Lyman decrement is 32-37). This range
(3.0-3.1) corresponds to that proposed by various authors (Gaskell \& Ferland 1984; 
Halpern \& Steiner 1983) for the intrinsic NLR Balmer decrement except that it is here
more the result of the presence of internal dust rather than of a hard ionizing
continuum or of deficiencies in the gas metallicity relative to solar. Binette et al.
(1990) pointed out the apparent lack of Seyfert 2 with H$\alpha$/H$\beta <$ 3.4 if one
considers only objects with good S/N in the H$\gamma$ line. If this high value of 3.4
was exclusively caused by internal dust present in all Seyfert 2's, then large values
of $\mu$ ($\sim$1) are implied from Figure 8. Alternatively, a portion of the
extinction may be the result of {\it intervening} gas containing dust, in which case
the amount of {\it internal} dust implied is lower. There are two possibilities for how
this intervening gas/dust may arise. In one case this dust is contained in a detached
screen. Such models, however, may not be physically plausible as discussed in
$\S$4.2. Another possibility is that the intervening dust+gas arises from a
large covering factor of the NLR clouds themselves. This case is presented in
$\S$2.4.3.

	Radiation-bounded calculations with our selected parameters are known to over
predict by a factor $\approx$2 the  [OI]$\lambda$6300/H$\beta$ ratio. Using a
moderately matter-bounded cloud such as the one shown in Figure 8, one can resolve
this discrepancy by cutting the cloud at some depth to make the [OI] fit the observed
value.
 As previously shown in Figure 7, the matter-bounded model
$\mu$=1 results in a lower Ly$\alpha$/H$\beta$ (than radiation-bounded calculations)
somewhat below case B and in a moderately reddened Balmer decrement. The procedure
of cutting the clouds to adjust [OI] to the observed value is,
however, quite arbitrary. A more realistic approach is the dual gas components model
of Viegas \& Prieto (1992) which was proposed as a solution to explain the observed
high HeII/H$\beta$ ratio observed in the spatially extended NLR (ENLR) (see also
Morganti et al. 1991). This model consists of a matter-bounded (MB) and a
radiation-bounded (RB) component. The MB component has a higher volume filling factor
(to compensate for its much lower emission measure) and a higher ionization parameter
($U_f>$0.0015). Although we did not consider this level of complication in our modeling,
we can still establish some of its general properties in relation to dust using the
results of $\S\S$ 2.3.2 and 2.3.3. Under most instances, the effects of dust on
the HI lines cannot be larger in the case of the dual-component model than with our
simpler single RB component because the MB component, despite its higher $U_f$, is
nevertheless extremely thin in column density and therefore generally results in little
Ly$\alpha$ destruction \footnote{The dual component model of Viegas \& Prieto (1992) suggest a MB component with opacity $\tau_{H^0}^c$(912 \AA\ ) $<$ 1. Adopting $\tau_{H^0}^c$(912 \AA\ ) =0.55 and a similar $U^{MB}$ = 0.01 to what they use, we obtain $N^{slab}_H \cong$ 0.067 $\mu$. Even with high dust content $\mu$=1, our calculations indicate a Ly$\alpha$ destruction of only 0.13 dex with a difference in flux between ``f'' and ``b'' perspectives of 15\%.}. Furthermore, the RB component should have $U^{RB} <$
0.0015) (Viegas \& Prieto suggest $U^{RB} <$ 0.0006) and therefore much lower $\tau_V$ than in our single RB component. In conclusion, a working dual-component model would
in all likelihood be characterized by {\it smaller} line absorption effects due to
dust at equal values of $\mu$ than in our simplified single component model.

\section{Discussion}

	After comparing calculations with internal dust with recent Ly$\alpha$/H$\beta$ data on Seyfert 2's ($\S$2.4.1), we discuss various interpretations of the low values observed. We look at the possibility of pure line scattering or pure dust absorption ($\S$2.4.2) as well as that of mutual covering of Ly$\alpha$ clouds ($\S$2.4.3). On $\S$2.4.4, we consider the possibility of dust absorption of the UV continuum for explaining the apparent deficit of ionizing photons.

\subsection{Observational data of KAW3}

	The available data on the NLR Lyman decrement is still rather limited. The most reliable information on the NLR comes from Seyfert 2's since, as pointed out by FO86, only for these objects can one deduce  an optical to ultraviolet line spectrum which is totally free of contamination by BLR components. It is likely, however, that the physical information gathered from the NLR of Seyfert 2 is relevant to that of Seyfert 1 and quasars. One of the motivation of the current work was the recent publication by Kinney et al. (1991) of a Seyfert 2 data set for which the optical spectra were taken through an aperture which matched the large IUE satellite aperture. The matching of aperture size for the different lines is a crucial point in the study of the NLR Lyman decrement because the emitting line region is spatially extended and in the likely event that Ly$\alpha$, H$\alpha$, H$\beta$ present a different spatial distribution in surface brightness, the interpretation of ratios of lines derived from different apertures becomes very complicated, if not impossible. By comparing for instance objects common to both the KAW3 and FO86 data set, one finds many objects with significantly different line ratios between the two data sets presumably as a result of the much smaller aperture 
of the FO86 optical observations. For this reason we have limited our analysis to the data set of KAW3 which are presumed free of this problem.

	In Figure 9 (which is an expanded scale version of Fig.8), we present the data of KAW3 ({\it filled squares}) with their estimated error bars. The earlier model sequences of Figure 8 now appear as broken dash lines in the upper left. What is immediately striking is that internal dust absorption for an isotropic distribution of emitting clouds does not account for the low Ly$\alpha$/H$\beta$ ratio which is significantly below case B.

	In order to explain the overall low Ly$\alpha$/H$\beta$ ratio in Seyfert 2's, either an intervening absorption/scattering screen is required or we must modify our simple NLR geometry. The first possibility, discussed in detail in $\S$4.2, would require pure line scattering from the screen to explain the near-case B H$\alpha$/H$\beta$ objects. The second possiblity is that the covering factor of the emission clouds is not small as was assumed so far. This idea is explored in $\S$2.4.3 along with the possiblity that if the covering is extreme, a closed spherical geometry might be preferable to that of an isotropic distribution of slab-approximated clouds.

\begin{figure}
\includegraphics{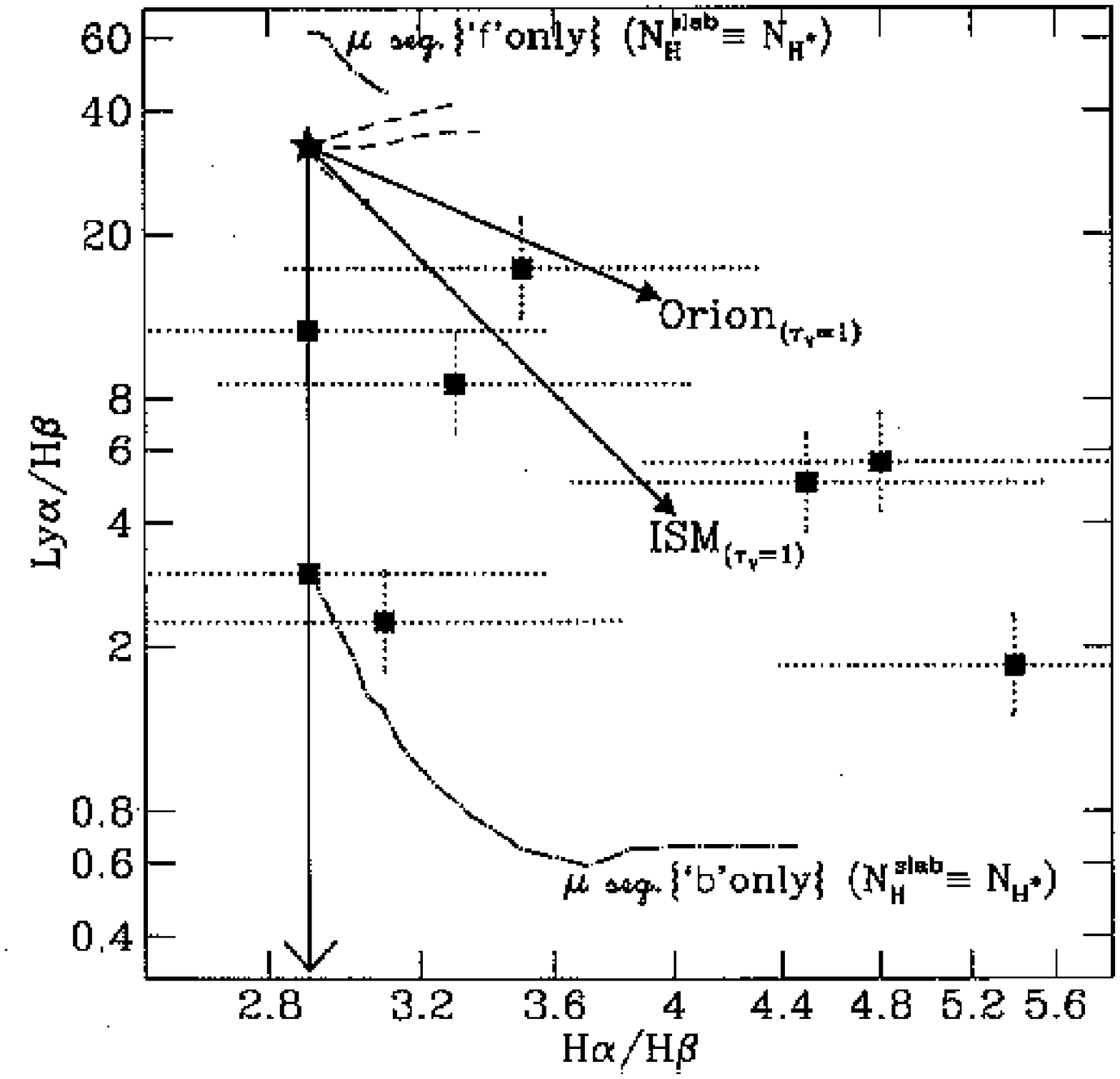}
\vspace{4in}
\caption[]{}{\small The squares represent the homogeneous data set of Seyfert 2 observations of Ly$\alpha$/H$\beta$ vs. H$\alpha$/H$\beta$ by Kinney et al. (1991) in which the aperture in the optical closely matches the large-aperture IUE Ly$\alpha$ observations. Two reddening vectors corresponding to the ISM and the Orion extinction curve, respectively, are drawn to indicate the expected displacement due to foreground {\it pure} dust. The vectors' length correspond to an extinction $\tau $ = 1 in the visible (5500 \AA\ ). The vertical 
solid line represents the effect of scattering of Ly$\alpha$ by a foreground 
neutral hydrogen screen at the same redshift as the NLR (see $\S$ 4.2). The dash-dotted lines represents the $\mu$ sequence with $U_f$=10$^{-2.8}$ and $N_{H^0}^0$ = 0 but as seen {\it separately)} from perspective ``f'' and ``b''. The three dash lines correspond to the same $\mu$ sequences of Fig.~8 (i.e., with ``f''+``b'') }

\end{figure}

\subsection{The pure line scattering and the pure dust models}

	There are broadly two types of Seyfert 2 objects in the KAW3 data: the objects with  near-case B H$\alpha$/H$\beta$ and the objects with reddened H$\alpha$/H$\beta$. The first category might be explained by pure Ly$\alpha$ scattering due to an intervening screen of H$^0$ which is supossed to be covering only the observer's view of the NLR but not the complete sky as seen from the NLR. Assume a slab geometry for the screen. Then for a pure scattering screen of thickness $\tau_{H^0scat}$, where $\tau_{H^0scat}=5.9\times 10^{-14}N_{HI}^{screen}(T_{HI}/10^4 K)^{-1/2}$ is the Ly$\alpha$ resonant scattering optical depth at line center, the fraction of photons impinging on the screen from the side facing the source which penetrates to the opposite side (the side facing the observer) is
$\propto \tau_{H^0scat}^{-1/2}$ for $\tau_{H^0scat} \gg 1$ (Slater, Salpeter, \& Wasserman 1982). Owing to the large magnitude of $\tau_{H^0scat}$, only modest values of $N_{HI}^{screen}$ are required to efficiently suppress the Ly$\alpha$ flux reaching the observer. Thus, for instance, for $N_{HI}^{screen} \sim 10^{15}$ cm$^{-2}$ (10$^{17}$ cm$^{-2}$), only 10\% (1\%) of the incident Ly$\alpha$ photons reaches the observer; the rest are reflected back toward the source. At such column depths, the dust opacity within the screen is entirely negligible, and so the extinction vector+ shown in Figure 9 is purely vertical.

	The lack of objects with near-case B H$\alpha$/H$\beta$ in the FO86 sample has led these authors to propose pure dust absorption (ignoring the destruction of Ly$\alpha$ after multiple resonant scatters) since their objects all broadly lie along the ISM reddening vector (see Fig.6 FO86). Within the sample of KAW3, many objects still lie in the general direction of the ISM reddening vector. The pure dust screen (without any line scattering; otherwise, the reddening vector would be steeper than shown above) remains a possible interpretation for the observed ratios, but only for the reddened H$\alpha$/H$\beta$ objects. Several limitations exist, however with the picture of the pure dust screen which we now proceed to analyze.

	 In the pure dust screen model, a layer of dust and gas is presumed to lie outside the NLR toward the line of sight to the observer. Photons that pass through this layer are assumed to suffer solely absorption by the dust, that is, interaction between the photons and the gas are ignored. While this is a good approximation for non resonant line photons such as H$\alpha$ and H$\beta$, it is a poor approximation for resonant line photons such as Ly$\alpha$. Owing to the large resonant scattering cross section of Ly$\alpha$ photons, resonant scattering by these photons with neutral hydrogen atoms as they traverse the layer cannot be ignored in general. For example, to obtain a reddening of $A_V$ = 1 require $N_H = 2\times 10^{21}$ cm$^{-2}$ for $\mu=1$. For a gas at $T_e$=10$^4$ K, this corresponds roughly to $N_{HI} \sim$ 10$^{20}$ cm$^{-2}$, which implies a line center resonant scattering optical depth fort Ly$\alpha$ photons of $\tau_{H^0scat} \sim 6\times 10^6$.

	One possible way to avoid resonant scattering so that only pure dust absorption is operative is to have the layer move with sufficiently large bulk velocity so that the line is shifted out of the Doppler core. How large a column density in neutral hydrogen can be tolerated without significant resonant scattering then depends on how fast the layer is moving away from the line source. We plot in figure 7 this critical neutral hydrogen column density, $N_{HI}^{critic}$, against the layer bulk velocity for gas at $T_e$ = 10$^2$, 10$^4$, and 10$^6$ K  where $N_{HI}^{critic}$ is defined as the $N_{HI}$ (measured in the comoving frame) that gives $\tau_{H^0scat}$=1 (see Appendix B, $\S$ 1). Thus, resonant scattering may be ignored for layers with a neutral hydrogen column density lying below the curves. The "steps"  in the curves correspond to where the line is shifted from the core into the wings. It is evident that $N_{HI}^{critic}$ is very low $<$ 10$^{14}$ cm$^{-2}$) unless the bulk velocity is very high. In order to have $A_V \geq 1$, we require $N_{HI} \geq 10^{20}$ cm$^{-2}$) (assuming $T_e=10^4$ K and $\mu \approx 1$), which requires $V \geq 500$ km s$^{-1}$) from figure 7. Aside from the issue of how to accelerate the layers to such high bulk velocites, there remains the problem that these clouds must be constantly replenished from some unknown source. This replenishing must be very efficient and continuous because the narrow-line gas in the dust screen is at $T_e$=10$^4$ K) and has $N_{HI} \geq 10^{20}$ cm$^{-2}$, the screen itself should have significant emissivity in lines such as H$\alpha$ and H$\beta$, unless the 
screen is placed sufficiently far outside the NLR so that geometric dilution reduces the impinging ionizing flux to insignificant levels. Otherwise, one should see highly blueshifted lines (from the screen) in addition to the usual lines from the NLR. This is not observed.

\begin{figure}
\vspace{3.in}
\caption[]{}{\small The critical neutral hydrogen column density, $N^{crit}_{HI}$, of the dust screen vs. the screen's bulk velocity, $V$. The quantity $N^{crit}_{HI}$ is defined as the column density (measured in the comoving frame) giving Ly$\alpha$ resonant scattering optical depth of unity ($\tau_{H^0scat}$ = 1). The region below the curves have $\tau_{H^0scat} <$ 1. The curves correspond to a dust screen gas temperature of 10$^2$ K ({\it dashed}), 10$^4$ K ({\it solid}), and 10$^6$ K ({\it dot-dashed}). The "step" in each of these curves correspond to where the line is shifted from the core into the wings. The impinging Ly$\alpha$ line is assumed to be monochromatic at the line center (as measured in the line source frame). }

\end{figure}

	Without invoking bulk motion, another way to make the pure dust screen viable is to make the gas in the dust screen very cold relative to the width of the line impinging on the screen. In this manner, only a vey narrow band about line center will suffer resonant scattering within the screen. In Figure 11 we plot the fraction of the impinging energy in the Ly$\alpha$ line (assumed to have a gaussian profile) that suffers interactions with neutral gas atoms in the screen as a function of the neutral hydrogen column density of the screen (see Appendix B, $\S$ 2). The horizontal line shows where 1\% of the incoming line energy suffers interactions. The curves are labeled by the width of the impinging line (expressed in velocity untis) relative to the width (in velocity units) across which the line will interact with the gas in the screen. Typical line widths for the NLR Ly$\alpha$  emision line is $\sim$ 10$^3$ km s$^{-1}$. It is evident from Figure 11 that even for (screen) gas at $T_e$=10$^2$ K so that $\kappa = \delta v_s/\delta v_a$ = 10$^3$, the critical neutral hydrogen column density (to have 1\% interaction or less) is only $\sim 10^{16}$ cm$^{-2} \sim N_H$, corresponding to $A_V \sim 5 \times 10^{-6} \mu$. Thus, a cold screen is unfeasible because the low hydrogen column densities it requires (to avoid resonant scattering) imply negligible absorption by dust internal to the screen.

\begin{figure}
\vspace{3.5in}
\caption[]{}{\small The fraction of energy, $f$, in the Ly$\alpha$ line impinging on the dust screen that suffers interactions with the screen's gas vs. the neutral hydrogen column density, $N_{HI}$, of the screen. The horizontal dotted line correpond to $f$=0.01. The curves correspond to $\kappa = \delta V_S/\delta V_a$ = 10 ({\it solid}), 10$^2$ ({\it dashed}), and 10$^3$ ({\it dot-dashed}), where $\delta V_S$ is the width (in velocity units) of the impinging Ly$\alpha$ line and $\delta V_a$ is the width of the Ly$\alpha$ resonant scattering line profile for the gas in the dust screen. }

\end{figure}

	We conclude that either with bulk motion or with a very cold cloud, the pure dust screen model is not physically plausible. The only scenario we can conceive of where a pure dust screen might work is a "coronal" model wherein the dust is embedded in a {\it very} highly ionized, and hence very hot ($T_e \sim$ 10$^6$ K), and tenuous ($n_e \sim$ 10 cm$^{-3}$) medium. Assume for now that dust can exist in significant quatities in this coronal gas. To have $A_V \sim$ 1 require $N_H \sim $ 10$^{21}$ cm$^{-2}$ for $\mu$=1, or $N_H \sim$ 10$^{22}$ cm$^{-2}$ for $\mu$=0.1. With $n_e \sim$ 10 cm$^{-3}$, this correpsonds to a coronal region with linear size $L \sim$ 10$^2$ pc ($\mu$=1) or $L \sim$ 10$^3$ pc ($\mu$ = 0.1). In the former case, the size is comparable to the size of the NLR. While the existence of such a region external to the NLR cannot be ruled out, it is somewhat ad hoc. In the latter case, the large volume required for the coronal region almost surely implies that the corona would cover the whole NLR. In this case, one would expect the same dust in the coronal gas that 
affects the narrow lines to redden the starlight from the central galactic buldge. This reddening is generally not observed.

	The properties of the coronal gas is probably not all that different from those of the intercloud medium in the NLR, if we assume a two phase medium for the NLR with the two phases (cloud and interclouds) being in pressure equilibrium. If so, then postulating a separate corona outside the NLR seems superflouos; it would be more natural to assume dust to be present in the NLR intercloud medium itself. If, however, one is to consider dust in the intercloud medium, then one should also consider dust in the clouds themselves. The narrow lines produced within the NLR cloud would then be affected in general by dust in both the cloud and intercloud medium. With regard to dust survival, it is 
likely that both evaporation (owing to the high gas temperature) and sputtering
 will render the dust life-times to be shorter in the coronal-like  intercloud medium than in the clouds. The main dust destruction process inside the clouds is sputtering which gives a dust lifetime of $\sim$ 10$^5$ yr (assuming $n_e \sim$ 10$^3$ cm$^{-3}$ and $T_e\sim$ 10$^4$ K). In the NLR model we adopt, we assume that dust is able to be replenished on such time scales inside the clouds to enable a significant dynamic steady state abundance of dust to be present. This replenishment is probably more difficult to achieve in the coronal-like intercloud medium and in our model, we assume that dust is not present in any significant quantities in the intercloud medium.

	\subsection{The partially covering models}
	
	One obvious advantage of a simple interpretation for the position of all the objects in Figure 9 rather than the two separate pure dust and pure line scattering interpretations is that it might be more informative of the NLR geometry. Although the Ly$\alpha$/H$\beta$ observed is much below that predicted by the ``f''+``b'' models, some of the objects are not that far from the pure perspective ``b'' models (see dot-dash line in Fig.~9). This naively suggests a nonnegligible covering factor for the emission region in which individual clouds partly cover one another in the line of sight to the external observer. Although plane-parallel calculations do not allow us to replicate very accurately such a geometry, a first order estimate of the effects involved can nevertheless be arrived at. For instance, we can safely expect the Ly$\alpha$ luminosity to be dominated by the few uncovered clouds which are directly seen from the front (i.e., ``f'' clouds). The Balmer lines in our simplified scheme would essentially come from three different components: the few ``f'' clouds seen directly, the many ``b'' clouds seen from the back, and finally the ``f'' clouds seen through other clouds and which we postulate are seen through a depth statistically equivalent to a single cloud (we neglect multiple cloud covering). By assuming that the transmitted spectrum of each covered ``f'' cloud is approximately equivalent to that of a ``b'' cloud \footnote{A zero-order aproximation based on the fact that the fully ionized zone is only a small fraction of the whole cloud and therefore any ``b'' spectrum is really seen through a screen of dust of depth comparable to that given by the whole slab.}, we can readily estimate the main effects of a partially covered geometry by simply giving more weight to the ``b'' clouds, i.e., ``f''+$\gamma$``b''. 

\begin{figure*}
\includegraphics{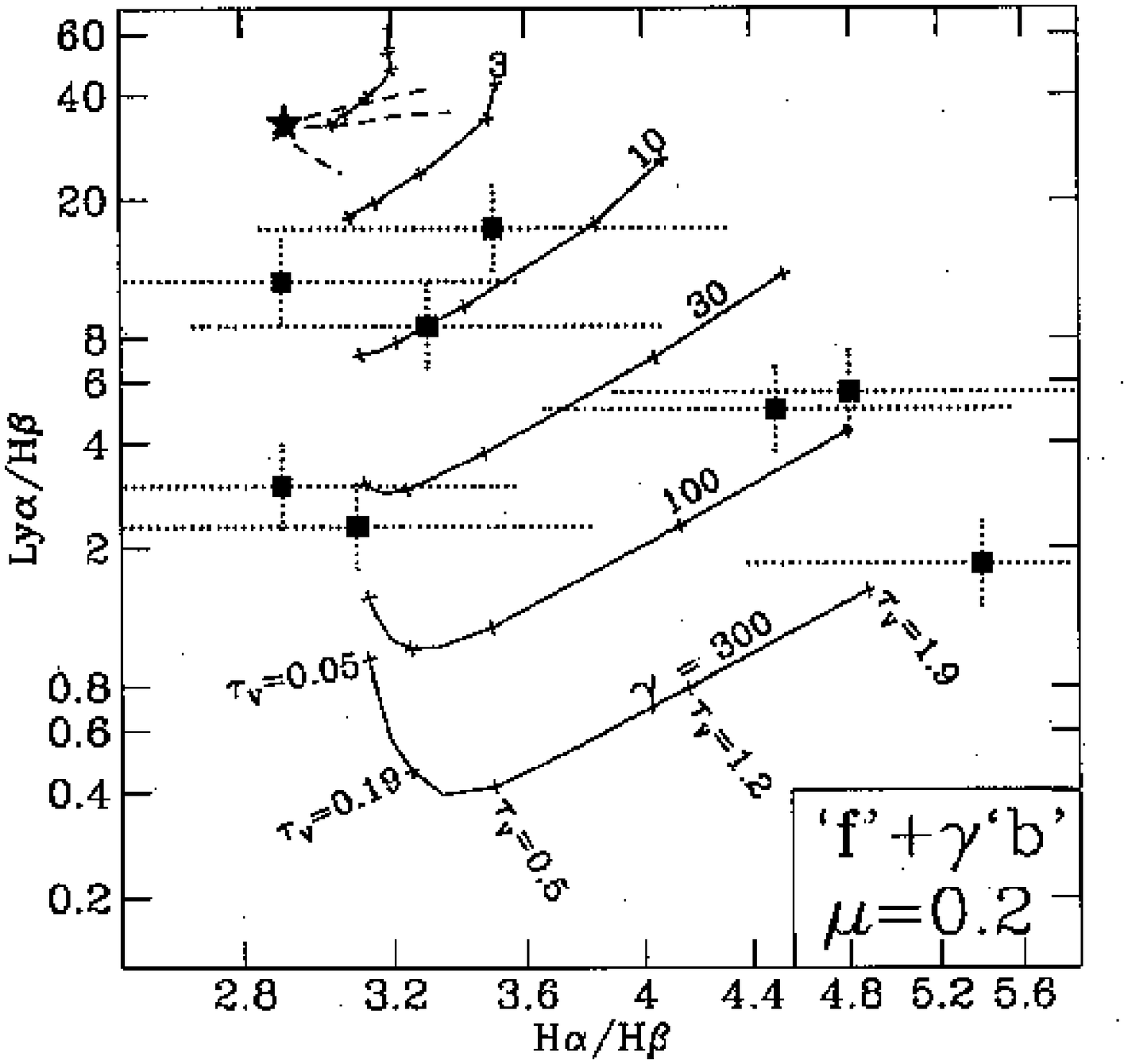}
\vspace{5in}
\caption[]{}{\small Same observations as in Fig.~9 but compared with the partially covering model using $\mu(H^*)$ = 0.2 (see eq.[3]) and with the line fluxes obtained by giving a higher weight $\gamma$ to perspective ``b'' (``f'' + $\gamma$``b''). For a given value of $\gamma$, the solid line illustrates the effect of increasing the neutral zone depth, $N_{H^0}^0$, characterized by a dust content $\mu(H^0)$ =1.0. The tick marks from left to right correspond
to opacities of the  $N_{H^0}^0$ zone of $\tau_V$ = 0.05, 0.19, 0.5, 1.2 and 1.9 [for example, with $\mu(H^0)$ = 1.0 these correspond to $N_{H^0}^0$ = 10$^{20}$, 4 $\times$ 10$^{20}$, 2.5 $\times$ 10$^{21}$ and 4 $\times$ 10$^{21}$ cm$^{-2}$, respectively]. The three dash lines correspond to the same $\mu$ sequences drawn in Fig.~8 ({``f''+``b''})}

\end{figure*}

In Figure 12, we present the loci of ``f''+``b'' calculations in which the weight $\gamma$ takes on the values of 1, 3, 10, 30, 100, and 300 using photoionization calculations in which the dust content $\mu(H^*)$=0.2. For a given value of $\gamma$, the solid line tracks result from varying the depth of neutral gas $N_{H^0}^0$ (beyond the PIZ). Since the dust content within the neutral zone $\mu(H^0)$ (see eq.[3]) might be plausibly be different from $\mu(H^*)$, we specify $N_{H^0}^0$ in Figure 12 in terms of the dust opacity in the visible, $\tau_V(H^0)$ [leaving $\mu(H^0)$ unspeciefied; see tick marks]. Considering the size of error bars ({\it dotted lines}) which characterizes the data, we find very encouraging that the tracks pretty much cover the whole region occupied by all the Seyfert 2 observations.

	Even if the tracks are only crude estimates and are not unique in terms of the parameter $\mu(H^*)$, they succeed in illustrating that a very simple and self-consistent mixture of internal line scattering and dust absorption can explain the observed ratios without the need for ad hoc intervening screens. We infer from this exercise that the NLR HI line ratios points toward a significantly self-covered distribution of gas which is very patchy and where all the Ly$\alpha$ originates from ionized gas in direct view of the observer.

	If this conclusion is valid, it is possible to study one limiting case of this picture by supposing the covering factor to be so high  that spherical geometry becomes a preferable representation of the NLR (and which at least addresses the effects of multiple scattering inside the geometry, a possibility excluded when considering only independent slabs).

	Adopting spherical geometry, we will represent the NLR as a Str\"omgrem sphere uniformily surrounded by a layer of neutral gas of opacity $\tau_V(H^0)$ except for a a small area where the ionized sphere is not covered as depicted in Figure 13. As in the simple partial covering model dicussed above, Ly$\alpha$ escapes exclusively from the region which is uncovered. The Balmer lines, on the other hand, emerge through both the uncoverd area and the neutral dust outer layer.

	To compute in spherical geometry and yet retain an adequate treatment of the transfer through dust, we integrate the emissivity of volume cells within either the near-side hemisphere ``b'' or the far-side hemisphere ``f''. For the line transfer we use the dust opacity occurring between each cell and the surface of the sphere along the direction of the observer's line of sight. The ionization and thermal structure are calculated, as usual, radially from the inner gas radius $R_0$ up to the limit of the photoexcited region $R_*$. For the line opacity, the integral extends out to include the outer neutral layer. The line emissivity cells are defined by the intersection of the photoionized spherical layers (used in the calculation of the radial ionization structure) of either hemispnere :``f'' or ``b'' with concentric cylinders oriented along the lines of sight to the observer. In this scheme, we are in effect simulating the use of 10 concentric ring apertures of an observer which is looking at the spherical NLR. The global model is the sum of 10 ``f''+ ``b'' cylindrically integrated contributions.

\begin{figure}
\includegraphics{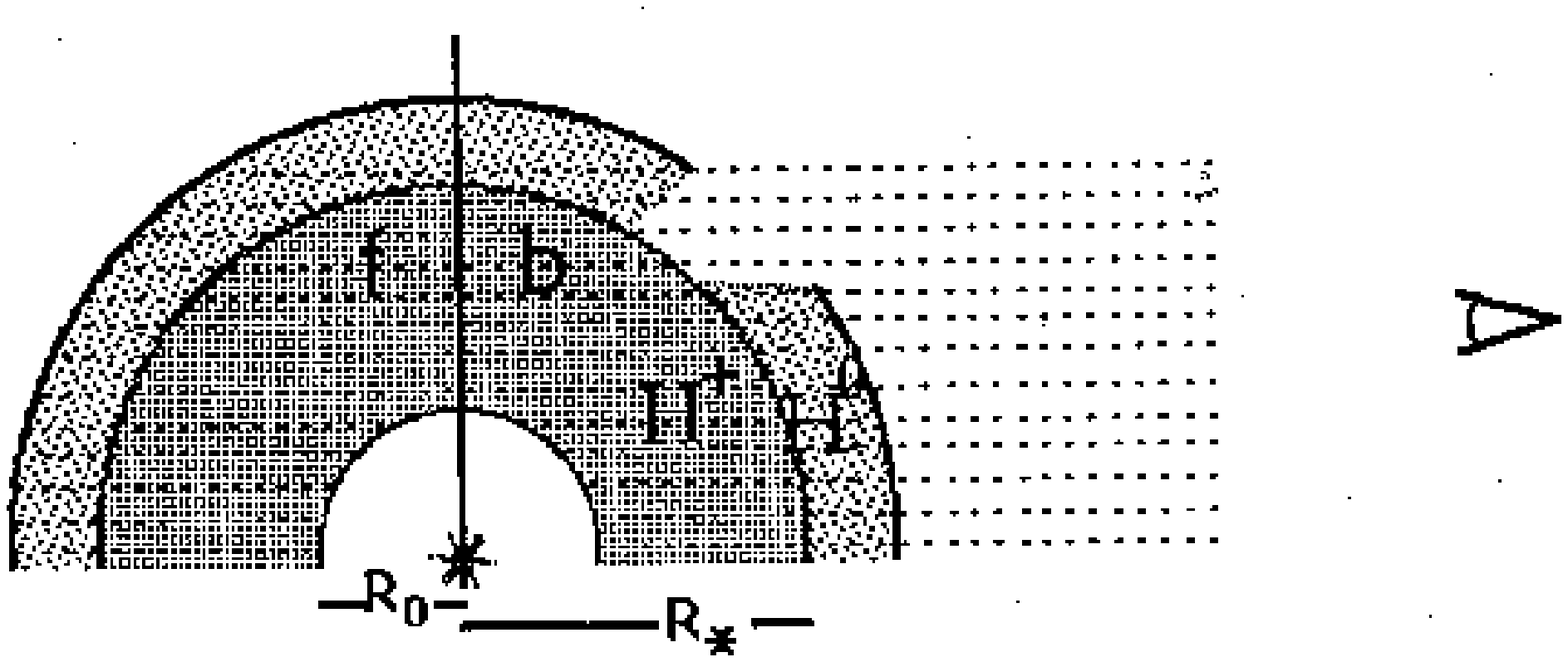}
\vspace{2in}
\caption[]{}{\small Schematic depiction of the spherical geometry assumed for the NLR. The photoexcited sphere of radius $R_*$ and dust content $\mu(H^*)$ is divided into two hemispheres ``f'' and ``b''. The integral of the emissivity of each hemisphere is integrated along 10 concentric cylinders whose axis defines the observer's perspective. The photoexcited sphere is surrounded by a neutral shell of gas of dust content $\mu(H^0)$ and opacity $\tau_v(N_{H^0}^0)$. A small area of the sphere is uncovered from which some Ly$\alpha$ escape probability is determined radially starting from the inner gas radius $R_0$. }

\end{figure}

	Supposing no external neutral gas layer, we find that for the same amount of dust $\mu(H^*)$ and column density of ionized gas $N^S_{H^+}$, the spherical geometry results in a reduced Ly$\alpha$/H$\beta$ compared to the slab ``f''+``b'' case, resembling in fact more the ``b'' perspective of the slab case. This is understandable given that the direction of the gradient ionization 
structure ($n_{H^0}/n_H$ increasing with radius) seen by an observer external to the H$^+$ sphere is similar to that of a slab looked up from perspective ``b''. We emphasize that the effect of clumpiness of the ionized gas (which we do not consider in the transfer) would be more consistent with the geometry of a large number of clouds superposed along the line of sight. This would allow a larger fraction of Ly$\alpha$ to escape the ionized region  as argued by Neufeld (1991), although the implementation of such effects is beyond the scope of this work.

	To derive the emergent spectrum for the partially uncovered case, we simply combine the spectra of the fully covered sphere -excluding the contribution of a certain number of aperture rings- with the correponding missing aperture rings of a spherical model calculated without any external gas layer covering the ``b'' hemisphere. Note that in the dust-free case, such procedure would underestimate the emerging Ly$\alpha$ by a factor of 2. With the presence of internal dust, the treatment is only approximate and rests on the reasonable assumption that most of the Ly$\alpha$ scattering is local to the point of emission an approximation which is increasingly valid as $\mu$ increases since a photon which would scatter too many times over a length scale of order $R_*$ will likely be destroyed.  Furthermore, the resonant photon does not know a priori where the hole in the outer layer is and will likely be absorbed in the much larger neutral outer zone which covers the ionized sphere.

	The results are presented in Figure 14 where the amount of uncovered area is expressed as a fraction of the total projected emitting area $\pi R^2_*$. The parameters of the models were derived as follows. As for the slab case, we adjusted the ionization parameter of the spherical radiation-bounded models until [OIII]$\lambda$5007/H$\beta \sim$ 10 (by varying the source luminosity in the center and/or the volume filing factor of the infinitesimal gas filaments). The ionizing continuum is the same as before, the density at $R_0$ is 5000 cm$^{-3}$ and the gas pressure is maintained constant with radius. The resulting photoexcited sphere in our calculations has $R_*/R_0 \sim$ 2.4 and $N_{H^*} \sim$ 6.3 $\times$ 10$^{20}$ cm$^{-2}$. The electron column density in $N_e$ = 2.1 $\times$ 10$^{20}$ cm$^{-2}$ which is exactly the same as for the reference slab model $\bar{U}$
=0.0013. Because more Ly$\alpha$ is destroyed in spherical geometry, we have assumed smaller values for ($\mu(H^*)$, [See, for instance, the long dash line in Fig.~14 which corresponds to models with $\mu(H^*)$ = 0.4, a clearly excesive value.]

	The solid lines in Figure 14 correspond to models with a small internal dust content ($\mu(H^*)$ =0.1). The dust opacity of the external layer takes on the values of $\tau_V(H^0)$ = 0.5 ({\it leftmost solid line}) and $\tau_V(H^0)$ = 2.4 ({\it rightmost solid line}). From the position of the near-case B objects on the left, we infer that their NLR is probably less deeply embedded in dust 
[$\tau_V(H^0) \sim$ 2-3]. The fraction of the uncovered area runs from 30\% to as little as 3.0\%-0.3\%. Overall, the models bracket quite well the region occupied by the Seyfert 2 observations.

\begin{figure}
\includegraphics{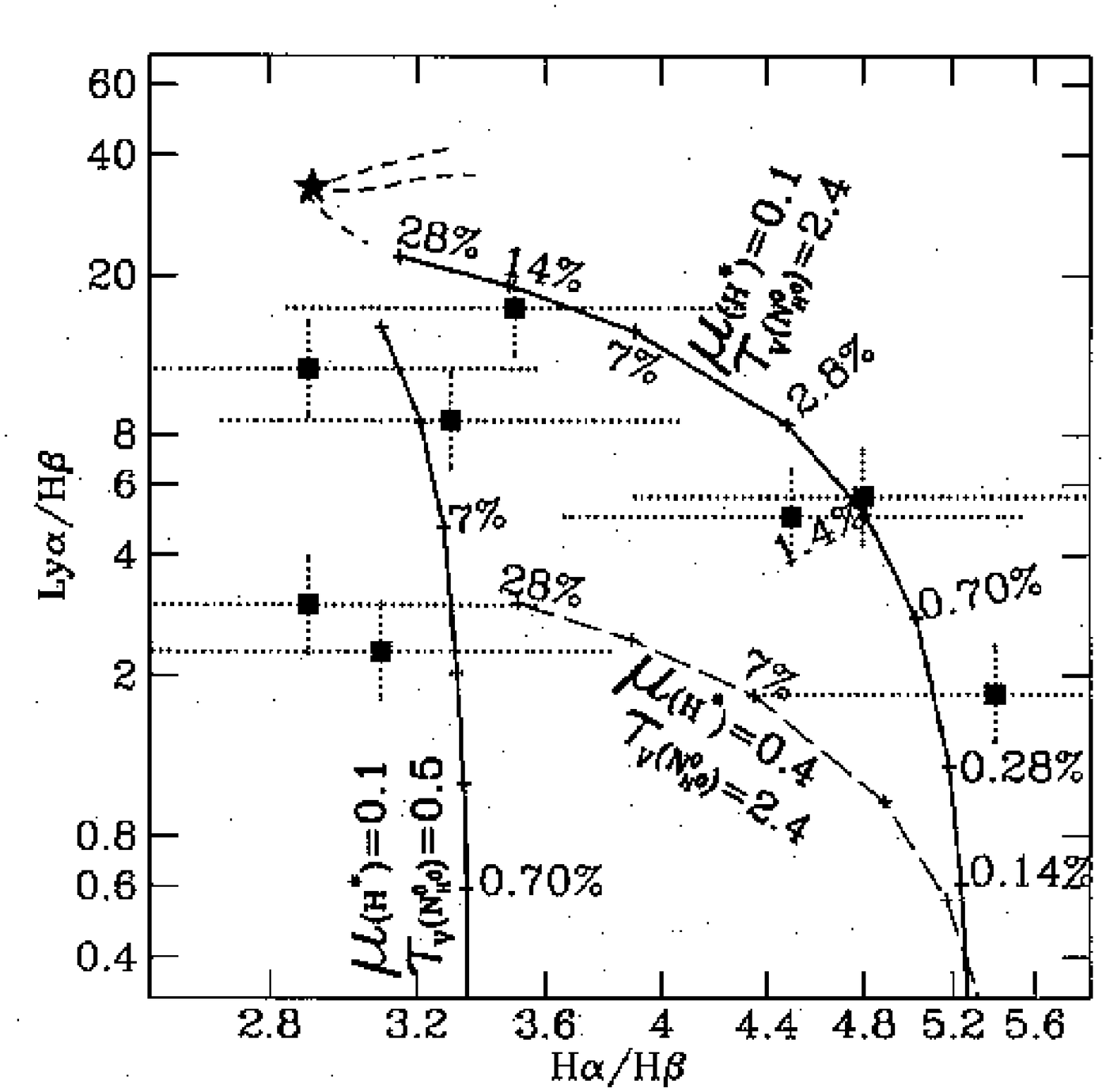}
\vspace{4in}
\caption[]{}{\small Same observations as in Fig.9 but compared with the spherically closed geometry calculations. The ionized sphere is surrounded by an outer shell of neutral gas $N_{H^0}^0$ [with dust content $\mu(H^0)$; see eq.[3]) which we express in terms of dust opacity $\tau_V(H^0)$. The curves represent the effect of having a hole in this shell which directly exposes part of the Str\"omgren sphere and allow some Ly$\alpha$ to escape out. Models with internal dust content $\mu(H^*)$=0.1 and either $\tau_V(H^0)$ = 0.5 or 2.5 are shown as solid lines. The long dash line correponds to a model with a higher internal dust content in the ionized region with $\mu(H^*)$=0.4. Tick marks along each curve represent the percentage of the uncovered area reltive to the sphere's projected emitting area ($\pi R^2_*)$.}

\end{figure}
\subsection{Dust and the UV continuum}

	It is not clear how representative of Seyfert 2 the sample of KAW3 is since as many as five objects of the 12 objects in Figure 9 show evidence of properties assigned to Seyfert 1. These include either a very weak BLR either seen directly as in Mrk 348 and NGC4388, or seen in polarized light as in Mrk 348, Mrk 3, and Mrk463 (also NGC4388), or variability of the featureless continnum as in Mrk 477. Unification models of Seyferts have been proposed in which Seyfert 2's present a very thick torus which hides direct view of the BLR as a result of its inclination (see, for instance, Antonucci \& Miller 1985; Krolik \& Begelman 1988). Our conclusion from $\S$2.4.3 is that the ionized gas regions responsible for the emission lines are  at least partly covered by large amounts of gas and dust with Ly$\alpha$ emerging from the least covered regions. According to this picture, we would expect the nuclear continuum to be reddened as well. We will present briefly some arguments which favor continuum reddening, a possibility earlier studied by Boisson \& Durret (1986) (see also Carleton et al. 1987), although the idea of an extinction free continuum is more commonly accepted (e.g., FO86, KAW3).

	KAW3 performed an important test on the photon budget of the nuclear engine by extrapolating the power laws which they had fitted to their IUE spectra and deriving the number of ionzing photons $N^{obs}_{ion}$ which originates from the nucleus (assuming an isotropic source). After comparing with the number of recombination photons deduced from the H$\beta$ luminosity $N^{der}_{rec}$ (using the {\it dereddened} H$\beta$ flux and assuming a covering factor of unity), they found that most objects showed a clear deficit of ionizing photons ($N^{der}_{rec}/N^{obs}_{ion} >$ 1), which they concluded as favoring the ``occultation/reflection picture'' of Antonucci \& Miller (1985). According to this picture, the ionizing photons stream out freely only along the axis of the torus  axis and are sufficently numerous to account for the luminosity of the HI recombination lines (although the continuum observed in our direction [off-axis] is too weak to account for them).

	Although the ``occultation/reflection picture''  is supported by other independent arguments, there remains, however, an unresolved paradox (see Binette, Fosbury \& Parker 1993): if H$\beta$ of the large NLR is so much reddened (and considering a closed geometry for the NLR), why is the nuclear continuum of comparatively pointlike size not reddened at all? Interestingly, using other IUE Seyfert observations, Boisson \& Durret (1986) as well as Carleton et al. (1987) concluded that there is no deficit of ionizing photons if dereddening of the nuclear continuum is taken into account. The imporant argument against continuum reddening given by KAW3 is that the 2175 \AA\ (10$^{15.14}$ Hz) dust absorption feature was not present in their featureless continua. Although this seems to be clearly the case for objects like NGC 1068, we are not strongly convinced that this feature is totally absent from some of their spectra given  the noise level present. In particular, objects like Mrk 3 and possibly NGC 4388 show plausible hints of such a feature. Furtermore, the abscence of the 2175 \AA\ feature is not so compelling if we consider the possiblity that the properties of the dust in AGNs might differ from that in the Galaxy. After all, we know that dust grain properties depend strongly upon the environment in which they form (Whittet 1992).

\begin{figure}
\includegraphics{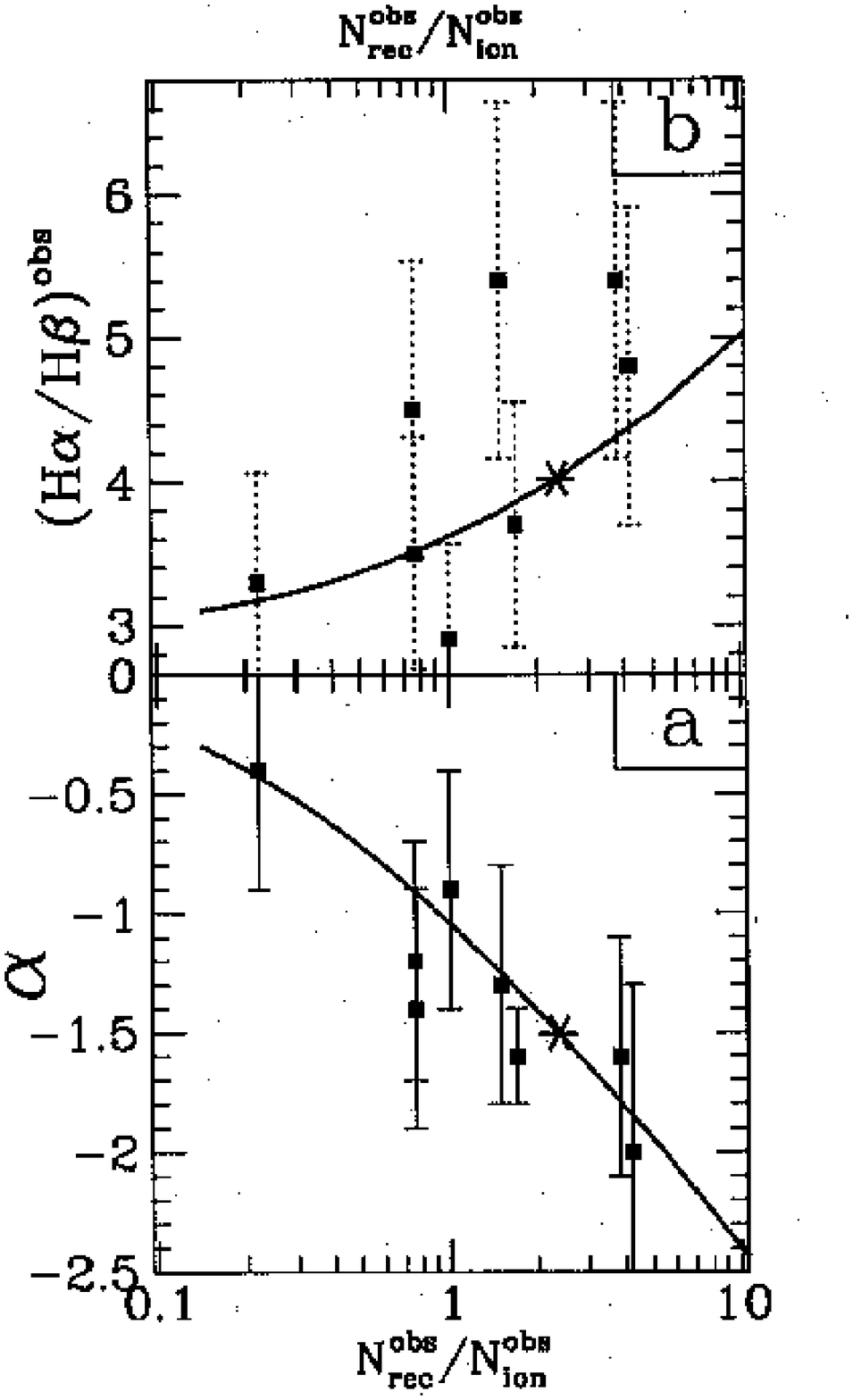}
\vspace{4in}
\caption[]{}{\small In panel $(a)$, the power-law index $\alpha$ vs. the ratio $N^{obs}_{rec}/N^{obs}_{ion}$. The spectral index $\alpha$ corresponds to the power-law fit by KAW3 of the observed continuum energy distribution between 2600 \AA\ and 1300 \AA\ . In panel {\it (b)}, the observed Balmer decrement is plotted as a function of $N^{obs}_{rec}/N^{obs}_{ion}$. The ratio $N^{obs}_{rec}/N^{obs}_{ion}$ differs from KAW3 in that the {\it observed} value of the H$\beta$ flux is used to derive the number of recombination photons instead of using the reddening-corrected H$\beta$ flux.  The solid line is a model representing the effect of increasing dust extinction on an intrinsic power-law ionizing continuum of index $\alpha=-$0.3 assuming a covering factor of 15\% (panel $[a]$) and an intrinsic Balmer decrement of 3.1 (panel $[b]$). The apparent index $\alpha$ of the calculated transmited spectrum is derived by simply using the flux ratios at 2450 \AA\ and 1310 \AA\ . The star represents the locus of a model with a shell dust extinction of $\tau_V$=1.3.}

\end{figure}

	One way of addressing the issue of extinction of the nuclear ``nonstellar'' continuum is to look at possible hints of correlation of the inferred deficit of ionziing photons with the IUE spectral index as well as with the amount of reddening inferred from the Balmer decrement. We present in Figure 15 the results of such an attempt based upon the KAW3 data set. Note that we plotted in abscissa the quotient $N^{der}_{rec}/N^{obs}_{ion}$ which differ from KAW3 in that it is based on the observed H$\beta$ {\it not} corrected for reddening (since we want to model all the observed quantities as a function of extinction). Although the correlations of Figures 15$a$ and 15$b$ are not statistiaclly significant, given the magnitude of the errors, the data is not inconsistent with a simple reddening model ({\it solid line}) described below and applied to an intrinsic power law of index $\alpha=-$0.3 (assuming the gas covering factor and the continuum's shape to be intrinsically the same in {\it all} the objects).

	To obtain the theoretical results represented by the solid line in Figure 15, we illuminate isotropically a slab of dust with opacity $\tau_V$ with an $\alpha$=-0.3 power-law continuum and a one-parameter sequence of slab models in $\tau_V$. For eah model, we compute the emergent reddened $\alpha$ (defined by the flux ratios at 2450 \AA\ and 1310 \AA\ ) in the calculated transmitted dust extinction ($\tau_V$ on the observed $N^{der}_{rec}/N^{obs}_{ion}$~ starting with an unreddend value ($N^{der}_{rec}/N^{obs}_{ion})_0$=0.15, corresponding to a gas covering factor of 15\% (varying the covering factor simply slides the curve horizontally). In figure 12$b$ we present the intrinsic ratio of 3.1. The standard extinction curve has been assumed in our model. Details on the transfer solution for the continuum and for the lines are given in Appendix C.

	We have assumed the impinging flux to be isotropic in order to approximate a shell layer of dust illuminated by a pointlike central source; a more appropriate treatment which distinguishes the isotropic optical lines from the radio continuum flux will be the subject of subsequent work. Because both the continuum and H$\beta$ are reddened by the same amount of dust, the equivalent widht of H$\beta$ is independent of $\tau_V$ in our slab treatment. Although the index $\alpha=-$0.3 is rather flat, it is not at odds with the range -0.08 to +0.11 denoted by Shuder (1981) (for the case of a covering factor of 10\%) in his study of the proportionality between continuum and H$\alpha$ luminosity in various clases of AGNs (see also Binette, Fosbury \& Parker 1993).

	Comparing Figures 15$a$ and 15$b$, the slope of the model in the case of the Balmer decrement panel seems too low. This could be a consequence of an 
inhomogeneous covering of the nuclear dust as proposed in $\S$2.4.3 for the HI lines (Figs.~13 and 14). Since $N^{der}_{rec}/N^{obs}_{ion}$ and $\alpha$ depend mostly on the UV part of the spectrum which is very sensitive to small amounts of dust, these quantities are expected to be strongly weighted towards the lest obscured regions, while the optical Balmer lines may merge from more obscured regions.

	Despite the recognized validity of the ``occultation/reflection picture'' in explaining cones of ionization and hidden polarized BLR in some Seyfert 2's, it remains important to test further whether the nuclear continuum is as reddening free as generally assumed and, therefore, whether part of the apparent deficit in the ionizing radiation does not simply result form dust extinction.

\vspace{0.5cm}

\newpage

\addcontentsline{toc}{part}{Bibliography}
\parskip=1ex
{\Large \bf References}

\vspace{0.5cm}

\noindent Anders E., Grevesse N., 1989, Geochim. Cosmochim. Acta, 53, 197 

\noindent Antonucci R.R., Miller J.S., 1985, \apj 297, 621 

\noindent Baldwin J.A., Ferland G.J., Martin P.G., Corbin M.R., Cota S.A., Peterson B.M.,
Slettebak A., 1991, \apj 374, 580 

\noindent Barvainis, R., 1992, \apj 400, 502 

\noindent Binette L., Dopita M.A., Tuothy I.R., 1985, \apj 297, 476

\noindent Binette L., Robinson A., Courvoisier T.J.L., 1988, {\it A\&A}, 194, 65

\noindent Binette L., Calvet N., Cant\'o J., Raga A.C., 1990, \pasp 102, 273

\noindent Binette L., Magris C.M., Martin P.G., 1993a, {\it Ap\&SS}, 205, 141

\noindent Binette L., Wang J.C.L., Zuo L., Magris C.M., 1993b, \aj 105, 797 (BWZM)

\noindent Binette L., Fosbury R.A.E., Parker D., 1993, \pasp 105 1550

\noindent Boisson C., Durret F., 1986, {\it A\&A}, 168, 32

\noindent Bruzual A., Magris C.M., Calvet N., 1988, \apj 333, 673

\noindent Carleton N.P., Elvis M., Fabbiano G., Willner S.P., Lawrence A., Ward M., 1987, \apj
318, 395
 
\noindent Chandrasekhar S., 1960, Radiative Transfer (New York: Dover)

\noindent Cimatti A., di Serego Alighieri S., Fosbury
R.A.E., Salvati M., Taylor D., 1993, \mnras 264, 421

\noindent Clavel J., Wamsteker W., Glass I.S., 1989, \apj 337, 236

\noindent Cohen R.D., 1983, \apj 273, 489

\noindent Crosas M., Weisheit J.C., 1993, \mnras, 262, 359

\noindent De Zotti G., Gaskell C.M., 1985, {\it A\&A}, 147, 1

\noindent DeRobertis M.M. \& Osterbrock D.E., 1984, \apj 286, 171

\noindent DeRobertis M.M. \& Osterbrock D.E., 1986, \apj 301, 727

\noindent Draine B.T., 1978, {\it ApJS}, 36, 595

\noindent Ferland G.J., Netzer H., 1979, \apj 229, 274

\noindent Ferland G.J., Netzer H., 1983, \apj 264, 105

\noindent Ferland G.J., Osterbrock D.E., 1985, \apj 289, 105

\noindent Ferland G.J., Osterbrock D.E., 1986, \apj 300, 658

\noindent Filippenko A.V., 1985, \apj 289, 475

\noindent Fosbury R.A.E., 1993, in The Nature of Compact Objects in AGN, ed. A. Robinson \& R.J.
Terlevich (Cambridge: Cambridge Univ. Press)

\noindent Gaskell C.M., 1984, {\it Ap.Letters}, 24, 43 

\noindent Gaskell C.M., Ferland G.J., 1984, \pasp, 96, 393

\noindent Halpern J.P., Steiner J.E., 1983, \apj 269, L37

\noindent Hummer D.G., Kunasz P.B., 1980, \apj 236, 609

\noindent Hummer D.G., Storey P.J., 1987, \mnras 224, 801

\noindent Johnson L.C., 1972, \apj 174, 227

\noindent Kinney A.L., Antonucci R.R.J., Ward M.J., Wilson A.S., Whittle M., 1991, \apj 377, 100

\noindent Krolik J.H., Begelman M.C., 1988, \apj 329, 702

\noindent Kwan J., Krolik J.H., 1981, \apj 250, 478

\noindent Lilly S.J., 1988, \apj 333, 161

\noindent Magris C.G, 1985, in senior physics thesis, Universidad Sim\'on Bol\'\i var and CIDA,
Venezuela

\noindent  Magris C., G., Binette L., \&
Martin. P. G., 1993, in proc. The Nearest Active Galaxies, ed. J. Beckman;
Astrophysics and Space Science, 205, p. 141

\noindent Malkan M.A., 1983, \apj 264, L1

\noindent Martin P.G., Rouleau F., 1991, in Extreme Ultraviolet Astronomy, ed. R.F. Malina \& S.
Bowyer (Oxford: Pergamon), 341

\noindent  Morganti R., Robinson A., Fosbury
R.A.E., di Serego Alighieri S., Tadhunter C.N. Malin D.F., 1991,
\mnras 249, 91 

\noindent Netzer H., ., 1993, in Proc. Madrid Meeting on The Nearest Active Galaxies, ed. J.E. Beckman, H. Netzer \& L. Colina, 219

\noindent Netzer H., Laor A., 1993, \apj 404, L51

\noindent Neufeld D.A., 1990, \apj 350, 216

\noindent Neufeld D.A., 1991, \apj 370, L85

\noindent Osterbrock D.E., 1989, Astrophysics of Gaseous Nebulae and ACtive Galactic Nuclei
(Mill Valley: University Science Books)

\noindent Pelat D., Alloin D., Fosbury R.A.E., 1981, \mnras 195, 787

\noindent Pier E.A., Krolik J.H., 1992, \apj 401, 99

\noindent Puetter R.C., Hubbard E.N., 1987, \apj 320, 85

\noindent Sanders D.B., et al. 1989, \apj 347, 29

\noindent Shuder J.M., 1981, \apj 244, 12

\noindent Shull J.M. \& van Steenberg M., 1985, \apj 298, 268

\noindent Slater G., Salpeter E.E., Wasserman I., 1982, \apj 255, 293

\noindent Tadhunter C.N., Robinson A., Morganti R., 1989, in ESO Workshop on 
``Extranuclear Activity in Galaxies'', ed. E.J.A. Meurs, Fosbury R.A.E. Fosbury, ESO Conf. and Workshop Proc. No.32, Garching, p.293

\noindent Viegas S.M., Prieto A., 1992, \mnras 258, 483 

\noindent Whittet D.C.B., 1992, Dust in the Galactic Environment (Bristol:IOP)

\chapter{Calcium depletion and the presence of dust in large scale nebulosities in radio galaxies (I).}
\centerline{Villar-Mart\'\i n \& Binette 1995, A\&A, in press}
\vspace{0.3cm}
\pagestyle{myheadings}

\markright{Internal dust in Large Scale Nebulosities in RGs}

\vspace{0.5cm}

{\LARGE  \bf Abstract}
 
\vspace{0.5cm}

	We show that the study of the Calcium depletion is a valid and
highly sensitive method for investigating the chemical and physical
history of the very extended ionized nebulae seen around radio
galaxies (EELR), massive ellipticals and `cooling flow' galaxies.  By
observing the near IR spectrum of nebular regions characterized by low
excitation emission lines (LINER-like), we can use the intensity of
the [CaII]$\lambda\lambda 7291,7324$\AA\ doublet --relative to other
lines, like H$\alpha$-- to infer the amount of Calcium depletion onto
dust grains. The presence of dust in these objects --which does not
necessarily result in a measurable level of extinction-- would favour
a `galactic debris' rather than a `cooling flow' origin for the
emitting gas. Before applying such test to our data, we study four
possible alternative mechanisms to dust depletion and which could have explained
the absence of the [CaII] lines: a) ionization of Ca$^+$ from its
metastable level, b) thermal ionization of Ca$^+$, c) a high
ionization parameter and/or a harder ionizing continuum than usually
assumed and d) matter bounded models associated to a hard ionizing continuum. 
We show that none of these alternative mechanisms explain the
absence of the [CaII] lines, except possibly for the highly ionized
EELR where  a high ionization parameter is required combined with a soft power
law. We thus conclude that for the other low
excitation emission regions (cooling flows, liners, low excitation EELR), the
absence of the CaII lines {\it must} be due to the depletion of Calcium onto 
dust grains.

\section{Introduction}

The study of the interstellar medium (ISM) of external galaxies
provides important information about the global kinematic (inflow,
outflow) of such gas, its chemical composition and the implied star
formation history, its mass distribution, etc.  This gas forms a vital
part of the record of the formation of the parent galaxy, and the
evolutionary processes involved.

	How can this material be studied in details? One way is to
have it illuminated or excited by a powerful AGN, gi\-ving rise to the
phenomenon of extended emission line region (EELR). The
drawback of course is that it only allows us to look at a restricted
class of galaxy. The large scale EELR phenomenon is observed in a
majority of the most powerful radiogalaxies with EELR extending to
radial distances of up to 100 kpc from the nuclei (Tadhunter 1986;
Baum et al. 1988), much larger radii than the stellar population
distribution of the parent galaxy. The morphologies and kinematics of
such regions cover the full range from regular disc/ring systems to
chaotic systems for which no pattern can be discerned.  Their spectra
show strong emission lines, covering a wide range in ionization.  It
is generally accepted that the EELR are ionized by some mechanism
connected with the nuclear activity, but there is no full consensus on
the excitation mechanism.  Some objects show evidences for an
interaction between the radio jets, which transport ener\-gy to the
outer radio lobes, and the gas in the outer region. Maybe this
interaction is responsible of the excitation of the gas through some
kind of shocks (Sutherland, Bicknell \& Dopita 1993).  For other
objects, that do not show evident spatial coincidence between the
radio structures and the EELR, the excitation might be due to direct
photoionization by the nuclear ionizing radiation field (Robinson
et~al. 1987: hereafter RBFT87). A reduced scale version of the EELR is
the one observed in many Seyfert galaxies (Haniff, Ward \& Wilson
1988) where the ionized gas may extend up to a few kpc although it is
brightest within the central 100--300pc. The EELR morphology tends to
be conical (Wilson \& Tsvetanov 1994). Its detailed observation is
complicated by the presence of the very luminous stellar background of the
bulge. In normal bright ellipticals, quite weak extended nebulosities
of low excitation are a common phenomenon (Buson et~al. 1993, Goodfroij
1994). The very large scale gas around radio-galaxies, on which we
focus here, presents the advantage that the lines are observed against
the sky background rather than against the bright parent galaxy bulge.

	The origin of the gas making up these {\it large scale}
extended nebulosities remains unknown. Furthermore, the distinction
between these  and the filamentary ne\-bulae seen in some clusters and around
some massive ellipticals and often identified with cooling flows, is
unclear.  We do know, however, that this gas in every case is
chemically enriched as compared to primordial gas. Emission line
analyses (RBFT87) show common element abundances to be
within a factor of a few ($\le$) of Solar and also to be rather
uniform over all the objects observed.

	The two most likely explanations for the origin of the material  are:

	a) debris from recent tidal interactions and mergers

	b) gas cooling from the hot ($\sim$ virial) phase which from X-ray
observations (Forman, Jones \& Tucker 1985) has been shown to exist around massive
ellipticals and inside galaxy clusters

The main arguments behind these explanations are the following:

	a) Heckman et~al. (1986) showed that a large fraction of powerful radiogalaxies
have morphological features --shells, tails, loops, etc-- similar to those produced in
numerical si\-mulations of galaxy interactions ({\it e.g.} Toomre and Toomre 1972, Quinn
1984). This could indicate that the activity has been triggered either because fresh
gas has been accreted from outside or because preexisting gas in the galaxy has been
caused to collapse to the core as a result of the interaction. This interaction
scenario is also supported by observations of a few nearby radio   galaxies which,
apart from morphological peculiarities, show large misalignments between the stellar
and the gaseous rotation axes, indicative of an external origin for the gas.

 	b) Hot X-ray corona ($T \ge 10^4$K) are a common feature of bright early-type
galaxies.  Within some critical radius, radiative cooling becomes important, leading
to the deve\-lopment of the so-called `cooling flow' hypothesis (Nulsen, Stewart \&
Fabian 1984, Thomas et~al. 1986) Eventually, condensations or filaments
could be formed, dense and cool enough to radiate detectable optical emission lines.
Most powerful radio galaxies are too distant for the characteristic X-ray emission to
be currently detectable, but it is quite plausible that the EELR gas has condensed
out of a surrounding cooling flow.  


	Distinguishing between these alternative hypotheses has been attempted using
gas kinematic measurements (Tadhunter, Fosbury \& Quinn 1989) which show that the
radio galaxy EELR generally have a high specific angular momentum which is difficult
to reconcile with the cooling flow picture. An alternative approach is to look for
the presence of dust associated with the EELR gas. If the gas has cooled directly
from a hot phase, there will have been no opportunity for dust to form, according to
the standard cooling flow theory. Any dust introduced from galaxies into the hot
intracluster medium will be rapidly destroyed (Draine \& Salpeter 1979). We would
need a mechanism to produce this dust once the gas has cooled down (Fabian, Johnstone \& Daines 1994).  If, on the other hand, the material has fallen in during a merger, 
the dust/gas ratio is expected to have a value appropriate to the gas chemical
composition found in normal galaxies.

Determining the presence or absence of dust is important because:

-of its relevance in deriving chemical composition which takes into account the
effects of depletion of metals unto dust grains

-of the implication for the star formation history and the ISM evolution: when does
dust form? (and so stars?)

-of the effects of dust on the apparent morphology of continuum and line features:
pure absorption (reddening) and/or scattering (blueing/polarization)

We discuss below how the absence of the forbidden [CaII] lines can be
used to infer the presence of dust mixed in with the emission gas. We
assess in detail all the most plausible {\it alternative} explanations
to that of internal dust for explaining the absence of [CaII]
lines. As no acceptable alternative solution is found, we conclude in
favour of the validity of the method initially proposed by Ferland
(1993). We however adapt and optimize the [CaII] dust detection
technique to the context of the EELR studies in which we are
involved. The observational results and their interpretation will be
presented in a subsequent paper.

\section{Outline of the method}

To investigate how the forbidden Calcium doublet of [CaII] in the
infrared is affected by the physical condition encountered in
photoionized plasma, we outline first the computer code which we used
and then proceed to illustrate how the [CaII] lines might be used to
infer dust in the ISM of galaxies and thus our interest in securing
these conclusions by closing up the possibility of alternative
interpretations.

	To compute the emission lines used in our study, we have used
the multipurpose photoionization--shock code MAPPINGS described in 
Chapter 2. The Ca$^+$ ion is considered as a five level
atom. Its structure is shown in Fig.~16.  

\begin{figure}
\includegraphics{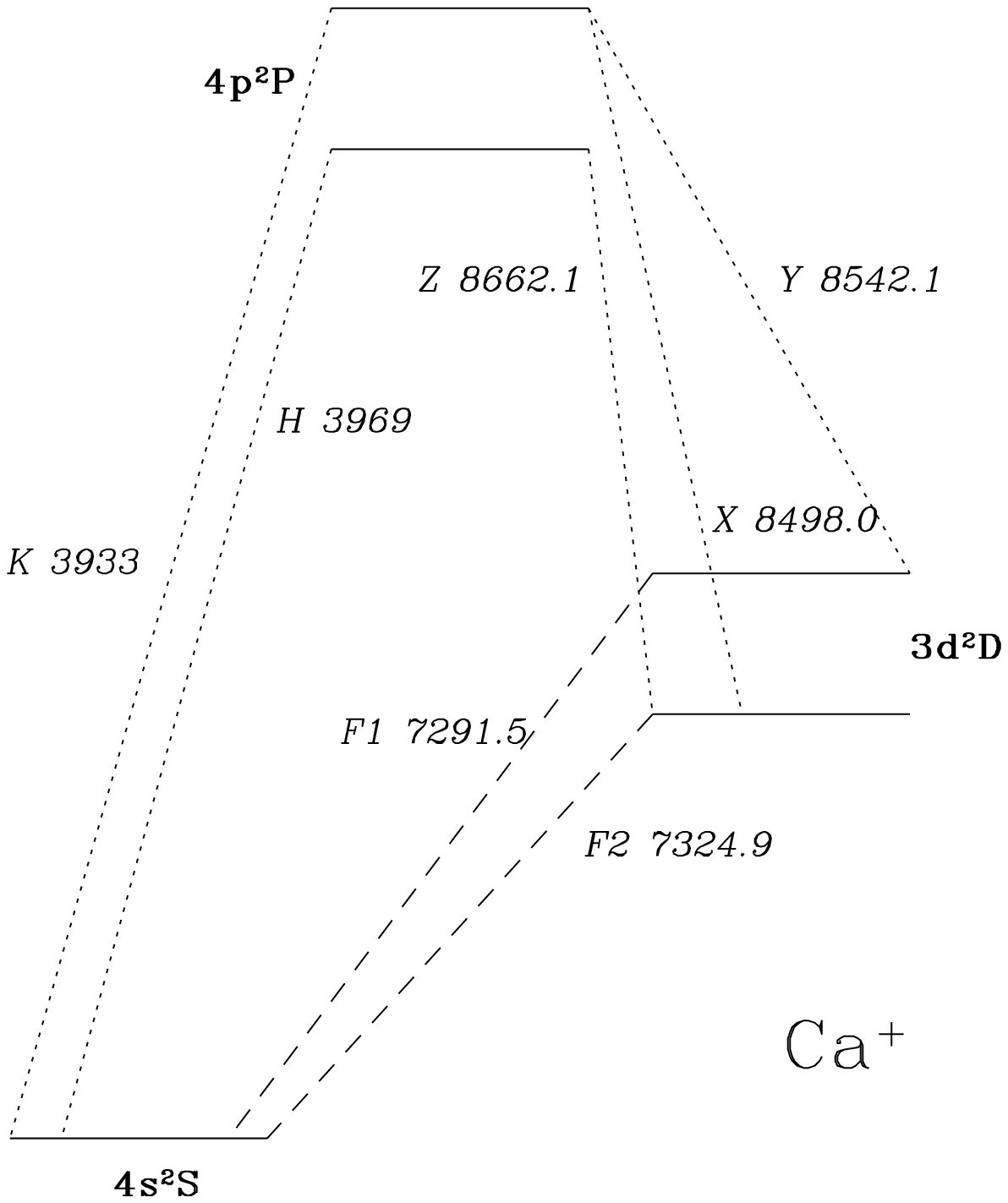}
\vspace{3in}
\caption[]{}{\small Energy level diagram for CaII showing the radiative transitions
among the five levels considered here. Permitted transitions are represented
with dotted lines and forbidden transitions with dashed lines. F1 and F2 are the 
forbidden doublet lines considered in this work.}
\label{fig-1}
\end{figure}

\subsection{The input parameters}

	We now proceed to define the parameters employed in the
calculations. Most of these are derived from our observational
knowledge of EELR although we recognize that it is incomplete and
can be biased by particular diagnostic tools which are employed.

\subsubsection{Metallicity}

For definiteness, we adopt a set of solar abundances (Anders \&
Grevesse 1989) for the trace elements, in line with the finding of
RBFT87 who indicated values for the EELR not radically
different from solar. The solar abundance of ({\it not} depleted) atomic
Calcium is $2.2 \times 10^{-6}$ by number relative to
hydrogen. Depletion into interstellar dust grains is known to reduce
this abundance in the local ISM by a factor of $\sim 5000$ (Whittet
1992) so that if depletion was taking place in the partially ionized
EELR plasma, even a small dust-to-gas ratio might be sufficient to
eliminate any detectable trace of atomic Calcium.

\subsubsection{Geometry}

	It's been accepted for a long time that the EELR
is formed by individual clouds that are ionized by the central
source.  In most models presented here, each emitting cloud is
considered  (see section 2.2.1) like a radiation bounded slab (optically thick to the Lyman
continuum) which comprises  a fully ionized region, closer to the
illuminated face and responsible for the high ionization lines and a
partially ionized zone (PIZ), where the low ionization lines like CaII
are emitted.   Matter bounded clouds will be considered in Sect.~3.3.4. As we want
to concentrate on the depletion phenomena, we will simplify the
calculations by not considering any neutral region (see Fig.~4, Chapter 2)
beyond the PIZ which would contain dust, the effect of which would be
to cause additional extinction for an hypothetical and unfavorably
placed observer looking from the back of the slab. We have verified
that the results reached here using line ratios of similar wavelengths
are not altered by the presence or not of this neutral absorbing zone.

\subsubsection{Gas density}

	We adopt a representative density for the EELR clouds of 300
cm$^{-3}$. Typical electron densities values measured vary between a
few tens (or less) to a few hundreds.  The densities derived from
forbidden line ratios might not apply to every subregion, but they are
sufficiently low to consider the low density limit as a generally valid
and very good approximation to the physical conditions affecting the
CaII lines of large scale EELR. In the low density regime, the effects
of density variations on the emission line ratios which are considered
in this paper are very small by comparison to the effects of other
parameters like the ionization parameter or the hardness of the
ionizing continuum.

The calculations consider the gas pressure to be cons\-tant within the
cloud (isobaric models) with the density behaviour modulated with
depth into the cloud by the behaviour of the temperature and by the
ionization fraction of the gas.

\subsubsection{The ionizing continuum and the ionization parameter U}

	We implicitly assume that the dominant ionization me\-chanism
of the ionized gas in radio galaxies is photoionization. It has been
shown by RBFT87 that a hard continuum extending well
down into the soft X-ray region, despite some discrepancies with the
observed spectra, can be considered to reproduce generally well the
measured line ratios. This continuum can take its source in the active
nucleus or be locally generated ({\it e.g.} fast shocks: Binette, Dopita,
Tuohy 1985). We have considered power law (PL) distributions of various
values of $\alpha$ ($F_\nu \propto \nu^{+\alpha}$) as well as hot
blackbodies (BB) of temperature T$_{bb} \approx 10^{5}K$ in order to study
the effects that hardness has on our conclusions.

	The ionization parameter, a measure of the excitation level of
the ionized gas, was defined in section 2.2.1.

	$$ U_f = \frac{1}{cn_H^f} \int {\frac{\phi^f_{\nu}}{h\nu} d\nu} = 
\frac{\phi^f_{\nu}}{cn_H^f}, $$

quotient between the density of impinging ionizing photons and the density of the gas
cloud.

	We find that the parameters having the strongest effect on the
line spectrum are the ionization parameter U and the mean ionizing
photon energy ({\it i.e.,} hardness of the continuum).

\subsection{How to detect dust}

	How can we detect dust within the gas associated with EELR? We here
summarize different techniques and compare their sensitivities. 

	a) Reddenning: given the nature of the gas distribution and
the fact that it is ionized externally (unlike HII
regions which are internally excited), the extinction may not 
necessarily be sufficiently high to be easily detected using optical
observations. Indeed, the EELR ratios studied by RBFT87 show little or
no reddening.  Furthermore a small enhancement of the Balmer decrement
over recombination case~B might be interpreted as resulting from
collisional excitation rather than from reddening.

	b) Scattering: polarization measurements of high redshift
radio galaxies (di Serego Alighieri et al. 1989, Januzi \&
Elston 1991, Tadhunter et al. 1992) show conclusive evidence
for scattered nuclear light over large volumes and, although there are
reasons to believe that the scattering medium is dust, it is difficult
to rule out entirely Thomson scattering by electrons. Detailed
studies of the low redshift galaxy PKS2152-69 do, however, show
polarized continuum radiation from highly excited extranuclear gas
cloud with an energy distribution which is so blue that it must arise
from dust scattering (Fosbury et al. 1990).

	c) Infra-red thermal dust emission: the IRAS satellite has
shown that many galaxies radiate significant fractions of their energy
in the far infrared sprectral region. Significant masses of dust at
temperatures of around 40K are responsible for this radiation at
wavelengths of 60$\mu$m and beyond. In many cases the FIR spectral
energy distributions is still rising at 100$\mu$m, out of the spectral
range detectable by IRAS. The cool dust can only be detected at
millimetre and submillimetre wavelengths. A strong limitation of this
technique is the very poor spatial resolution of the IRAS satellite.
Groundbased studies of the far IR emission of galaxies in the sub-mm
range also exist ({\it e.g.} Clements, Andreani \& Chase 1993), but
still with poor spatial resolution.  Although it is possible to infer
masses and temperatures of the warm and cool dust components
(dependent on models), the spatial distribution of the dust is not
known.

	d) Indirect effects on the line spectrum. There are se\-veral
ways this can happen: effects of dust on the gas temperature
(photoionization of dust grains may raise the temperature of the
plasma), effects of dust on the ionization structure (dust grains selectively
absorb ionizing photons of lower energies), and influen\-ce on apparent chemical
composition via the depletion of refractory elements onto dust grains.
The first effect does not provide a unique interpretation for the
unusually high temperatures seen in some EELR (Tadhunter, Robinson \& Morganti 1989) while the second effect cannot be discriminated
against reliably since even dust-free photoionization models are still
too uncertain to be used as absolute reference point. For these
reasons, the last effect is the only clearly promising one and is
looked into details below.

\begin{figure}
\includegraphics{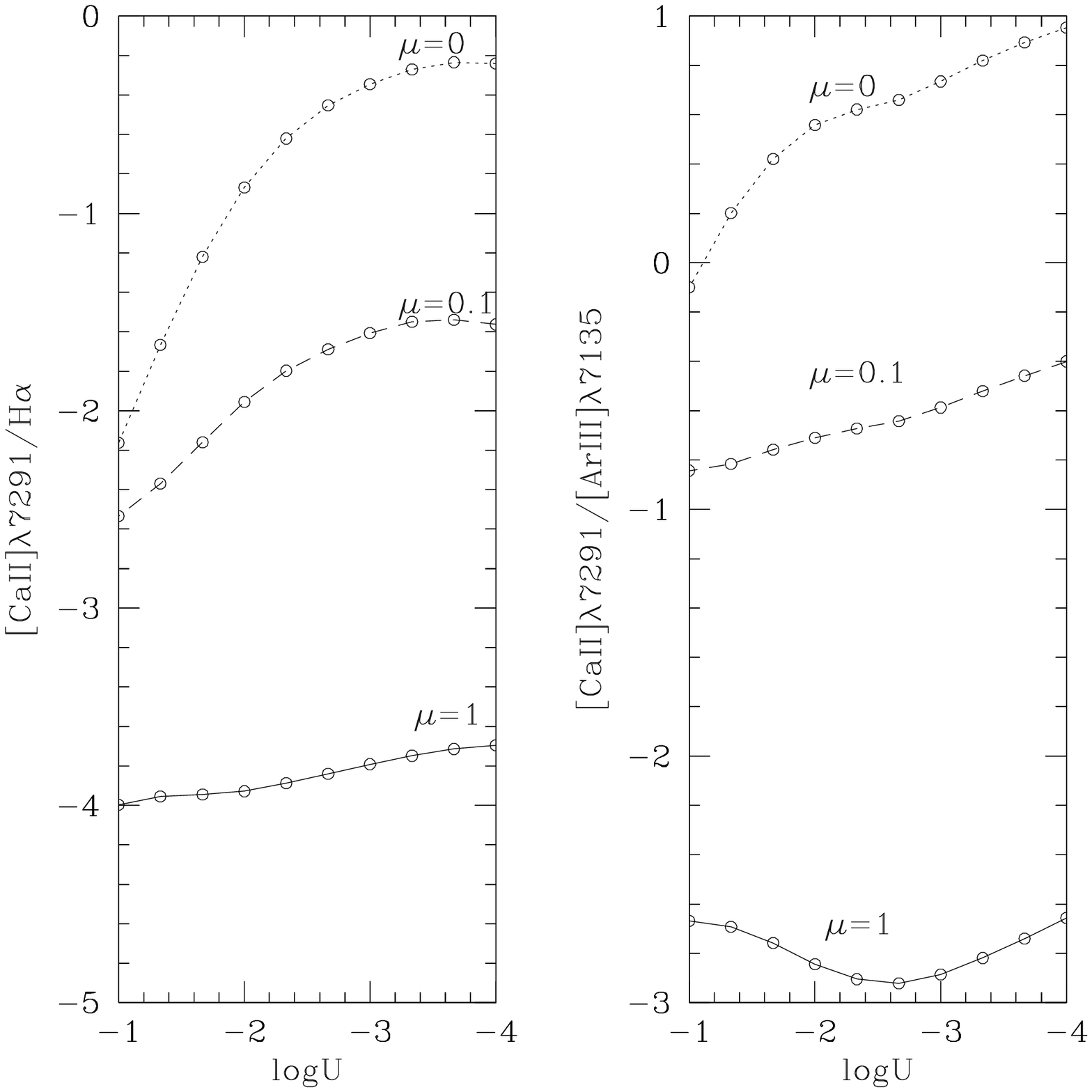}
\vspace{2.5in}
\caption[]{}{\small MAPPINGS photoionization models showing the strenght of [CaII]$\lambda7291$
relative to H$\alpha$ (left panel) and [ArIII]$\lambda$7135 (right panel)
as a function of the ionization parameter $U$. The models assume solar
abundances but with different amounts of dust-to-gas ratios $\mu$
expressed in solar neighborhood units.  The gas phase abundances of
Ca/H for the $\mu = 0$, 0.1 and 1.0 models are $2.2 \times 10^{-6}$
(no depletion), $9.2 \times 10^{-8}$ and $4.4\times 10^{-10}$ (Whittet
1992), respectively, after taking into account depletion. The hydrogen
density is 300 $cm^{-3}$ and the ionizing power law index is $\alpha =
-1.4$}
\end{figure}

Calcium is very sensitive to the presence of dust as it is always
found to be depleted in the interstellar medium ({\it e.g.} Crinklaw,
Federman \& Joseph 1994).  Photoionization calculations appropriate
to LINERS --a hard ioni\-zing spectrum with a relatively low
ionization parameter-- invariably predict the [CaII]$\lambda\lambda
7291,7324$\AA\ doublet (4$s^2$S-3$d^2$D) to be very strong (Ferland
1993).  These two forbidden lines (hereafter
F1 $\equiv\lambda7291$\AA\, F2 $\equiv\lambda7324$\AA\ ) have a high
critical density $\sim 10^6$ cm$^{-3}$. The latter line, 7324\AA\, is
the weakest of the doublet and is furthermore blended with the
[OII]$\lambda7325$\AA\ multiplet. The other line, $\lambda7291$\AA\,
lies some 30\AA\ shortward of [OII] and is therefore straightforward to
isolate given reasonable spectral resolution.  The fact that any of
these doublet lines are generally not seen in \-LINERS but are so in
some novae when their envelopes have reached the appropriate ionization
level can be interpreted as evidence of Calcium depletion onto dust
grains in the former objects. Since EELR which are the subject of our
investigation are often seen in the ionization parameter region of the
line ratio diagnostic diagrams occupied by LINERS ($10^{-4}\leq U\leq
10^{-3}$), we can similarly use the [CaII] doublet measurements to
infer whether or not there is depletion taking place in EELRs and thus
conclude whether dust is also mixed with the gas as is thought to be
the case in LINERs.

	 The above arguments have already been used for se\-veral
objects with `cooling flow' filaments by Donahue \& Voit (1993) to
infer the presence of dust mixed with the ionized gas. Ferland (1993)
has shown the great sensitivity of this method to the presence of dust
under NLR conditions.  We now show it to be also the case under EELR
conditions. We present two diagnostic diagrams in Fig.~17, the ratio
[CaII]/H$\alpha$ (7291/6563) and the ratio [CaII]/[ArIII] (7291/7135)
as a function of the ionization parameter $U$.  The variable parameter
distinguishing the three different sequences in U is $\mu$, the
dust-to-gas ratio of the plasma expressed in units of the solar
neighborhood dust-to-gas ratio. The dramatic difference in line ratios
between the grain depleted Ca/H ($\mu > 0$) and the undepleted case
($\mu = 0$), shows the sensitivity of this method, particularly for
low values of $U$.


\section{Possible alternative explanations to depletion.}

	 Before we carry the conclusions of the current analysis to
the interpretation of our observations (Chapter 4), we report first on our effort in investigating other possible
alternative mechanisms to dust depletion. If the warm Ca$^+$ region
predicted by standard models does exist, then Calcium depletion
becomes the only reasonable explanation for the non detection of the
doublet lines. What we consider in this section is the possible NON
EXISTENCE of the emitting [CaII] region by investigating different
mechanisms which could eliminate it. During
our investigation, we require however that successful models do not result in important
discrepancies with other observed line ratios. The mechanisms we have
considered to eliminate the [CaII] region are
	
	- Ionization of Ca$^{+*}$ by Ly$\alpha$ and soft continuum photons from the
metastable level of Ca$^+$

	- Thermal (collisional) ionization of Ca$^+$

	- Photoionization with a much harder continuum or a much
higher U than usually assumed

\subsection{Ionization by Ly$\alpha$ and soft continuum photons.}

	Wyse (1941) proposed that the ionization of CaII from the
me\-tastable level by Ly$\alpha$ photons, could explain the fact that
the IR lines of CaII at 8498, 8542 and 8662\AA\ appear in emission
near the maximum phase of Me variables, whereas the H and K lines only
occur in absorption.  Trapped Ly$\alpha$ photons could also play a
part in ionizing metastable Ca$^{+*}$ as suggested by Wallerstein
et~al. (1986).

	We investigate here if this process is important under the
conditions found in EELR clouds. In order to do this, we add two terms
to the ionization equilibrium equation of CaII. One which considers
photoionization of excited Ca$^{+*}$ by the impinging UV continuum. 
The other is photoionization of excited Ca$^{+*}$ by the nebular
Ly$\alpha$ photons.  The statistical equilibrium equations give the
relative population of the mestastable level, which turns out to be,
under EELR conditions, $\frac{n_{3d}}{n_{4s}}\sim 10^{-7}$, being
$n_{4s}$ the density of Ca$^+$ ions in the ground level. With such a
negligible population, the density of ionizing photons must be very
high to increase the ionization rate to a non negligible level as
compared to the ground state ionization rate.  A simple estimate
presented in Appendix D demonstrates this level to be out of reach.

	In summary, the very diluted radiation fields and the low
densities appropriate to the EELR implies an extremely small
population for the excited levels which prevents the ionization of
Ca$^{+*}$ by Ly$\alpha$ and soft continuum photons from being of any
significance.

\subsection{Thermal ionization of Ca$^+$.}

\begin{figure*}
\includegraphics{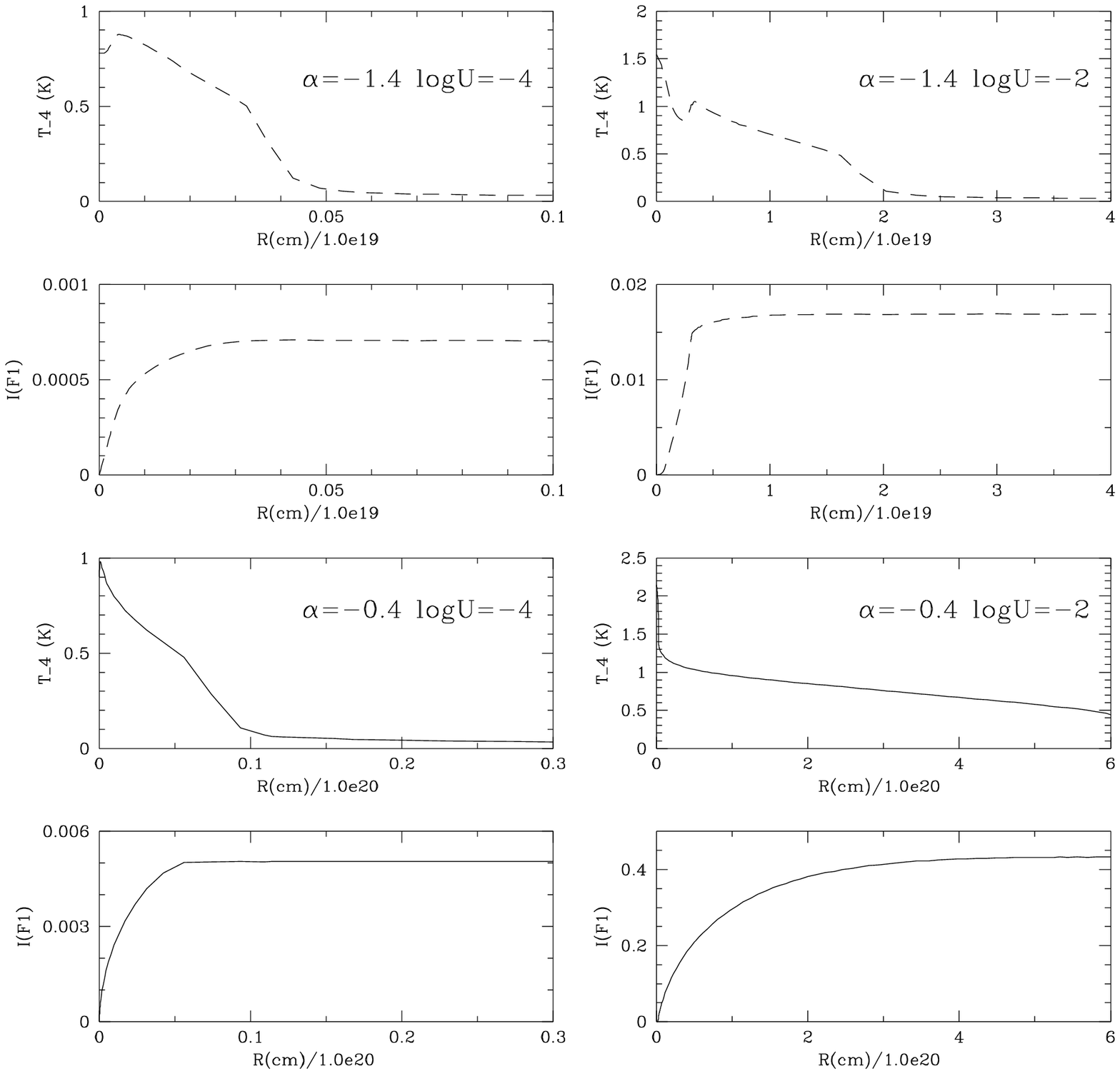}
\vspace{6.5in}
\caption[]{}{\small Distribution of temperature  and integrated intensity of F1
(in {\it ergs s$^{-1}$cm$^{-2}$}) line across an EELR cloud.  
 Dotted lines (top diagrams) correspond to a power law of index $-1.4$; solid
lines (bottom diagrams) to a power law of index $-0.4$. Each temperature diagram must be
compared with the one just below, to see the temperature that correspond to each point in
the nebula.}
\end{figure*}

	 We investigate here the possibility of collisional ionization
by thermal electrons of Ca$^+$ to Ca$^{++}$, a process which is
important when the electronic temperature becomes higher than 20000K
(Jordan 1969). In order to establish a comparison in U, we have
considered two extreme cases in our calculations, log$U=-4~ \& -2$. To
illustrate how a much harder continuum will result in much higher gas
temperatures, we also use two different PL of index $\alpha = -1.4$ and
$-0.4$.  Note that such a hard continuum as $\alpha =-0.4$ is probably
quite unrealistic. However, our intention here is simply to test
whether very high temperatures can be achieved with photoionization
models and specifically near the Ca$^+$ region. The results are shown
in Fig.~18 as a function of depth in the photoionized slab.  Of the
eight plots, the four upper ones correspond to $\alpha = -1.4$ while
the four at the bottom to $\alpha = -0.4$. The four plots on the
right, have logU$=-2$, and the four on the left, logU$=-4$. Two plots
therefore are shown for each pair of [U, $\alpha$] values: one is the
the temperature T4 in units of 10000K and the one immediately
underneath is the intensity of F1 ($erg.s^{-1}.cm^{-2}$), both as a
function of the depth in units of $10^{20} cm$ into the slab. These
plots allow us to see the corresponding electron temperature to the
position where the bulk of the [CaII]F1 emission takes place.

	For the traditional PL of index $\alpha=-1.4$, the electron
temperatures are not anywhere near high enough for the process of
thermal ionization to be relevant. Harder continua and increasing U
values do produce higher temperatures, but, unless U are
unrealistically high (even logU$=-2$ is not enough), the gas is never
hot enough. Thus, realistic photoionization models are {\it not}
able to heat the gas sufficiently to thermally ionize Ca$^+$. We might
conjecture that there could be an additional heating source like
shocks which could raise the temperature of the gas. This would be an
interesting point for further investigation.

\subsection{Effects on F1 of varying U and the continuum hardness. }

\begin{figure*}
\includegraphics{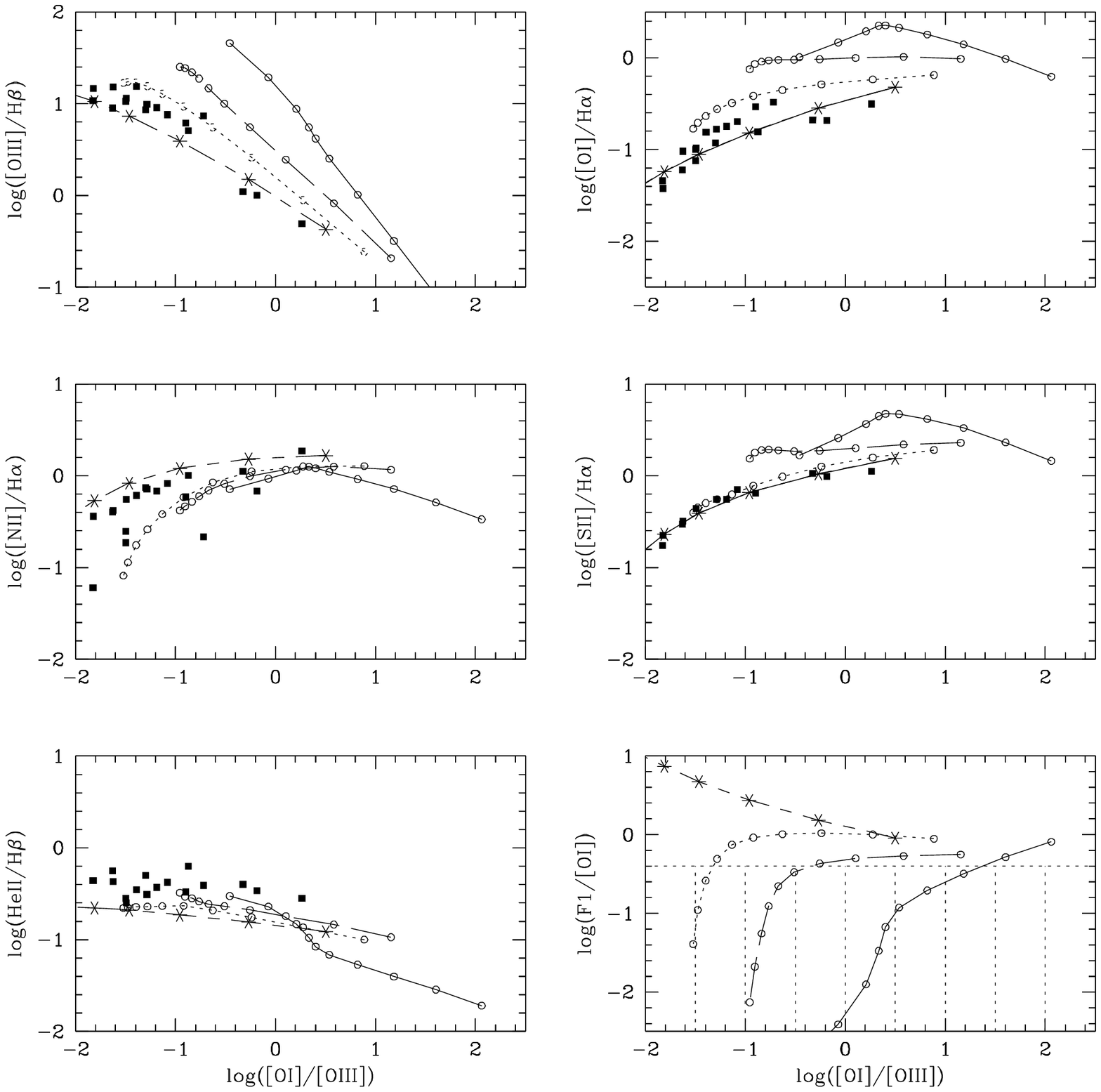}
\vspace{7in}
\caption[]{}{\small The curves correspond to ionization parameter
U model sequences for different ionizing continua: power laws (small
circles used as fiducial marks) of index $\alpha = -1.4$ (......),
$-1$ (---~---~), $-0.4$ ($\_~\_~\_~\_$) and a black body continuum of
$T_{bb}=1.2 \times 10^5K$ (*--*--*). The region represented by
vertical dotted lines in the last diagram shows the area where F1
would fall under the detection limit defined in Sect. 3.3.3. All the
models within this region are successful in accounting for the
absence of the [CaII] lines without requiring depletion by dust. Line
ratios of extended emission line regions observed by RBFT87
(uncorrected for reddening) are shown with solid squares. }
\end{figure*}

	We now investigate how the F1 line might become undetectable
by simply varying arbitrarily the ionization parameter  $U$ or the
continuum hardness. Fig.~19 shows six diagnostic diagrams with the abscissae
always representing the line ratio [OI]$\lambda6300$/[OIII]$\lambda5007$. The [OI]/[OIII] ratio
monotonically increases with
decreasing gas excitation (i.e., with decreasing U) and is therefore
a good measure of the
excitation level of the gas. Each dia\-gram
shows in the ordinate a different line ratio which can be related to a
given gas property. [OI]$\lambda6300$/H$\alpha$, for instance, might
measure the hardness of the continuum. In the last diagram, the
ordinate corresponds to the quotient F1/[OI]$\lambda6300$.  The three
sequences of models shown in each diagram differ by the slope of the
power law which takes on the values of $\alpha =-1.4, -1$ and $-0.4$. The
values of Log~U covered by each curve is in the range $-4$ to $-1$.
Our aim is to look for models which can decrease the F1 intensity
below the detection limit. Let's look at how we might define a
practical detection limit. The open squares in the diagrams of Fig.~19
represent line ratios measured by RBFT87 in several EELR. The faintest
line they measure is typically HeII$\lambda$4686. The mean ratio of
HeII$\lambda$4686/[OI]$\lambda$6300 observed is $10^{-0.4}=0.4$ for the
large scale nebulosities. We establish our 'artificial' detection
limit in the following way: since HeII is one of the weakest line
successfully measured by RBFT87, we will assume that any line fainter
than 0.4 below the [OI]$\lambda$6300 flux is not detectable. In
the last diagram, the region where F1 falls below this detection limit
is shown by a dash line grid. Any model found  in this area is deemed 
successful 
in explaining the non-detection of F1 without requiring depletion.

	We see that for the standard PL ($\alpha=-1.4$), only mo\-dels
with high U (log U$>-2$) decrease log(F1/[OI]) below --0.4.  These
models, as we can see in the diagrams, would therefore be valid only
for the high excitation EELR, but not for LINERs, cooling flow
filaments, or EELR of low and intermediate excitation. On the other
hand, increasing the hardness of the continuum (flatter power laws),
helps F1 to get fainter with respect to [OI]$\lambda$6300, but the
discrepancies with observed line ratios in other diagrams become
totally unacceptable (see top two diagrams of Fig.~19).

	It is interesting to compare a BB sequence ($1.2\times
10^5$K) with the canonical PL sequence $\alpha = -1.4$. We see that
both ionizing continua reproduce rather well the observed line ratios
as was earlier shown by RBFT87. From these line ratios alone, there
are no reasons to favour power laws over hot blackbodies. A similar
conclusion was reached by Binette, Robinson and Courvoisier (1988) for
the mean NLR spectrum of Seyferts.

	A BB produces a much stronger F1 compared to [OI]$\lambda$6300
than any of the power laws considered.  One reason for this is that
the fraction O$^0$/O$^+$ in the PIZ is completely controlled by the
charge exchange reactions of O$^0$ and O$^+$ with H$^+$ and H$^0$,
respectively, and not by direct photoionization of O$^0$. This is not
so for Ca$^+$/Ca$^0$ which is free to respond to the different amount
of hard photons (the only one to make it to the PIZ) available in a PL
or a BB.  We conclude from the last diagnostic diagram that the BB
models could not explain the absence of F1 from the observed spectra
without invoking depletion.

\subsection{Effects on F1 of truncating clouds }

\begin{figure*}
\includegraphics{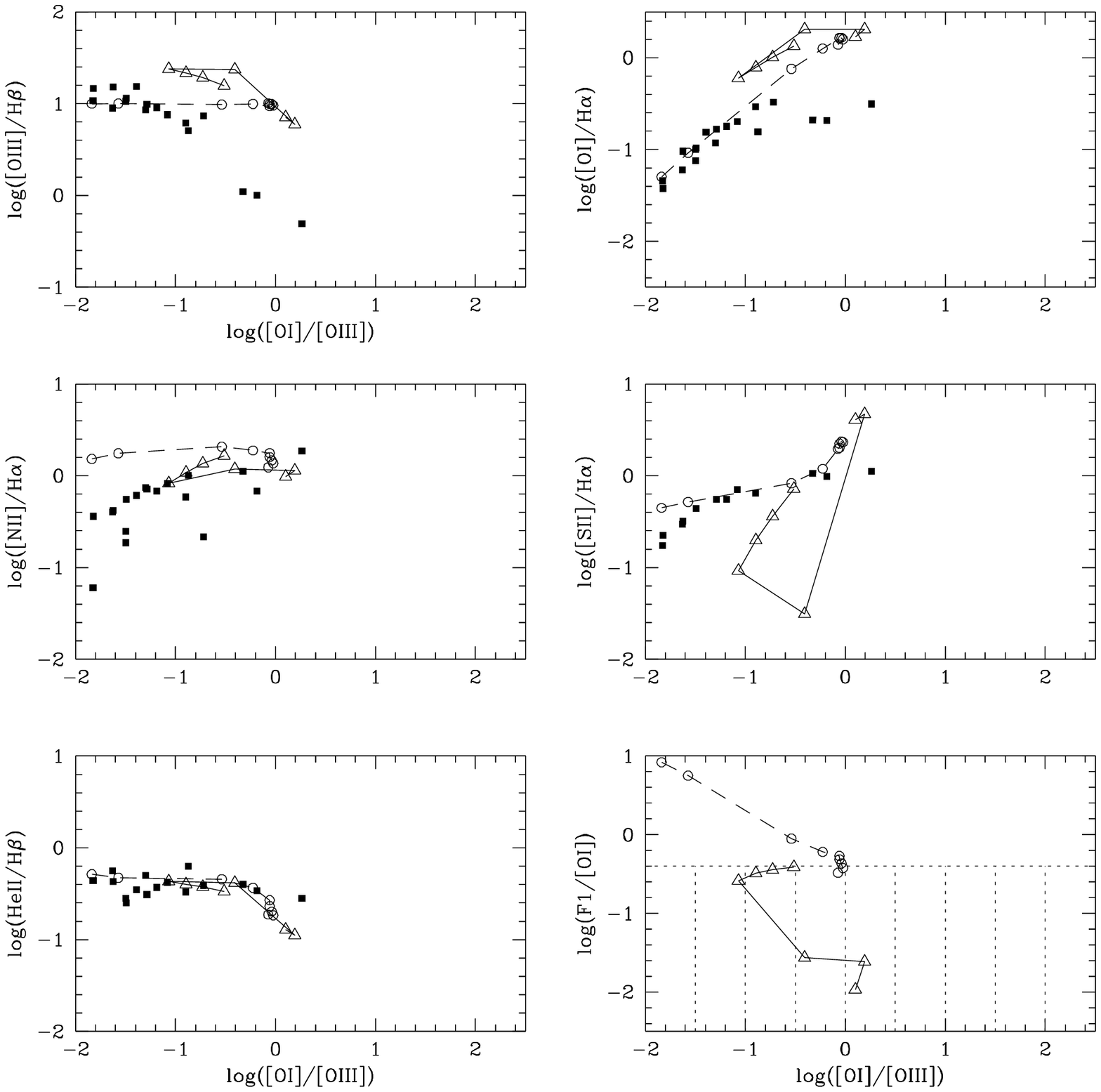}
\vspace{7in}
\caption[]{}{\small Same diagrams as in Fig.~19 but for truncated gas clouds and
only for hard power laws ($\alpha=-0.4$). The open circles (connected by the dashed line)
correspond to matter bounded clouds ($5.2\times 10^{-5} \leq U
\leq 5.2\times 10^{-4}$)
truncated at a depth which produce [OIII]/H$\beta =10$ (see text). 
The open triangles (connected by
the solid line) are similar models truncated at a
depth such that F1/[OI]$\leq$0.4 ($3.7\times 10^{-3} \leq U \leq 2.7\times
10^{-2}$). }
\end{figure*}

        We showed in the previous section that a harder continuum can
potentially bring F1 under the detection limit but results in important
discrepancies with the low excitation lines. The reason is that the
partially ionized zone (PIZ) where most of the low excitation lines
are generated gets larger and larger with increasing hardness of the
continuum. If the clouds were truncated, the smaller PIZ would
generate weaker low ionization lines, thus improving the overall fit.

        We have investigated models with $\alpha=-0.4$
which were truncated at a depth which satisfies a given
criterion based on a specific line ratio. This has been done in two
ways:

\noindent 1) The criterion in this case is to truncate the calculations
when OIII/H$\beta$ has reached a value of 10 which is the typical ratio for the
high excitation EELR. 

        The sequence of models shown in Fig.~20 are separated by a
factor of 0.14dex in U.  Altough there are still discrepancies, there
is a notable improvement compared to the radiation bounded models of
Fig.~19. The predicted line ratios are now located closer to the
observed data (same scale as in Fig.~19).  We conclude that models with
a hard ionizing continuum must be matter bounded in order to fit
acceptably most observed ratios.

        What happens now with F1/[OI]? In the last diagram of Fig~20, we see
that all these models produce F1 above the detection limits and cannot
therefore explain the non detection of Calcium.

\noindent 2) The second method consists on imposing that the models
produce F1 under the chosen detection limit (F1/[OI]=0.4) and check if
such models agree with the position of the observed line ratios in the
rest of the diagrams.

         The models in the sequence which satisfy this criterion are
found in the range $3.7\times 10^{-3} \leq U \leq 2.7\times 10^{-2}$.
They are represented in Fig.~20 as open triangles connected by a solid
line.  As we see in the top left diagram, the ratio [OIII]/H$\beta$ is
not any more defining a simple trend with excitation (which is
represented by the ratio [OI]/[OIII]).  Furthermore [OI]/H$\alpha$
remains a discrepant ratio as in Fig.~19. So although it is in
principle possible to satisfy the F1/[OI]=0.4 criterion with a hard
power law, the truncation must be done at a specific yet {\it ad~hoc}
depth and furthermore the previous trend of excitation with U has
disappeared.

Without rejecting the possibility of a more complicate mixture of
matter and radiation bounded models, we believe that simply truncating
clouds does not convincingly solve the problem of the weakness of the
CaII doublet and dust depletion remains the most likely
interpretation.
	
	It is interesting to note that truncated clouds adjust better the
HeII/H$\beta$ ratio (bottom left diagram) as proposed before by Morganti
et al. (1991) and Viegas and Prieto (1992).

\section{Conclusions}

	This work is based on the method proposed by Ferland (1993) to
investigate the presence of dust mixed with the gas of the Narrow Line
Region of active galaxies.  Because photoionization models predict
remarkably strong forbidden lines [CaII]$\lambda\lambda7291,7324$\AA\
assuming reasonable abundances of atomic Ca, the basic idea is to
infer a systematic depletion of Calcium onto dust grains whenever the
infrared [CaII] lines are observed very weak or undetected. This test
of the dust content was applied to cooling flow filaments by Donahue
\& Voit (1993) who concluded on the presence of dust.

	We have shown here that this sensitive method is also
applicable to the conditions found in the EELR of radiogalaxies. In
order to make more secure any inference about the presence of dust
based on [CaII] lines, we have investigated alternative explanations
for their absence: ionization of Ca$^{+*}$ to Ca$^{++}$ by Ly$\alpha$
photons and soft continuum photons from the metastable level of
Ca$^+$, thermal ionization of Ca$^+$, ionization of Ca$^+$ due to
either a very high U value (ionization bounded case) or to a hard
continuum (with truncated clouds). Except for the highly excited EELR
which might not possess any Ca$^+$ region due to their high ionization
level, the results are negative: none of the alternative mechanisms or
models studied can explain the absence of the [CaII] lines without
dust depletion.

	Our conclusion is that the dust content test appears generally valid for the
EELR of radiogalaxies (unless the excitation level of the gas is extremely
high). This will allow us to make important conclusions about the origin
of such gas, discriminating between galactic debris and the standard
cooling flow theory.

	A BB ionizing continuum characterized by a temperature of
$1.2\times 10^5$K can reproduce the observed line ratios of the EELR
at least as well as a PL of index $\alpha=-1.4$. On the other hand,
the F1/[OI] from a BB is higher than that of a PL so the case in
favour of depletion is even stronger.

	In a follow up paper, we will present long slit spectra of
EELR, cooling flow filaments and Seyfert~2 NLR, all taken in the
region of the [CaII] doublet.  The goal will be to apply the test of
the Calcium depletion described above in order to conclude whether or
not the gas in these nebulosities is mixed with dust. This will be our
starting point for deciphering the origin of the emitting gas.

\newpage

\vspace{0.5cm}

\addcontentsline{toc}{part}{Bibliography}
\parskip=1ex
{\Large \bf References}

\vspace{0.5cm}

\noindent Anders E., Grevesse N., 1989, Geochim. Cosmochim. Acta, 53, 197 

\noindent Baum S.A., Heckman T., Bridle A., van Breugel W., Miley G., 1988, {\it ApJS}, 68, 643

\noindent Binette L., Dopita M.A., Tuothy I.R., 1985, \apj 297, 476

\noindent Binette L., Robinson A., Courvoisier T.J.L., 1988, {\it A\&A}, 194, 65

\noindent Bruzual A., Magris C.M., Calvet N., 1988, \apj 333, 673

\noindent Buson L.M., Sadler E.M., Zeilinger W.W., Bertin G., Bertola F., Danzinger I.J.,
DeJonghe H., Saglia R.P., de Zeeuw P.T., 1993, {\it A\&A}, 280, 409

\noindent Clements, D.L., Andreani, P. and Chase, T., 1993, \mnras 261, 299

\noindent Crinklaw G., Federman S.R., Joseph C.L., 1994, \apj 424, 748

\noindent di Serego Alighieri S., Fosbury R.A.E., Quinn P.J., Tadhunter C.N., 1989, {\it Nature}, 341, 307

\noindent Donahue, M., Voit G.M., 1993, \apj 414, L17

\noindent Draine B.T., Salpeter E.E., 1979, \apj 231, 77 

\noindent Fabian A.C., Johnstone R.M., Daines S.J., 1994, \mnras 271, 737

\noindent Ferland G.J., 1993, in Proc. Madrid Meeting on The Nearest Active Galaxies, ed. J.E. Beckman, H. Netzer \& L. Colina, p. 75

\noindent Forman W., Jones C., Tucker W., 1985, \apj 293, 102

\noindent Fosbury R.A.E., di Serego Alighieri S., Courvoisier T., Snijders M.A.J., Tadhunter C.N.,
Walsh C.N., Wilson W., 1990, in Evolution in Astrophysics, Toulouse,
ESA SP-310 

\noindent Goudfrooij P., 1994, Ph.D. Thesis, University of Amsterdam, The Netherlands

\noindent Haniff C. A., Ward M.J., Wilson A.S., 1988, \apj 368, 167

\noindent Heckman T.M., Baum S.A., van Breugel, W.J.M., Miley G.K., Illingworth G.D., Bothun
G.D. \& Balick B., 1986, \apj 311, 526 

\noindent Januzi B.T., Elston R., 1991, \apj 366, L69 

\noindent Jordan C., 1969, \mnras 142, 501

\noindent Kingdon J., Ferland G.J., Feibelman W.A., 1995, \apj 439, 793

\noindent  Morganti R., Robinson A., Fosbury
R.A.E., di Serego Alighieri S., Tadhunter C.N. Malin D.F., 1991,
\mnras 249, 91 

\noindent Nulsen P.E., Stewart G.C., Fabian A.C., 1984, \mnras 208, 185

\noindent Quinn P.J., 1984, \apj 279, 256

\noindent Robinson A., Binette L., Fosbury R.A.E., Tadhunter C.N., 1987, \mnras 227, 97 (RBFT87)

\noindent  Sutherland R.S., Bicknell G.V., Dopita M.A.,
1993, \apj 414, 510  
  
\noindent  Tadhunter
C.N., 1986, D.Phil. Thesis, University of Sussex

\noindent Tadhunter C.N., Fosbury R.A.E., Quinn P.J., 1989, \mnras 240, 255

\noindent  Tadhunter C.N., Robinson A., Morganti, R., 1989.
In: {\it ESO Workshop on Extranuclear Activity in Galaxies}, p. 293,
eds Meurs, E.J.A., Fosbury, R.A.E., ESO Conf. and Workshop Proc. No.
32, Garching 

\noindent Tadhunter C.N., Scarrot S.M., Draper P., Rolph C.,1992, \mnras 256, 53p

\noindent Thomas P.A., Fabian A.C., Arnaud K.A., Forman W., Jones C., 1986, \mnras 222, 655 

\noindent Toomre A., \& Toomre J., 1972, \apj 178, 623

\noindent Viegas S.M., Prieto A., 1992, \mnras 258, 483 

\noindent Villar-Mart\'\i n M., Binette L., 1995, {\it submitted to A\&A}

\noindent Wallerstein G., Bolte M., Whitehill-Bates P., Mateo M., 1986, \pasp 98, 330

\noindent Whittet D.C.B., 1992, Dust in the Galactic Environment (Bristol:IOP)

\noindent Wilson A.S., Tsvetanov Z., 1994, \aj 107, 1227

\noindent Wyse A.B., 1941, \pasp 53, 184

\chapter{Calcium depletion and the presence of dust in large scale nebulosities in radio galaxies (II).}
\centerline{Villar-Mart\'\i n \& Binette 1995, A\&A, submitted }
\vspace{0.3cm}
\pagestyle{myheadings}

\markright{Internal dust in Large Scale Nebulosities in RGs (II)}

\vspace{0.5cm}

{\LARGE  \bf Abstract}
 
\vspace{0.5cm}

We investigate here the origin of the gas observed in extended emission line
regions surrounding AGNs. We use the technique of calcium depletion  as a test to
prove or disprove the existence of dust in such a gas in order to discriminate
between two main theories: (1) a cooling process from a hotter X-ray
emitting phase surrounding the galaxy, (2) merging or tidal interaction
between two or more components. We have obtained long slit spectroscopy
of a sample of objects representative of different galaxy types although
our main interest focus on radio
galaxies.  The spectral range always includes the
[CaII]$\lambda\lambda$7291,7324 doublet.  The faintness or absence of
such lines is interpreted as due to the depletion of calcium onto the
dust grains and, therefore, is a proof of the existence of dust mixed with
the gas in the EELRs.

\newpage

\section{Introduction}

	In Chapter 3, we have demonstrated that the test of proof of the existence of internal dust 
based on the depletion of calcium is valid (and very sensitive)
under the conditions found in EELR studied
here. We studied in detail all the most plausible {\it alternative}
mechanisms to that of internal dust for explaining the absence of
[CaII] lines. No acceptable alternative solution was found and we
concluded in favour of the validity of the method. The observational
results and their interpretation are presented in this chapter. We
have applied this test to a sample of objects of different types:
radiogalaxies with extended emission line regions, cooling flows 
(hereafter CF) and starburst galaxies. 

	We describe the observations, data reduction and analysis of
the spectra in $\S$4.2. In $\S$4.3 we present the results of the
comparison between measurements and model predictions.  The
implications on the origin of the gas are discussed in $\S$4.4. The
conclusions compose section $\S$4.5.

\section{Observations and data reduction}

The observations were carried out on the nights 21-23 August on 1993. All the spectra were
obtained at La Silla Observatory, Chile,  with the 3.6 m telescope, using the EFOSC 1
spectrograph with a CCD detector (TEK\#26) of 512 x 512 pixels$^2$ of 27 $\mu$m$^2$. The
slit width  
was 1.5". The grism, R150, with a dispersion of 120  \- \AA\ /mm, a wavelength bin of 3.3
\AA\ /pixel and covering a spectral range of
$\sim$ 6870-8560 \AA\ . The observing conditions where photometric. Table 1 gives a log of the observations.

\begin{table*}[htb]
\footnotesize
\caption{Observing Log}
\begin{tabular}{lllllll} \hline

Name & RA(1950) & Dec(1950) & z & Comment &  Exp time (s) &  PA \\ \hline
NGC1052 & 02 38 37.33 & -08 28 09 & 0.005 & Liner & 1800 & 270\degr \\ 
NGC6215 & 16 46 47.0 & -58 54 30 & 0.005 & Liner & 2700 & 240\degr  \\ 
NGC7552 & 23 13 25.0 & -42 51 24 & 0.005 & Liner & 1200 & 270\degr  \\ 
NGC7714 & 23 33 40.59 & 01 52 42 & 0.009 & Starburst &  2700 & 270\degr  \\ 
PKS1404-267 (nuc)& 14 04 38 & -26 46 51 &  0.021 & RG  & 2700 & 270\degr  \\ 
PKS1404-267 (5" S) & & &       &  & 2700 & 270\degr  \\
PKS2014-55  &  20 14 06 & -55 48 52 & 0.061 & RG & 3600 & 190\degr  \\ 
PKS2152-69 (nuc) & 21 52 58 & -69 55 40 & 0.028    & RG & 2700 & 270\degr  \\ 
PKS2152-69 (cloud) &  & &       &  & 3600 & 290\degr  \\
PKS2158-380 & 21 58 17 &  -38 00 51  & 0.033 &  RG & 2700 & 270\degr  \\ 
PKS2356-61 & 23 56 29 & -61 11 42 & 0.096 & RG & 3300 & 285\degr  \\ 
PKS2300-18 & 23 00 23 & -18 57 36 & 0.129 &    RG & 2700 & 240\degr  \\ 
2A 0335+096  &  03 35 52 & 09 48 10 & 0.035 & CF & 5400 & 147\degr  \\ 
A2029   &  15 08 30 & 05 57 00 & 0.077 & CF & 2700 & 270\degr  \\ 
A2597   &  23 22 42 & -12 23 00 & 0.085 & CF+RG & 2700 & 197\degr  \\ 
\hline
\end{tabular}

\end{table*}

\subsubsection{a) Basic data reduction}

The reduction of the data was done using standard  methods in IRAF. The spectra were bias subtracted
and divided by a flat-field frame (dome flat-field). Illumination
corrections along the slit were found to be negligible.

 In general, we obtained three frames for each object and each slit
position, allowing the
direct removal of cosmic ray events. For a given object, all 
the frames corresponding 
to the same slit position were averaged together. In the cases where only two
frames were available, the cosmic rays were removed visually, replacing the affected
pixels by the mean of the surrounding region.
	The spectra were calibrated in wavelength using comparison spectra of an  HeAr arc
taken at the beginning and end of each night, and additionally before and after each
object. The  wavelength calibration  was done very carefully in order to later 
subtract the sky as accurately as possible.
 The IRAF routine ``background'' was used to substract the contribution of the sky by interpolation of the background detected in windows close to and on both sides of the object. With this method the sky features were successfully substracted.

\subsubsection{	b) Atmospheric extinction}

\begin{figure}
\includegraphics{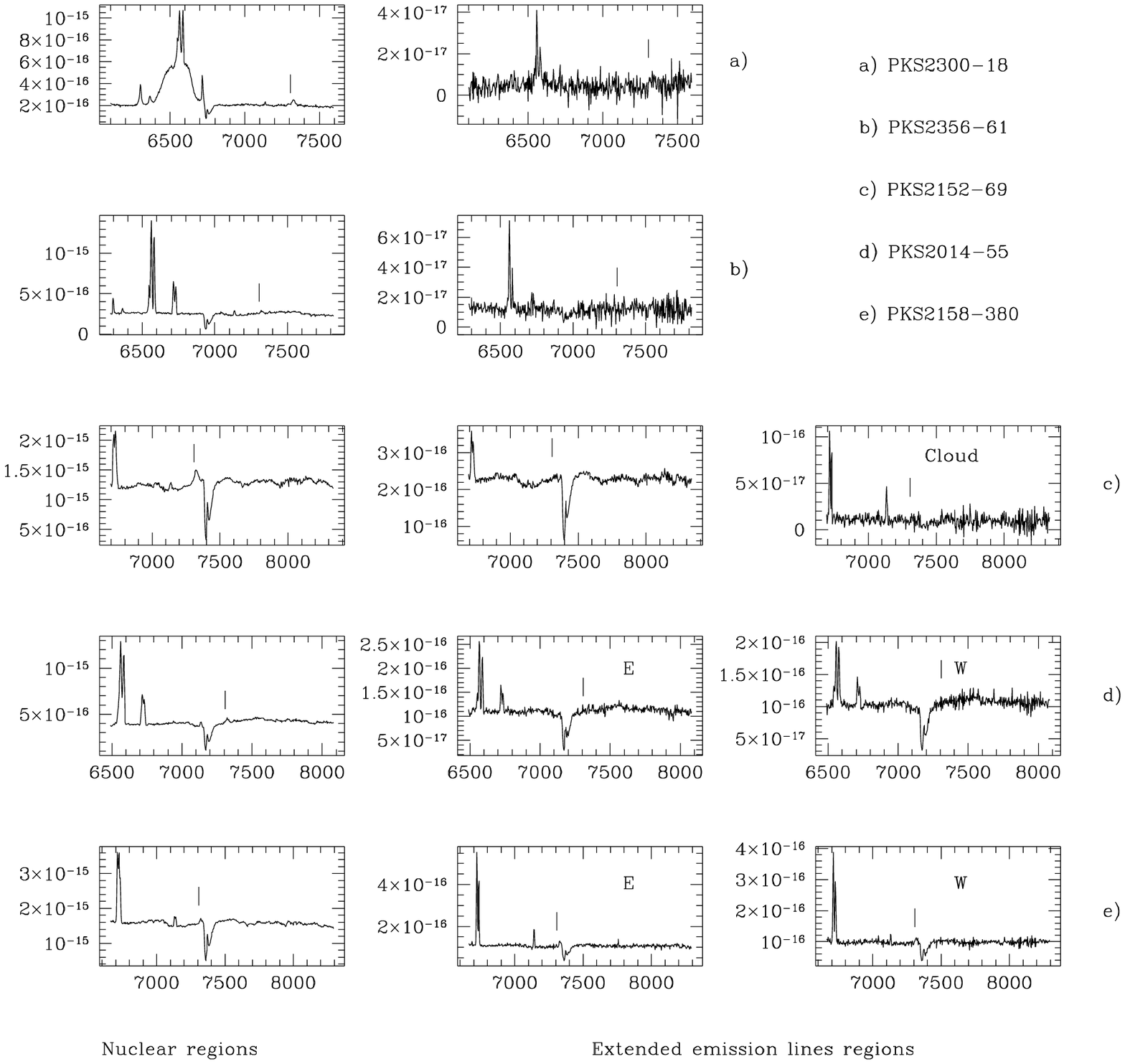}
\vspace{8in}
\caption[]{}{Spectra of observed Radiogalaxies, nuclear (first column) and extended emission line regions
(second and third columns). The expected
position of the F1 line is indicated with  $\mid$ . Flux is given in units of $erg~ s^{-1} cm^{-2} \AA ^{-1}$.}
\end{figure}

\begin{figure}
\includegraphics{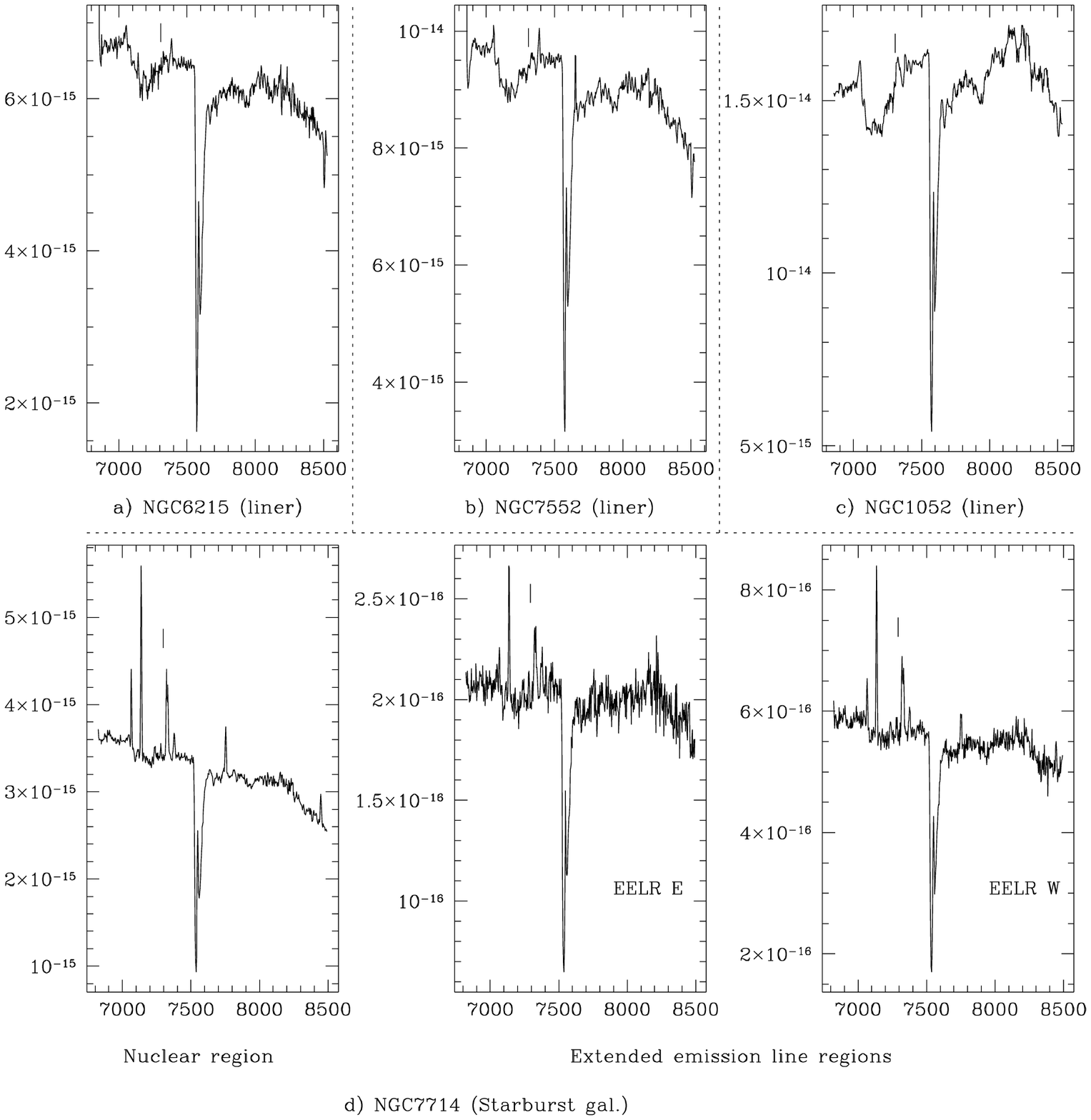}
\vspace{3.5in}
\caption[]{}{Spectra of observed Liners and Seyfert 2 galaxies. The liner spectra have been extracted
from the whole spatial extension of the object. The expected
position of the F1 line is indicated with $\mid$ . Flux is given in units of $erg~ s^{-1} cm^{-2} \AA ^{-1}$.}
\end{figure}

\begin{figure}
\includegraphics{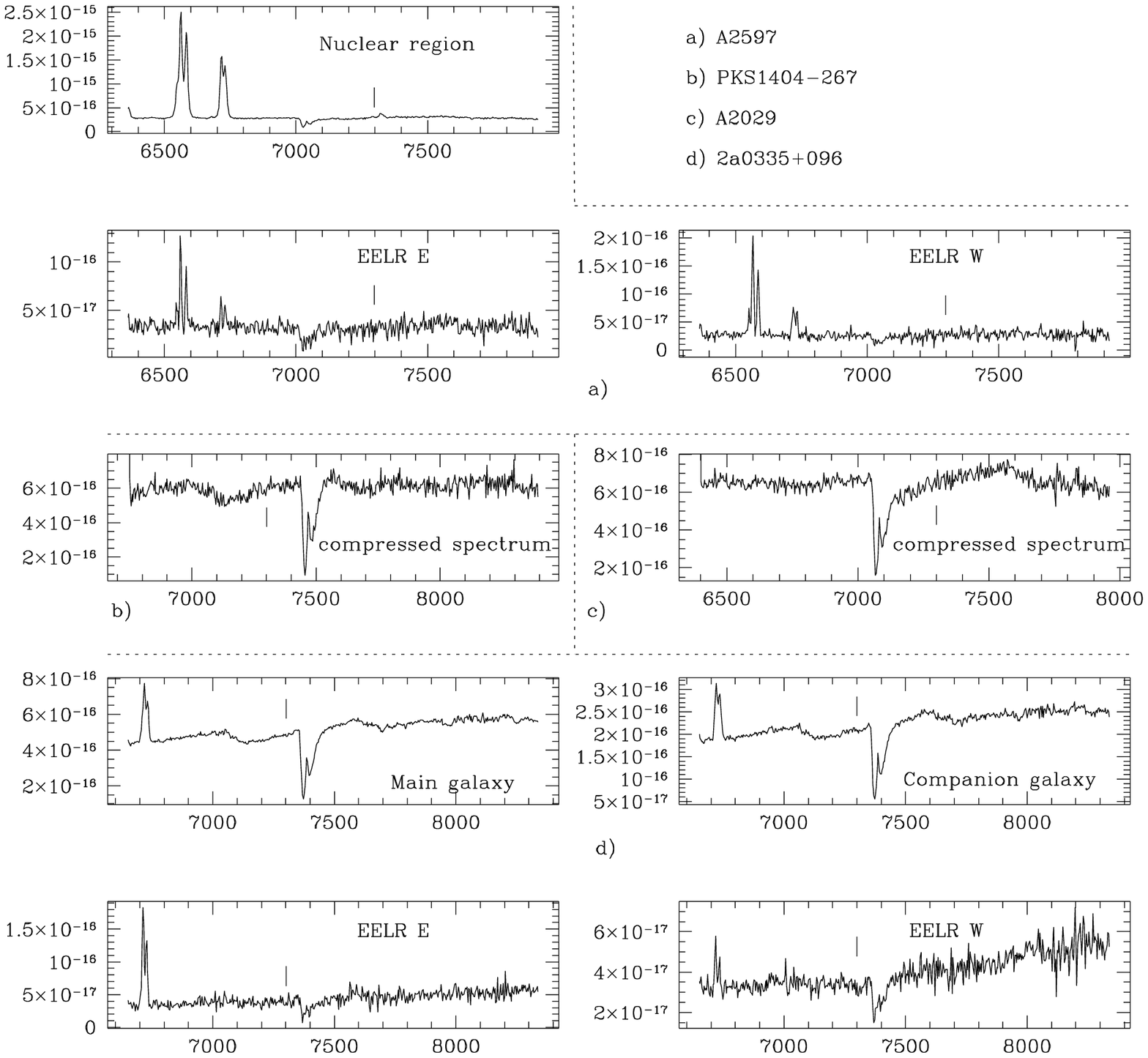}
\vspace{8in}
\caption[]{}{Spectra ot objects with properties associated with cooling flows. The spectra of
PKS1404-267 and A2029 have been extracted from the whole spatial extension of the object.
The expected
position of the F1 line is indicated with $\mid$ . Flux is given in units of $erg~ s^{-1} cm^{-2} \AA ^{-1}$. }
\end{figure}
 
	The spectra were corrected for atmospheric extinction with the aid of mean
extinction coefficients for La Silla. Molecular absorption bands of O$_2$ (the A
band at 7620 \AA\ and the H$_2$O bands, bands in the region 7100-7450 \AA\ and
8100-8400 \AA\ ) were also evident in the spectra. They were removed separately from
all of the observed spectra. These absorption bands are composed of many closely
spaced absorption features, unresolved in our spectra. Many of the features which
comprise the A band are optically thick and therefore nearly independent of zenith
distance, but most of those in the H$_2$O bands are not independent of time and
zenith distance. In principle, the best method to remove these lines is to observe standard
stars as close as possible in zenith distance and in time to that of the program
object to obtain the best correction  (c.f. Osterbrock, Shaw \& Veilleux, 1990)
. But as these authors pointed out
 the star closest to the objects's observing time and zenith distance is not always the one
producing the best correcting spectrum.

	The procedure we used was the following: we observed three different standard
stars whose frames were reduced in the same way as described before. After modeling the
atmospheric bands with them, we decided that the best results were obtained with Feige 110, a dwarf with
no intrinsic absorption features in the spectral range observed. An 1-dimensional integrated spectrum
of the  standard
star was obtained by adding all the light along the spatial direction. A smooth fit to the
continuum was done and then the original spectrum was divided by this. The result was a
normalized spectrum keeping the features due to the atmospheric absorption. Due to the dependence
on zenith distance and time, we had to model the bands for each object, i.e. to
construct an specific restoring spectrum for each object, multiplying the normalized spectrum by
appropriate factors. 

\subsubsection{	c) Flux calibration}

	The atmospherically corrected spectra of the standard stars were used to
obtain the flux calibration. For each night we built a mean response curve from the two
standard stars observed that night. 

\subsubsection{	d) Template galaxy subtraction.}

We checked that effects of stellar absorption features intrinsic to the galaxy could
be neglected: the calcium doublet is on top of the raising part of a molecular band 
whose narrow line components are not
resolved. Its only effect is to change the slope of the continuum under the doublet
and we can easily correct for this effect, fitting a
smooth continuum of certain slope in such region. H$\alpha$, a line that has been used 
as a reference to
predict the flux in the F1 line, can be underestimated due to the underlying H$\alpha$ absorption line. But in our objects  the H$\alpha$ emission is so strong that the difference is less than 5\% of the emitted flux.  
This was computed by comparing the EW
of the emission line in our objects and the EW of the absorption feature in a
template elliptical galaxy.

\subsection{Data analysis}

\subsubsection{	a) Extraction of the spectra}

	For those objects for which no extended emission lines were detected,
we compressed the whole spatial extension to extract a 1-dimensional spectrum.
 For those objects showing extended emission lines, we separated it into two
different components: the nucleus, and the extended emission line spectrum. In order to
do this,  we subtracted from 
the spatial profile in the brightest spectral line, the spatial
profile in the nearby continuum. In this way, we are left with the spatial profile of the line emitting 
regions. In
some cases  blobs appear well detached from the nuclear emission. But
sometimes the extended and nuclear emission were mixed. In order to avoid contamination of the EELRs by the nuclear spectrum, we fit a Gaussian (PSF) to the nuclear region. The residuals after
subtracting this Gaussian were considered to be due to the EELR.  The resulting spectra
are shown in Fig.~21, 22 and 23.

	It is possible that we may be venturing close enough
to the nucleus to sample higher densities, more typical of the classical narrow line
region (with densities higher than 10$^3$ cm$^{-3}$). To verify this, when it was possible, we used the [SII]
doublet to estimate the density.  The values were always below 100
cm$^{-3}$, which indicates that the spectra are dominated by low
density gas.  In any event, even if the NLR dominated the emission, as a result of its high critical density
the line F1 (in the dustfree case) was
calculated to remain
stronger with respect to other emission lines (like H$\alpha$ or [ArIII])
than in the EELR. Our conclusions about inferring dust or not
are therefore not affected by high densities. (Higher density models
would imply a
stronger calcium depletion).

\subsubsection{	b) Line measurements}

	IRAF routines were used to measure the emission line fluxes. For the blends, 
decomposition procedures were used, fitting several Gaussians at the expected positions
of the components. The fluxes were the values estimated for the Gaussians that better
fitted the given profile. 

	The upper limits on [CaII] emission were measured directly from the 
noise level in the data.

\section{Observed and predicted F1 fluxes.}

\subsection{Prediction of F1 fluxes}

	The second component of the [CaII] doublet (F2) is the weakest and is
blended with the [OII]$\lambda\lambda$7320,7330 multiplet. The first component is the
strongest and lies well aside, some 30 \AA\ shortward of [OII]. It is straightforward
to isolate it given a reasonable spectral resolution. Therefore, for simplicity, 
we have based our study on the measurement of this line. The observed values are compared
with the flux that photoionization models predict. MAPPINGS (Binette et al. 1993a,b) is
the multipurpose photoionization-shock code that we used for the modeling. It considers
the effects of dust mixed with the ionized gas: extinction of the ionizing continuum
and of the emission lines, scattering by the dust, heating by dust photoionization and
depletion of heavy elements. The appropriate input parameters for the
models were justified in Chapter 3.

	The predicted value of the F1 flux has been computed in the following way: we used as 
a reference line ($ref~line$), the strongest and/or the simplest to measure, like H$\alpha$,
 [SII]$\lambda\lambda$6716,6731, the 
[OII]$\lambda\lambda$7320,7330  blend and/or [ArIII]$\lambda$7136. 
The appropriate photoionization model
allows the prediction of $R=\frac{F1}{ref~line}$, from which we can
easily deduct the F1 flux:

$ Predicted~~F1~~flux = R \times Measured~~flux~~in~~the~~reference~~line. $
	In table 2, we show the reference lines used for each object
and for each spatial region, the measured flux, the predicted
R=$\frac{F1}{ref~line}$ ratio, the measured F1 flux and its
theoretical value. We have used different lines for different objects
and different zones. For the nuclear regions, we always base our predictions on
the blend [OII] $\lambda\lambda$7320,7330.  The density in the nuclear
narrow line region spands a wide range up to 10$^6$
cm$^{-3}$. The critical density of the [CaII] forbidden lines is very
high, and the flux of the lines under the nuclear conditions are
predicted to be far stronger than the [OII] lines (Ferland 1993). [OII] can be de-excited due to high density so
that if CaII remains much smaller than [OII], it is actually a more
stringent test since high densities would have actually helped to
increase the CaII/OII ratio.
To measure the flux
in the [OII] blend, we assume that there is a negligible contribution from the F2
component. This assumption is reasonable: it is fainter that F1 and
this line is near or under the detection limit in all cases. We can
therefore assume that the [OII] multiplet is not contaminated by F2.

\begin{table*}
\footnotesize 

\caption{Emission line fluxes for the sample of galaxies observed (in units of $erg~ s^{-1} cm^{-2}$). The superscripts $^M$ and $^P$ indicate measured and predicted
values respectively. The last two columns show the predicted and measured values for the $F1$ line. For those objects where the calculations were possible, the observed line is clearly fainter than photoionization models predict.  }

\vspace{0.5cm}

\begin{tabular}{llllclll} \hline
\centering

  Name & Spatial reg. & $ref~line$ & $Flux^{M}$  & ${\frac{F1}{ref~line}}^{P}$ & $Flux(F1)^{P}$ &  $Flux(F1)^{M}$  \\ \hline \hline
NGC1052 & Total  & None & --- & ---  & --- &   $\leq$~6.99e-16   \\ \hline
NGC6215 & Total & None & --- & --- & --- &   $\leq$~ 3.34e-16 \\ \hline
NGC7552 & Total  & None & --- & ---  & --- &   $\leq$~3.95e-16  \\ \hline
NGC7714 & Total  & [OII]$\lambda$7325 & 1.59e-14 & $\geq$1.00 & $\geq$ 1.59e-14 & $\leq$~6.84e-17   \\ \hline
PKS1404-267 & Nucleus & None & --- & ---  & ---  & ~ $\leq$~ 1.79e-16 \\ 
            & 5" South     & None   & ---  & ---  & ---    & ~ $\leq$~ 5.23e-17 \\ \hline
PKS2014-55 & Nucleus & [OII]$\lambda$7325 & 1.54e-15 & $\geq$1.00 & $\geq$1.54e-15 & $\leq$~3.34e-17 \\ 
        & EELR (E)      & H$\alpha$ &   1.47e-15 & $\sim 0.45$ & 6.62e-16 & $\leq$~ 2.60e-17   \\
        & EELR (W)      & H$\alpha$ &   1.06e-15 & $\sim 0.45$ & 4.77e-16 & $\leq$~  3.18e-19  \\ \hline
PKS2152-69 & Nucleus & [OII]$\lambda$7325 & 7.48e-15 & $\geq$1.00 & $\geq$7.48e-15 & 1.92e-16
  \\
         & EELR & [SII]$\lambda$6731 & 1.33e-15 & $\sim 0.40$ & 5.32e-16 & $\leq$~3.03e-17  \\
        &  Cloud & [ArIII]$\lambda$7136 & 3.86e-16  & $\sim 4.50$  & 1.74e-15 & $\leq$~2.24e-17   \\ \hline
PKS2158-380 & Nucleus &  [OII]$\lambda$7325 &   4.03e-15 & $\geq$1.00 & $\geq$4.03e-15 &
3.15e-16   \\
        & EELR (E) & [ArIII]$\lambda$7136 & 7.28e-16 & $\sim 5.0$ & 3.64e-15 &  $\leq$~2.55e-17  \\
        & EELR (W) & [ArIII]$\lambda$7136 & 3.00e-16 & $\sim 5.0$ & 1.5e-15 & $\leq$~5.00e-17 \\ \hline
PKS2356-61 & Nucleus & [OII]$\lambda$7325 &  9.39e-16 & $\geq$1.00 & $\geq$9.39e-16& 
2.78e-16 \\
        & EELR & H$\alpha$ & 5.36e-16 & $\sim 0.18$ &  9.6e-16 &  $\leq$~ 1.03e-17 \\ \hline
PKS2300-18 & Nucleus & [OII]$\lambda$7325 & 1.80e-15 & $\geq$1.00 &  $\geq$1.80e-15 & 4.37e-16 &    \\
        &  EELR & H$\alpha$ & 3.61e-16 & $\sim 0.40 $ & 1.44e-16 &   $\leq$~1.97e-17   \\ \hline
2A 0335+096  & cD galaxy & [SII]$\lambda$6725 & 7.46e-16  & $\sim 0.40$  & 2.98e-16 & $\leq$~ 2.72e-17 \\ 
        & EELR (E) & [SII]$\lambda$6725 & 2.24e-15 & $\sim 0.40$   & 8.96e-16 & $\leq$~1.29e-17  \\
        & EELR (W) & [SII]$\lambda$6725 & 4.22e-16 & $\sim 0.40$  & 1.68e-16 & $\leq$~1.40e-17   \\
        & companion  & [SII]$\lambda$6725 & 7.89e-15 & $\sim 0.40$  & 3.16e-15 & $\leq$~ 1.69e-17  \\ \hline
A2029   & Total  & None & --- & --- & ---  & $\leq$~1.43e-16  \\ \hline
A2597   & Nucleus & [OII]$\lambda$7325  & 1.79e-15 & $\geq$1.00 &  $\geq$1.79e-15  & 5.14e-16  \\ 
        & EELR & H$\alpha$ & 2.32e-15 & $\sim 0.50$&  1.16e-15 &  $\leq$~5.86e-17 \\ \hline
\hline

\end{tabular}
\end{table*}

\subsection{Comparison with the observations}

	The comparison between the measured and predicted F1 flux values shown in
table 2 demonstrates that, whenever the calculations were possible (sometimes there
was no a reference line available, however),
it is always fainter (non detected in most cases) than
expected. This result is common to all  the regions considered here: nuclear,
EELRs and cooling flow filaments. It demonstrates that calcium is depleted and,
therefore, {\it the gas is mixed with dust}, both in the nuclear as in the extended
emitting gas in radio galaxies and cooling flow filaments.
 
Donahue \& Voit applied this same test to emission line nebulae in cluster
cooling flows (1993). They interpreted the observed lack of [CaII] emission as  depletion of calcium onto dust grains in the ionized filaments.
Our work support these results, in a different sample of objects: the NLR in  AGNs and
the filaments in objects classified as 'cooling flows' are well mixed with dust.

The most important result is the confirmation that {\it dust exists mixed with the ISM in low $z$ radio galaxies}.

\section{Discussion}

\subsection{Implications on the origin of the gas}

As explained in Chapter 3, there are some pieces of evidence indicative of dust {\it in the extended ionized gas} in
active galaxies, mainly derived from polarization measurements that show the existence of scattered nuclear light over large spatial scales, although it is difficult to discriminate between dust and electrons as the scattering agent.
Our results confirm that dust exists {\it mixed with the ISM in low redshift radio galaxies}.

As mentioned before, this produces discrepancies with the traditional
cooling flow theory, which would now be required  to explain the formation of dust in a
shorter time than the cooling time! According to this theory (see Fabian 1994 for a review), the galaxies, groups and clusters of galaxies were formed out of gas that collapsed gravitationally. During this process, gravitational energy was released which heated  the clouds and a hot X-ray atmosphere would remain. Inhomogeneities in the gas would cause matter from the hot atmosphere to cool down and fall towards the center of the galaxy or cluster. The resulting filaments would eventually be visible at optical wavelengths and would emit strong lines. For massive galaxies and cluster, the cooling process would have been slower than for normal galaxies and the hot atmosphere would still exist, with typical temperatures of several million K and emitting strongly in the X-ray band.

If this is actually the origin of the gas in the EELRs (and cooling flow filaments), it should clearly be devoid of dust.
Any  dust introduced  in the hot intracluster medium 
would be sputtered and rapidly destroyed (time scale of the order of 10$^7$ years) 
(Draine \& Salpeter 1979), much before the filaments cooled down and became visible.
Is there a way to introduce dust in the filaments during the cooling process? The
common place where dust is formed is stellar atmospheres. Do stars exist in the accreted
gas?
Such gas must be deposited in some form and it has been generally thought that the
more plausible fate for most of the gas is the formation of new stars (e.g.,
O'Connell \& McNamara 1989; Fabian, Nulsen \& Canizares 1991). 
Although there are some indications of star formation in cooling flow galaxies,
the latter seems to be taking place in the inner
parts (over the central few kpc) (Cardiel, Gorgas \& Arag\'on Salamanca 1995, Fabian 1994).   There is a lack of evidence
of star formation  (or any kind of stars) at large distances, out of the main
body of the galaxies, where the EELRs filaments are still visible.

In high z RG there is an extended blue continuum aligned with the radio axis
that has often  been interpreted as due to young stars. However, as I mentioned already before, further
results (e.g., Tadhunter, Fosbury \& di Serego 1988; Cimatti et al. 1993) proved that this continuum is polarized and the
contribution of a hidden quasar continuum scattered by dust and/or electrons in the
ISM can well be an important component (in fact, this extended  polarized UV continuum has been used as a proof of the existence of dust in the EELRs of high z radio
galaxies). This effect has also been
detected in some radio galaxies at low $z$ (Cimatti \& di Serego Alighieri S. 1995).  Altough young stars may exist, we don't have a clear idea about
their overall contribution.

	There is another mechanism which can form dust: if most of the
cooled gas from a flow does not form stars with normal IMF, maybe it
remains as cold clouds or as low-mass stars.  There are
evidences of X-ray absorption (White et al. 1991; Mushotzky 1992;
Allen et al. 1993) in cooling flows in cluster of galaxies.  The
absorbing material could be in the form of cold gas embedded in the
hot intracluster medium of the cooling flow (White et al. 1991;
Daines, Fabian \& Thomas 1994). This cold gas, very slightly
photoionized by X-rays from the surrounding hot corona, can become
molecular (Ferland, Fabian \& Johnstone, 1994). Fabian, Nulsen \&
Canizares (1994) propose that the conditions suitable for dust to form
may occur in this cool gas, through the condensation of gaseous
particles. The authors propose that the gas which is cooling towards the center
could be a mixture of both the molecular gas and the very hot gas and,
therefore, could contain the existing dust. However, even if such a scheme was
possible, the dust grains will not remove the calcium which already existed in the hot gas; the temperature is too high for the condensation of calcium 
onto the dust grains. Therefore, we should observe the CaII lines from the hot gas when it cools down, even if it contains dust.

	On the other hand, the interaction scenario  (gas  accreted from
outside the galaxy, as a result of recent tidal interactions or mergers
between two or more components) predicts the existence of dust mixed with the  gas, the one already 
existing in the interacting objects. Some morphological evidences
and theoretical ideas (see Chapter 3) support this scene.  Heckman et al. (1986) showed
that a large fraction of powerful radio galaxies have morphological features (shells,
tails, loops, etc) similar to  those produced in numerical simulations of galaxy
interactions (e.g., Toomre and Toomre 1972, Quinn 1984). Kinematic measurements 
show that the radio galaxy EELR generally have a high specific angular momentum which
is difficult to reconcile with the cooling flow picture (Tadhunter, Fosbury \& Quinn
1989).

	Therefore, if there is a common origin for the EELRs of all radio galaxies, our results 
suggest that it is mergers or tidal interactions. However, we don't exclude the
possibility of an origin which is {\it not} universal, that is, which may differ from object to object.

\section{Conclusions.}

	We have confirmed that the gas in extended emission line
regions in radio galaxies at low $z$ is mixed with dust. 

	Our results support the existence of dust mixed with the gas in the
Narrow Line Region and in the cooling flow filaments.
	
	If there is an universal origin for the EELRs, the existence of internal dust favours mergers
or tidal interactions as the most plausible scenario.

\vspace{0.65cm}

\addcontentsline{toc}{part}{Bibliography}
\parskip=1ex
{\Large \bf References}

\vspace{0.5cm}

\noindent Allen S.W., Fabian A.C., Johnstone R.M., White D.A., Daines S.J., Edge A.C., Stewart G.C., 1993, \mnras, 262, 901

\noindent Cimatti A., di Serego Alighieri S., Fosbury
R.A.E.,
Salvati M., Taylor D., 1993, \mnras 264, 421

\noindent Cimatti A., di Serego Alighieri S., 1995, \mnras 273, L7

\noindent Cardiel N., Gorgas J., Arag\'on-Salamanca A., 1995, \mnras 277, 502

\noindent Daines  S.J., Fabian A.C., Thomas P.A., 1994, \mnras 268, 1060

\noindent Donahue, M., Voit G.M., 1993, \apj 414, L17

\noindent Draine B.T., Salpeter E.E., 1979, \apj 231, 77 

\noindent Fabian A.C., Nulsen P.E.J., Canizares C.R., 1991, {\it ARA\&A}, 2, 191

\noindent Fabian A.C., 1994, {\it ARA\&A}, 32, 277

\noindent Ferland G.J., 1993, in Proc. Madrid Meeting on The Nearest Active Galaxies, ed. J.E. Beckman, H. Netzer \& L. Colina, p. 75

\noindent Heckman T.M., Baum S.A., van Breugel, W.J.M., Miley G.K., Illingworth G.D., Bothun
G.D. \& Balick B., 1986, \apj 311, 526

\noindent Mushotzky R.F., 1992, in Fabian A.C., ed., Clusters and Superclusters of Galaxies. Kluwer, Dordrecht, p.91 

\noindent O'Connell R.W., McNamara B.R., 1989, \aj 98, 2018

\noindent Osterbrock D.E., Shaw R.A., Veilleux S., 1990, \apj 352, 561

\noindent Quinn P.J., 1984, \apj 279, 256

\noindent  Tadhunter C.N., Fosbury R.A.E., di Serego
Alighieri S.,
1988, in Maraschi L., Maccacaro T. \& Ulrich M.H., eds., "BL Lac Objects",
Springer-Verlag, Berlin, p.79

\noindent Tadhunter C.N., Fosbury R.A.E., Quinn P.J., 1989, \mnras 240, 255

\noindent Toomre A., \& Toomre J., 1972, \apj 178, 623

\noindent White D.A., Fabian A.C., Johnstone R.M., Mushotzky R.F., Arnaud K.A., 1991, \mnras 252, 72

\newpage 
~
\newpage

\chapter{Effects of dust and resonant scattering on the UV spectrum of radio galaxies.}
\centerline{Villar-Mart\'\i n, Binette \& Fosbury 1995, A\&A, submitted}
\vspace{0.3cm}
\pagestyle{myheadings}
 
\markright{Resonance scattering and dust in HZRGs}

\vspace{0.5cm}

{\LARGE  \bf Abstract}
 
\vspace{0.5cm}

In the powerful, high redshift radio galaxies, it is believed that the dominant
source of ionization for the  gas is the hard radiation field
associated with the active nucleus. The photon source is generally external to
the clouds being ionized and so the geometrical perspective from which the gas
is observed and the presence and distribution of dust must be properly
accounted for in the diagnostic process. In this paper, we examine the
formation of three strong lines, CIV$\lambda$1549, Ly$\alpha$\ and
CIII]$\lambda$1909 which are often observed in the nuclear and extended
emission from these sources. We find that the observed trends, in particular
the high CIV$\lambda$1549/Ly$\alpha$ ratio, are often better explained by
geometrical (viewing angle) effects than by the presence of large quantities of
dust either within or outside the excited clouds. We show that 
condensations of neutral gas along the line-of-sight can increase the observed CIV/Ly$\alpha$ ratio, by reflecting photons near the
wavelength of Ly$\alpha$,. The
existence of HI absorption clouds (i.e., mirrors) external to the emission
region leads also to the presence of large, diffuse haloes of what appears to
be pure, narrow Ly$\alpha$\ emission.

\newpage 

\section{Introduction}

The strong, spatially extended, rest-frame ultraviolet emission lines
observed in high redshift radio galaxies provide one of the principal
diagnostics in establishing the state of the interstellar medium in
galaxies at early epochs.  The presence of a blue continuum and
emission lines from regions aligned with the radio axis (McCarthy
et~al. 1987; Chambers, Miley \& van Breugel 1987) warned us that much of the observed
optical radiation might be associated with the nuclear activity and so
may not be giving us a clear picture of the stellar processes which
are of great interest in studies of galaxy formation and
evolution. Subsequent work has shown, indeed, that much of the blue
light is scattered, polarized nuclear radiation (e.g., Tadhunter
et~al. 1989; di Serego Alighieri et~al. 1993; Cimatti et~al. 1993) and
that the emission lines have a high ionization state and cannot result
from  photoionization by normal stars (McCarthy 1993). It is clearly necessary,
therefore, to reach a clear understanding of the physical processes
involved in the formation of the various lines and continua to be able
to disentangle the stellar and the AGN-related sources.

For objects at high\,$z$, the UV rest-frame lines are shifted into the optical
band and the spectrum is generally dominated by Ly$\alpha$, CIV$\lambda$1549,
HeII$\lambda$1640 and CIII]$\lambda$1909. The strength of the high ionization
lines suggests the presence of a hard photoionizing continuum which could
originate at the AGN itself (Robinson et~al. 1987) or be associated with fast
shocks generated in extranuclear regions by the radio jets (Sutherland,
Bicknell \& Dopita 1993). The strong radio/optical asymmetries observed in
these objects which exhibit the `alignment effect' (McCarthy, van Breugel \& Kapahi 1991) may
simply result from a one-sidedness in the distribution of material. It it is
clear, however, that correlated line and continuum asymmetries could be
produced by dust scattering and line fluorescence for sources where the radio
axis falls significantly away from the plane of the sky.

In this work, we concentrate on modeling the high excitation lines for which
rather extreme ratios relative to Ly$\alpha$ have recently been reported. The
presence of dust has been universally invoked to explain the weakness of
Ly$\alpha$ which is a resonance line and therefore, due to multiple scattering,
more susceptible to absorption. We explore the fact that any resonance line
will be extremely sensitive to geometrical factors, an aspect of the problem
which has so far been overlooked in modeling the UV lines. If in
radio-galaxies the distant gas clouds are photoionized from the outside by
partially collimated UV radiation emitted by the nucleus, the line formation
process --- particularly for the resonance lines --- is very different from that of the 
internally ionized HII regions. The escape of resonance line photons is
strongly influenced by the presence of voids between the line emitting clouds.

We have collected from the literature the observed line ratios for a number of
 radio-galaxies at high $z$ ($z>$1.5) in which no contribution from any nuclear BLR is
apparent. We have built a diagnostic diagram consisting of the lines
CIV$\lambda$1549/Ly$\alpha$ {\it vs.} CIV$\lambda$1549/CIII]$\lambda$1909, in
which we compare the position of the objects with photoionization models which
not only consider the effects of internal dust but also those of the viewing
perspective --- the angle between the incoming ionizing radiation and the
observer's line of sight. Our concentration on the particular class of radio
galaxies is purely for pragmatic reasons. It is these objects, which we presume
to harbour a powerful quasar which is hidden at optical/ultraviolet wavelengths
to our line of sight, which are most readily found and studied at high
redshifts where we have access to the ultraviolet spectrum from groundbased
observations. Our conclusions should be equally applicable to other classes of
AGN. 

For some objects, Ly$\alpha$ is observed to be fainter with respect to CIV than
predicted by dust-free photoionization models. The explanation previously proposed to explain the weakness of Ly$\alpha$ with respect to H$\alpha$ or
H$\beta$ has been dust destruction of resonant Ly$\alpha$ photons. This is {\it
not} borne out by our calculations in which we have used arbitrary amounts of
dust and found that this cannot simultaneously weaken Ly$\alpha$ while leaving
the CIV/CIII] ratio relatively unchanged since resonant CIV suffers also from
dust absorption. Alternatively, by varying the proportions of the illuminated
and the shadowed cloud faces which contribute to the observed spectrum, we are
better able to match the data. 

As we find that geometry alone (with or without internal dust) can in
principle explain most of the specific line ratios observed (fainter
Ly$\alpha$ compared with either CIV or HeII), we also discuss the
possibility of a patchy outer halo of neutral gas to account for the
diffuse Ly$\alpha$ seen in some cases to extend much beyond the CIV
emitting region and even the outermost radio lobes.  Reflection by
cold gas of the brighter Ly$\alpha$ emitting side of the ionized
clouds would lead to a narrower profile for such a diffuse
component. Another possibility is that part of the beamed nuclear {\it
continuum and BLR} radiation might be reflected at the wavelength of
Ly$\alpha$ by thin matter-bounded photoionized gas at very large
distances from the nucleus leading to a diffuse Ly$\alpha$ component
aligned with the radio axis. It appears to us that geometrical
perspective effects are an essential component of the interpretation
of the UV spectrum of radio-galaxies whether or not dust is
present. Furthermore, a spectrum in which only Ly$\alpha$ appears does
not necessarily imply starburst activity, other lines must be observed
before the existence of HII regions can be inferred.

\newpage

\section{Data sample and modeling procedure}

\subsection{The data}

We have constructed a data sample containing galaxies at high~$z$ for
which the CIV, Ly$\alpha$ and, in most cases CIII], emission lines
have been measured.  Since very high densities such as those
encountered in the BLR alter significantly the line formation and
transfer processes, we have excluded those objects which show evidence
of a BLR. In Table~3, we list the object names, the line ratios of
interest to us here, the redshift and the reference to the
observations. The larger fraction of the data are taken from the
recent thesis by van Ojik (1995) which includes objects selected on
the basis of a very steep radio spectrum. Probably by virtue of the
radio selection, these sources populate the region of the line ratio
diagram (Fig.~29) with lower CIV/CIII] ratios (lower ionization
parameter) than the previously published objects. The line
measurements refer to the integrated emission from the object
collected with a long slit aligned with the major axis.

\begin{table}
\centering
\small
\caption{Observed UV line ratios for several high~$z$ RG with not apparent broad component }
\vspace{0.5cm}
\begin{tabular}{lllll} \hline

Name &  CIV/CIII] & CIV/Ly$\alpha$ &  $z$  & ref. \\ \hline
  Average RG &  2.054 &  0.118  & 0.1$<z<$3 & McCarthy (1993) \\ \hline	
MG1019+0535A & 2.12 &  1.24 & 2.76  & 	 Dey et al. 1995\\
F10214+4724 &  3.68 & 8.75 & 	2.29 & 	Elston et al. 1994	\\ 	
TX0211-122 &  3.33  &	   0.91 &  2.34 	& van Ojik et al. 1994 \\ 	
3C294   &    0.83  &      0.10     &   1.79 &  McCarthy et al. 1990a \\
0902+34   &    -- & 0.11 & 3.40 & Lilly 1988 \\
3C256-3C239 &	1.90 & 	  0.14 & 1.82 \& 1.78  & Spinrad et al. 1985	\\ 
0200+015 &  1.05 & 0.241 & 2.23	& van Ojik 1995 \\ 
0214+183 &  1.67  &	--     &   2.13 	& ~~~~~~ " \\ 
0355-037 &  1.17  &	     0.24 &  2.15 	&  ~~~~~~ " \\ 
0417-181 &  -- & 0.40 &  2.73	&  ~~~~~~ " \\ 
0448+091 & 	0.44	&     0.10  & 2.04 & 	~~~~~~ " \\
0529-549 &  0.22	 &    0.05 & 2.58 & 	~~~~~~ " \\
0748+134 &  1.29	 &    0.29 & 2.42 & 	~~~~~~ " \\
0828+193 &  0.95	 &    0.14 &  2.57 & 	~~~~~~ " \\
0857+036 &   -- & 0.39 & 2.81 & 	~~~~~~ " \\
0943-242 &  1.70	 &    0.19 & 2.92 & 	~~~~~~ " \\
1138-262 &  0.62   &     0.06 & 2.16  & 	~~~~~~ " \\
1357+007 &  -- & 0.19  & 2.67 & 	~~~~~~ " \\
1410-001 &  1.58	&     0.16 & 2.36 & 	~~~~~~ " \\
1545-234 &  0.61	 &    0.17 & 2.76 & 	~~~~~~ " \\
1558-003 &   2.25	 &    0.18 & 2.53  & 	~~~~~~ " \\
2202+128 &  -- & 0.25 & 2.71 & 	~~~~~~ " \\
2251-089 &  2.20  & -- & 1.99 & 	~~~~~~ " \\
4C23.56 &  3.40  & -- &  2.48 &	~~~~~~ " \\
4C24.28	 &   -- & 0.23 & 2.88  & 	~~~~~~ " \\
4C26.38	 &  3.71 & -- & 2.61 & 	~~~~~~ " \\
4C28.58	 & 0.17 & -- & 2.89 & 	~~~~~~ " \\
4C40.36	 &  1.05 & -- &  2.27 & 	~~~~~~ " \\
4C41.17	 &  -- & 0.05  & 3.80 & 	~~~~~~ " \\
4C48.48	 &  2.18  & -- & 2.34 & 	~~~~~~ " \\
4C60.07	 &  -- & 0.27 &  3.79 & 	~~~~~~ " \\ \hline
\hline
\end{tabular}
\end{table}

\subsection{The model and its parameters}

The data are compared to photoionization models computed using the code MAPPINGS. 
The version described in Chapter 2 is particularly suited to this study
since it considers both the
effects of the observer's position with respect to the emitting slab and the
ionizing source (see Fig.~4, Chapter 2), distinguishing between the spectrum seen from the
back and from the front of the slab. Phenomena related to dust
are also properly considered. The treatment of the escape of resonant CIV and 
Ly$\alpha$ 
photons in a dusty
medium (a fundamental proces that needs to be well understood) is described in Appendix~B of Binette et~al. (1993a) and is based on the
results of Hummer \& Kunasz (1980).
The presence of dust implies 
depletion of refractory trace
elements and this effect has also been considered, as described in Apendix A.
The calculations consider the gas pressure to be constant (isobaric models) and
so the density behaviour with depth in the cloud is determined by the behaviour
of the temperature and the ionization fraction of the gas.

\subsection{Adopted physical conditions}

The detailed studies of the optical emission lines of low~$z$ radio galaxies
provide our basic reference for the properties of the emitting gas and the
ionizing continuum (e.g., Robinson et~al. 1987, hereafter RBFT87). As a guide
in choosing the input parameters, we have assumed that the excited gas of {\it
very} high~$z$ radio galaxies ($z>2.5$) has similar properties to that of the low
redshift ($z<0.1$) objects. As explained in Chapter 3, for the extended (EELR) as 
well as the (narrow) nuclear emission lines,
photoionization by a hard continuum appears to best explain the various line
ratio diagnostic diagrams: a power law of index 
$\alpha \simeq -1.4$
($f_{\nu } \propto \nu^{+\alpha}$), that has produced
remarkably good agreement with the optical line ratios at low $z$,  will be 
adopted as ionizing energy distribution.

Concerning metallicity, RBFT87 indicated that the abundances cannot be much
higher than solar values. They could however be lower by a factor of a few, a
distinct possibility in the case of the extranuclear gas in very high~$z$
galaxies. Unless specified otherwise, we have adopted $Z=1$ in our calculations
and have verified that lower abundances do not in any way affect the
conclusions reached here.

The red [SII] doublet ratio in low~$z$ radio galaxies indicates densities for the
extended emission line regions which are lower than 100  cm$^{-3}$.
We therefore adopt the low density regime, specifically $n_H = 100
{\rm cm}^{-3}$,
since the {\it extranuclear} gas generally dominates the line luminosities.

\section{Models and comparison with observed UV lines}

In this section we first extend the assumption of photoionization to modeling
the line emission in the ultraviolet rest-frame of high~$z$ radio galaxies. We
distinguish the effects of having a back and a front view of an externally
photoionized slab (section 5.3.2). In 5.3.3, we introduce internal dust and discuss
how, alone, it is insufficient to explain the low value of the Ly$\alpha$/CIV
ratio in some objects. In 5.3.4 we distinguish between the resonance line
`mirror' intrinsic to our photoionized slab and the possibility of having
external cold gas --- which is shadowed from the ionizing source --- acting as
a reflector of Ly$\alpha$ photons.

\subsection{The photoionization assumption at low and high~$z$}

At very high redshift, observational access to the normal optical plasma
diagnostic lines is restricted and the available data set is small. Also, where
measurements are available, few lines are measured in any given source. It is
nevertheless interesting to compare the [OIII]$\lambda$5007/H$\beta$ value
observed at very high~$z$ to that of low~$z$ radio galaxies as plotted in Figure~24.
The curved line corresponds to the sequence of photoionization models of RBFT87
which reproduce reasonably well the optical (rest frame) line ratios of the
extended emission line regions and their associated narrow nuclear line
emission. The range in $U$ is [$10^{-4}, 10^{-1}$]. The three lower horizontal
lines correspond to high~$z$ radio galaxies while the upper one corresponds to
the ``average'' radiogalaxy spectrum as defined by McCarthy (1993). The high~$z$
objects are not substantially different from the low~$z$ sample although the
apparent trend towards weaker [OIII]/H$\beta$, if confirmed, might indicate
substantially lower metallicities: low enough to overcome the higher ratios
produced by the higher kinetic temperatures in moderately underabundant
objects. Such an effect could also be produced by insufficient spatial
resolution at high~$z$ to separate the high from the low excitation
regions. 


\begin{figure}[htb]
\includegraphics{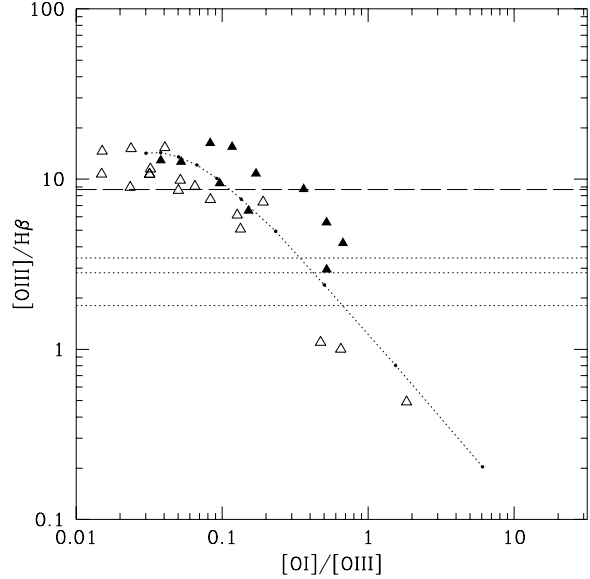}
\vspace{3in}
\caption[]{\small Excitation diagnostic for the ionized gas in low~$z$ radio-galaxies ([OIII]$\lambda$5007/H$\beta$ {\it vs.} [OI]$\lambda$6300/[OIII]$\lambda$5007).
Open triangles correspond to extended emission line regions (EELR)
while solid triangles correspond to nuclear regions.
The three lower horizontal lines indicate the observed [OIII]$\lambda
5007$/H$\beta$ ([OI] is not measured) ratio of very high~$z$ radiogalaxies 
(Eales \& Rawlings 1993) while
the long-dash line corresponds to the ``average'' radiogalaxy spectrum as
derived by McCarthy (1993). The dotted line correspond to the sequence of
photoionization models of RBFT87. $log~U$ is in the range [-4,-1].}
\end{figure}

\subsection{Effects of viewing direction on the UV lines}

To see if these models reproduce the observed CIV/Ly$\alpha$ and
CIV/CIII] ratios, we have presented the observed and predicted values
in Fig.~25. The line ratios shown correspond to calculations in which the
ionization parameter is varied, generating a sequence in $U$. The three sequences differ in density or metallicity and log$U$ is in the range [-4,-1].

\begin{figure}
\includegraphics{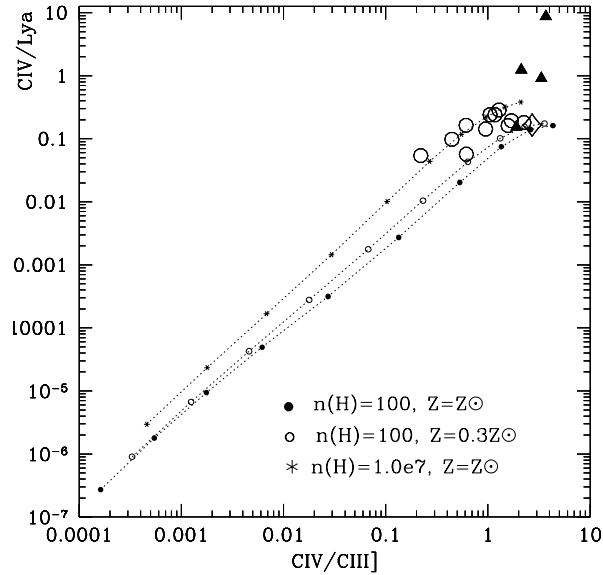}
\vspace{2.8in}
\caption[]{\small Observed and predicted UV line ratios. Filled triangles are the
ratios of objects observed by several authors. Open 
circles correspond to data taken from van Ojik's thesis,
selected on the basis of a very steep radio spectrum.  The open diamond is the average radiogalaxy
spectrum of McCarthy (1993). 
Dotted lines represent models in which the front and back spectra have
been summed. $log~U$ is in the range [-4,-1], as in Fig.~24. It is clear that high values of the ionization parameter
$U$ are required to reproduce the high CIV/CIII] value observed.}
\end{figure}	


We find that somewhat higher values of $U$ than are tipically used in the  optical are required to
reproduce the CIV/CIII] ratio observed. We have investigated possible effects 
due to variations in metallicity and density (see Fig.~25), but none of them is able to explain the high values of the line ratios, unles a higher ionization parameter than in the optical is considered. This is most probably a consequence of
the different way in which the objects at high and low~$z$ are selected: the
distant sources are all very powerful radio sources with luminous AGN. For a
power law of index --1.4, the optimum value is $U\simeq 0.1$. Hereafter,
diagrams will only cover the range of [0.01,0.1] in $U$.

A notable observation is the distribution of observed points above the model loci
in Fig.~25 even at the highest value of $U$. In the extreme objects at least ---
like F~10214+4724 and TX~0211-122 --- this results from a weakening of
Ly$\alpha$ rather than from atypical values of CIV/CIII]. It is usually claimed that
the destruction of Ly$\alpha$ photons by resonance scattering in the presence of
dust is the explanation for its faintness, but why does {\em not} the same process
reduce CIV which is also a resonance line? Is there an alternative explanation
for this selective dimming of Ly$\alpha$?

To answer this question, we examine the geometrical aspects of the line
formation process. Each emitting cloud is approximated as a plane parallel
slab which contains a fully ionized region and a partially ionized zone
where low ionization species co-exist (e.g., O$^0$, S$^+$, etc) with a
mixture of H$^0$ and H$^+$. In principle there can be an additional neutral
zone which does not contribute to the emission line intensities (see
Fig.~4, Chapter 2).



In this section we consider the dust-free case. For most lines (like CIII],
HeII, H$\beta$, [OIII], etc) line opacity is negligible and the line is emitted
isotropically with photons escaping freely in all directions. However, when the
line opacity is important as it is for CIV and Ly$\alpha$, line scattering
occurs which increases the path length. Another important effect of large
optical depths is that the line photon will not escape isotropically. A
resonance line photon following many scatterings must statistically escape in
the direction of highest escape probability which can be shown to be the front
for the photoionized slab depicted in Fig.~4 (Chapter 2). In the case of Ly$\alpha$, the
reason is that --- while Ly$\alpha$ photons are generated more or less
uniformly within the slab (except within the PIZ) by recombination --- the
neutral fraction and therefore the incremental line opacity $d\tau _L/dx$
increases monotonically as a function of depth as discussed in more detail by
Binette et al. 1993b (BWVM3). This means that for an {\it open} geometry like that shown in Fig.~4 (Chapter 2),
the zone of equal escape probability of front {\it vs} back occurs far beyond
the point where half the luminosity of Ly$\alpha$ is produced. For the
collisionally excited CIV line, the tendency to escape from the front also
exists although it is less pronounced. It arises mainly because the emissivity
of CIV is larger towards the front due to the temperature gradient across the
C$^{+3}$ zone. Note that while CIV is emitted and scattered within the rather
limited zone containing C$^{+3}$, Ly$\alpha$ remains subject to scattering
outside the region where it is produced. While most Ly$\alpha$ emission is
produced in the fully ionized zone, most of the line opacity occurs within the
PIZ. The presence of a layer of neutral gas beyond the PIZ will increase the
anisotropy of Ly$\alpha$ escape.

\begin{figure}

\includegraphics{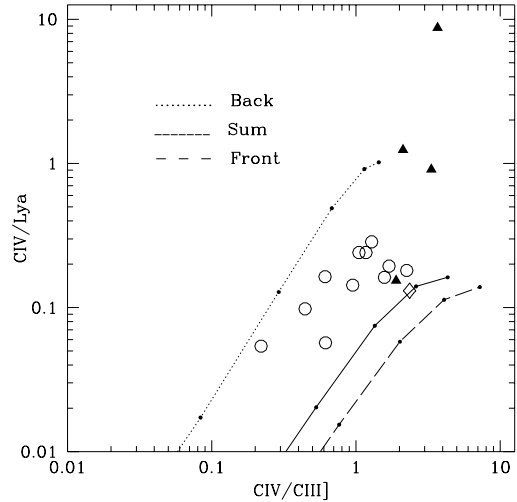}
\vspace{2.95in}
\caption[]{\small Influence of viewing geometry on the UV line ratios. The dotted 
line corresponds to the line spectrum seen from the {\it back} of the slab
(cf, Fig~4, Chapter 2) while the dashed line corresponds to the spectrum seen from the
{\it front} (UV irradiated face). It is apparent that perspective plays a
very important role on the CIV/Ly$\alpha$ ratio. The solid line represents
models obtained by summing the back and front spectra which would represent
symmetric case where equal numbers of clouds are observed with shadowed and
illuminated faces. The data are the same as in Fig.~25.}

\end{figure}

The effects described qualitatively above are shown in Fig.~26 using detailed
photoionization calculations. We present the same sequence of dust-free models
as in Fig.~25 but distinguish between the spectrum seen from the {\it back} ---
equivalent to observing the clouds through the PIZ --- from that seen from the
{\it front} --- equivalent to seeing the UV irradiated side. The differences
are striking: both Ly$\alpha$ and to a lesser extent CIV are fainter when seen
from the back. The CIII] line is isotropic in the dust-free case. The fact that
Ly$\alpha$ is more affected by perspective is due to the significant amount of
neutral hydrogen (i.e., large line opacity) in the PIZ which acts as a mirror.
Although we have considered a very simplified geometry in our calculations, the
method nevertheless treats properly the essential physical effects and
indicates how important the viewing direction is in this open geometry.

Fig.~26 suggests that perspective effects alone ({\it without any dust}) are
sufficient to explain the weak Ly$\alpha$ seen in some objects. Ionization
bounded calculations with $U=0.1$ imply total hydrogen column densities (H$^+$
region + PIZ) $N_H \sim 10^{22} \,{\rm cm}^{-2}$ (of which about 60\% is ionized).
Adding a modest neutral zone beyond the PIZ of $\simeq 4\, 10^{21} \,{\rm cm}^{-2}$
would double the CIV/Ly$\alpha$ ratio without affecting in any way the
CIV/CIII] ratio. It seems, therefore, that a geometry where we see
preferentially the ionized gas from the side of the PIZ gives us an explanation
for the weakness of Ly$\alpha$.

How would this apply to the EELR of  powerful radio galaxies? In a very
simplified scheme, we can imagine (see Fig.~27) that the clouds seen from the
nearside cone are seen from a direction which we approximate as the back
perspective in our slab calculations while clouds on the far side would be seen
from the front. The studies of McCarthy, van Breugel \& Kapahi (1991) which emphasized the
one-sideness of the line brightness distribution suggest that the observer with limited
spatial resolution at very high~$z$ will be biased towards either a back or front
dominated perspective depending on whether it is the near or the farside
illuminated cone which is intrinsically brighter. Our proposed interpretation
of the high CIV/Ly$\alpha$ ratio of some objects in Fig.~26 is that they
correspond to the case where the brightest clouds are seen from behind. Note
that, because of the intensity weighting, there is rather little difference
between the pure front perspective and the sum of equal numbers of front- and
back-clouds, a small effect which is not easily distinguished from that of
slightly reducing the value of $U$.

\begin{figure}[htb]
\includegraphics{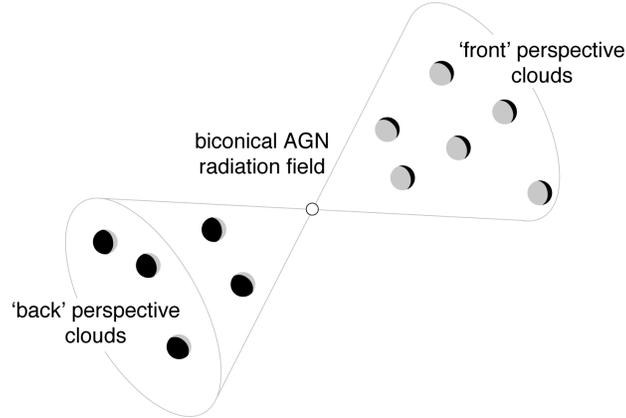}
\vspace{2.5in}
\caption[]{{\it Front} and {\it back} clouds within the ionized cones of a radio galaxy.}

\end{figure}

Our conclusion from this section is that perspective effects in an open
--- externally illuminated cloud --- geometry go a long way towards
explaining the behaviour of the CIV, Ly$\alpha$, CIII] line ratio diagram which implies that, when spatially resolved spectra become available, marked
asymmetries in the CIV/Ly$\alpha$ ratio could arise when the axis of the
illuminated cones forms a large angle with the sky plane.

\subsection {The effects of internal dust}

\begin{figure}
\includegraphics{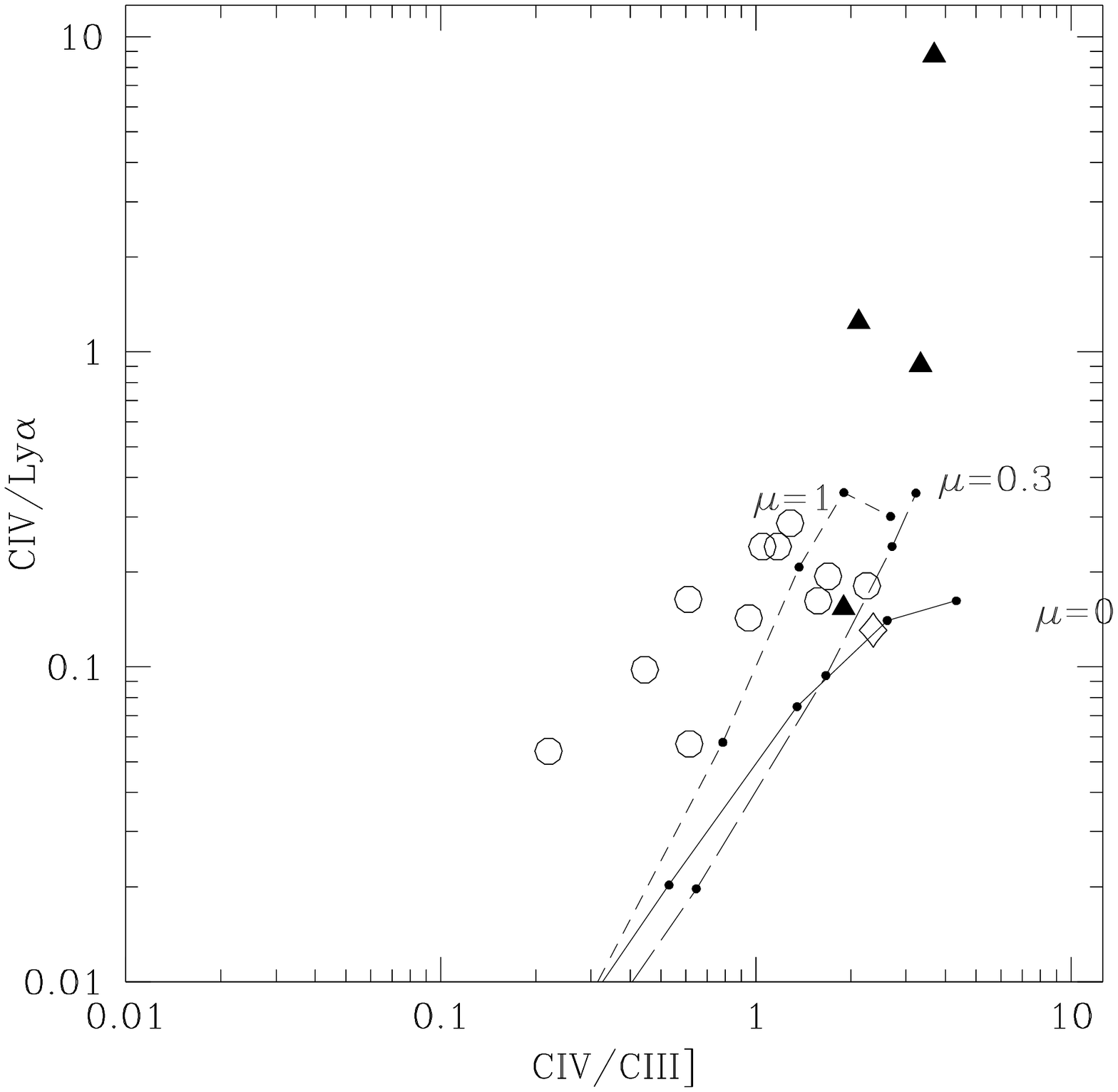}
\vspace{2.5in}
\caption[]{\small Effects of varying the amount of internal dust as seen from
the front perspective. As in the previous figures, the various lines correspond to sequences in $U$. Short-dashed line corresponds to models with $\mu=1.0$, the
long-dashed to models with $\mu=0.3$ and solid line to dust free
models.}

\end{figure}

\begin{figure}
\includegraphics{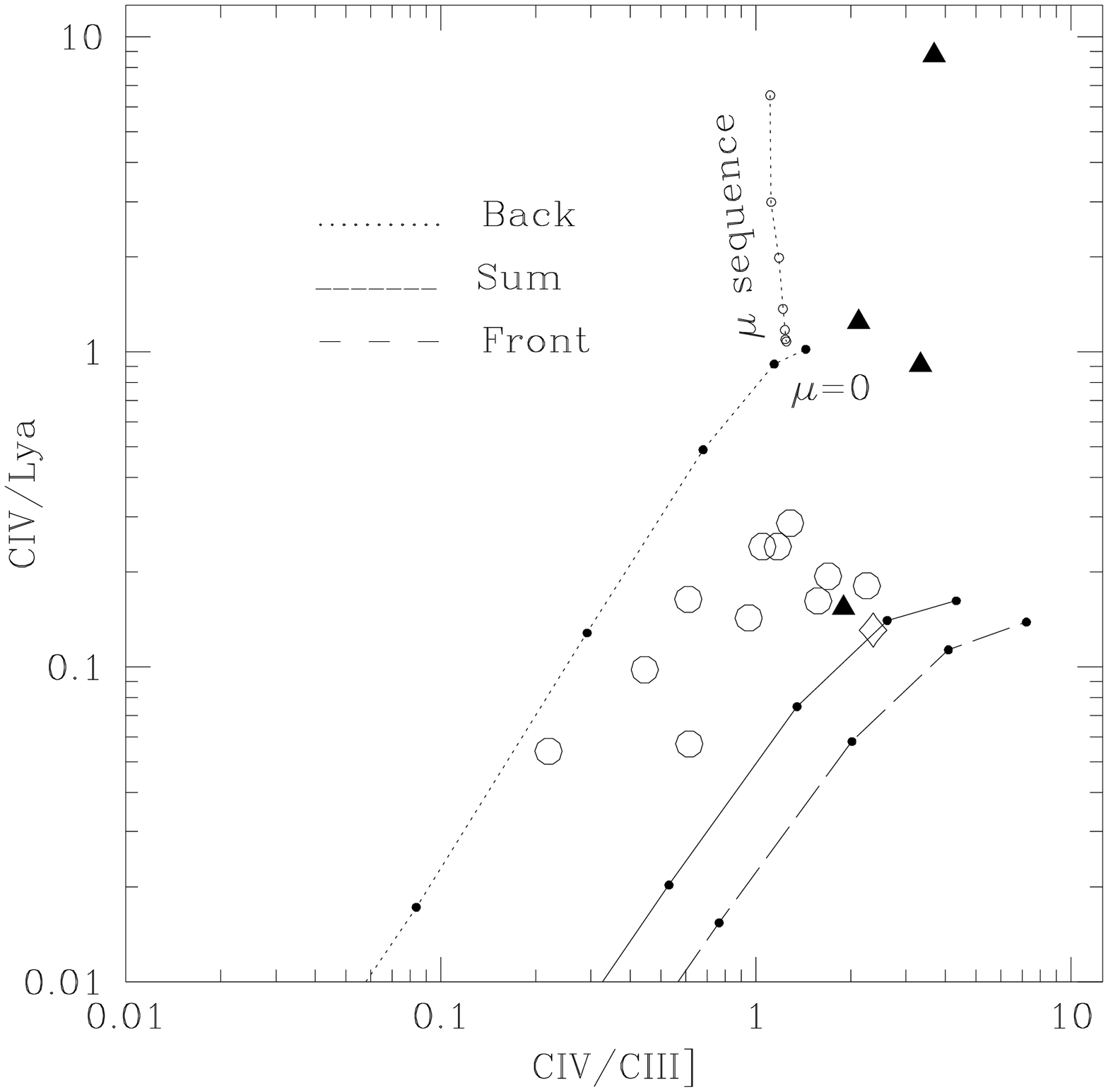}
\vspace{2.5in}

\caption[]{\small Effects of perspective combined with small quantities of
internal dust. The dust-free $U$ sequences of Fig.~26 are repeated in
this figure. The dotted line connected by open circles corresponds to
the last model with $U=0.1$ (but with metallicity $Z= 0.3$ in which
the dust content is increased in proportion up to $\mu=0.017$. Larger
quantities of dust do not improve these results because dust 
severely extinguishes both the
CIII] and CIV lines which are produced in the front layers. }

\end{figure}

There is evidence for the existence of dust in some high~$z$ radio
galaxies. The detection of 4C41.17 ($z=3.8$) and B2~0902+34 ($z=3.5$)
(Chini \& Kr\"ugel 1994; Dunlop et al. 1994) and 8C1435+635 ($z=4.26$)
(Ivison 1995) in the mm spectral range is attributed to warm
dust. Also, the IRAS galaxy F10214+4724 ($z=2.29$) has been shown,
from the far infrared flux, to contain $\sim 10^8 M_{\odot}$ of dust,
although this figure may be reduced by the gravitational lensing
amplification factor (e.g. Eisenhardt et al. 1995; Serjeant et~al. 1995). In addition, the mid- to
far-infrared measurements at intermediate~$z$ are explained as emission
from warm dust (Heckman, Chambers \& Postman 1992). We should
emphasize, however, that we have no direct measurement of how this
dust is spatially distributed. In the event that this infrared
emission arises from the reprocessing of higher energy photons from
the AGN by a dusty torus, it has no direct bearing on our modeling of
ionized gas at tens of kpc. However, there is considerable evidence
that aligned blue polarized continuum is the result of scattering of
the anisotropic nuclear radiation field by dust (e.g., Cimatti
et~al. 1993). In this case it is very likely that at least some of the
dust is internal to the extended line emission regions. In this
section we consider the effects of including dust within the gas
clouds which emit the UV lines.

We illustrate the effects of dust mixed with the emitting gas in
Fig.~28 where we plot sequences of models which correspond to the front
perspective for three different dust-to-gas ratios: $\mu=0$, 0.3 and
1. The effects on the line ratios are evident. The resonance
scattering suffered by CIV and Ly$\alpha$ increases their pathlengths
many times and, therefore, the probability of their being absorbed by
dust grains is much higher than for CIII]. In the case of Ly$\alpha$,
however, the geometrical thickness of the H$^+$ region exceeds greatly
that of the C$^{+3}$ since we observe more than one stage of
ionization of metals (e.g. C$^{+2}$). Any reasonable parameters for
the ionization structure of a photoionized slab with $Z\sim 1$
indicates that the opacity in Ly$\alpha$ greatly exceeds that of CIV,
which implies a larger pathlength increase for Ly$\alpha$ than for
CIV. This results in relatively more Ly$\alpha$ absorption by dust.
This effect explains how the ratio CIV/Ly$\alpha$ increases somewhat
with increasing $\mu$.  Dust absorption of resonant CIV on the other
hand causes a comparable decrease in CIV/CIII]. What is important in
these results is that, even with a concentration of internal dust as
high as $\mu =1$ (equivalent to that in solar neighbourhood cold
clouds), it is not possible to reproduce the high ratio CIV/Ly$\alpha
\ge 1$ which is observed in some objects and has been attributed to
dust.

When the clouds are viewed from behind, even with $\mu =0.3$ they are
sufficiently opaque that all the UV lines are severely absorbed. With $\tau_V=5 \, 10^{-22} \mu N_H$ $\sim 5$ and $\mu =1$, all of the high excitation UV
lines become absorbed within the PIZ. To produce an acceptable spectrum,
without reddened CIV/CIII] and CIV/HeII ratios, as seen from the back of an
ionization bounded slab with $U=0.1$, we have to use much smaller amounts of
dust like $\mu \simeq 0.017$ (2\% of local ISM) in models. In Fig.~29, we plot
the back and front sequences for such models. These can reproduce the weak
Ly$\alpha$ objects although this result is obtained only for the back spectrum,
emphasizing that perspective is the dominant factor. It should be noted that
the amount of extinction within the ionization bounded slab implied by $\mu =
0.017$ (i.e., $A_V \simeq 0.1$--0.2) is consistent with the quantity of small
dust grains needed to explain the extremely blue continuum of the ``detached''
ionized cloud in PKS 2152-69 (di Serego Alighieri et~al. 1988; Magris \&
Binette private communication), supposing that the continuum energy
distribution is the result of dust scattering of the nuclear radiation.

Note that because the scattering/absorbing dust is locally kinematically linked
to the emission line gas, we require much smaller column densities of HI than
the absorbing screen proposed by van~Ojik et~al. (1994) ($\sim 10^{23}
\,{\rm cm}^{-2}$) to explain CIV/Ly$\alpha \sim 1$. The line widths and centroids of
the PIZ and of the fully ionized gas are expected to be quite similar within
each of the emitting clouds, the ensemble of which could have a greater
`turbulent' velocity dispersion.

To reproduce the extreme case of the IRAS galaxy F10214+4724 with CIV/Ly$\alpha$ $\sim$ 10, we need additional neutral gas beyond the PIZ. For instance, it
requires only a column density of $N_{H^0} \sim 1.5 \, 10^{21} \,{\rm cm}^{-2}$
assuming $\mu=0.017$. Alternatively we might consider that $\mu$ within the PIZ
increases with depth as the degree of ionization decreases. In this case, no
additional HI gas would be required. If this IRAS galaxy is indeed an extreme
Seyfert~2 as argued by Elston et~al. (1994), then a closed, dust enshrouded
geometry as proposed by BWVM3 to explain the Lyman and Balmer decrements in
Seyfert~2 would be more appropriate than the open geometry adopted here which
may be applicable only to the truly extended large scale gas. It is likely that
objects with unusually strong NV$\lambda 1240$ emission (such as in F10214+4724
and TX0211-122) are cases where the NV originates predominantly from the inner
NLR. A high NV/CIV ratio indicates very enriched gas which is not unexpected
within the inner parts of an AGN (Hamman \& Ferland 1993) and is consistent
with the lack of convincing evidence that NV is spatially resolved in any of
these objects.

\subsection{Neutral gas mirrors}

So far we have considered the effects of scattering by gas which forms part of
the line emitting clouds: we refer to this as an `intrinsic' process. Similar
effects could be produced by a large-scale distribution of predominantly
neutral material surrounding the emitting regions: we will refer to this as the
`extrinsic' case. Any extrinsic neutral gas component with a non-negligible
covering factor could affect the observed spectra in a way which mimics that of
the back perspective described earlier. Let us suppose that this outer material
is broken up into cold gas clumps which are randomly distributed. In such a
case, some Ly$\alpha$ photons which leave the ionized cones will escape the
region through the holes between the external clumps while others will strike
the neutral clumps and be immediately scattered away to escape eventually
through another hole in a different direction (see Fig.~30). If observed with
sufficient spatial resolution, such a geometry would result in holes in the
Ly$\alpha$ brightness distribution due to reflection by intervening clumps as well as
diffuse sources corresponding to reflection from clumps on the far side of the
source. The bulk of the Ly$\alpha$ luminosity would be preserved but
redistributed on an apparently larger scale than the true line emitting clouds.
Only a more closed geometry would result in a significant destruction of
Ly$\alpha$, assuming that the interclump space does not contain pure dust
segregated from the gas phase.

The reflection efficiency of clouds will, of course, depend strongly on the
relative velocity fields of the emitting and the cold regions. If the extended
gas has a large scale ordered motion but small `microturbulence', the
reflection effects would be localized. Broad Ly$\alpha$ emission and continuum
from the AGN could, however, be scattered by any of the extranuclear
clouds (see below, 3C294). In
general, we would expect the diffuse, scattered Ly$\alpha$ to show a narrower
line profile than the integrated emission profile.

We will review the observational evidence for the existence of such large scale
mirrors.

\subsubsection{The radio galaxy PKS2104-242}

PKS2104-242 (McCarthy et~al. 1990b) shows very extended emission lines of
Ly$\alpha$, CIV and HeII associated with continuum knots which lie between the
radio lobes and are aligned with the radio axis. The Ly$\alpha$ image shows
emission resolved into three distinct clumps, two of them corresponding roughly
with the two continuum knots. In addition, there appears to be a low surface
brightness halo of emission surrounding the entire object. Is this halo a
consequence of reflection by neutral material of either Ly$\alpha$ emission or
of the intense nuclear {\it continuum}, or are these photons emitted locally by
H$^+$ recombination? A way of discriminating between these two possibilities is
to look for the detection of any other emission line in the halo. If Ly$\alpha$
is the result of reflection by cold HI, there will be {\it no} other lines
(except possibly resonant MgII). If it is instead produced by recombination,
other lines should be detected in the rest-frame optical band like H$\alpha$,
[OIII]$\lambda$5007, etc.

\begin{figure}
\includegraphics{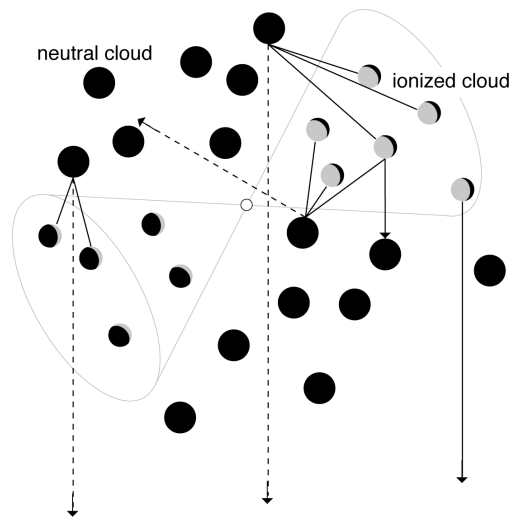}
\vspace{3in}

\caption[]{\small The extrinsic case. Neutral material external to the
ionized regions can strongly influence the appearance of the object in
Ly$\alpha$. In this figure, the neutral clumps (black clouds) cover a
significant fraction of the ionized regions (grey+black clouds). The solid
arrows represent Ly$\alpha$ photons emitted directly in the ionized
regions. Dashed arrows represent Ly$\alpha$ photons that after striking
neutral clumps are then reflected in other directions. The reflection by
the neutral H atoms is so effective that the dust has little chance to
interact with the photons before they find a hole to escape.
}

\end{figure}

\subsubsection{The radiogalaxy 3C~294}

From images of this object, McCarthy et~al. (1990a) report the
existence of Ly$\alpha$ emission extending over a region covering
170kpc. The Ly$\alpha$ emission is elongated and well aligned with the
inner radio-source. An intriguing aspect of the observations is the
one-sidedness of the CIV emission (obtained with an long slit aligned
with the axis of elongation). The decreasing linewidth of Ly$\alpha$
on the side where CIV is absent (South) suggests to us the possibility
that Ly$\alpha$ in the south corresponds to scattered {\it continuum
and BLR Ly$\alpha$ photons} by HI clumps lying along this direction
provided the ionizing radiation has already been reprocessed (filtered
out) within the nuclear regions into NLR or BLR emission.

Failing that (scattering by HI clumps), an alternative possibility
resides in very thin matter-bounded photoionized sheets of gas which
can be very efficient at reflecting the intense (beamed) nuclear
continuum as well as BLR Ly$\alpha$ photons. Due to its large
scattering cross section, very small column depths of H$^0$ (a trace
specie within the ionized phase) are sufficient to scatter effectively
the impinging flux within the core of the Ly$\alpha$ thermal profile.
If we suppose the existence within the cone of ionization of the
radio-galaxy of a population of clouds of similar physical conditions
to that thought to apply to Ly$\alpha$ forest clouds, the energy
reflected due to resonant scattering by HI typically exceeds that
generated within the clouds by reprocessing of the ionizing radiation
(i.e. by recombination). In effect, adopting $N_{H^0} = 10^{13.8}
\,{\rm cm}^{-2}$ as a typical column density of such cloud, we derive
an equivalent width in absorption (i.e. the scattered intensity) of

$$EW^{abs}_{Ly\alpha} \approx 0.15 \sqrt{{\rm ln}(4.2\,10^{-14}
N_{H^0})} \,{\rm \AA }=0.15{\rm \AA } $$

\noindent which corresponds to the saturated part of the Ly$\alpha$
curve of growth. Adopting for
definiteness $U=0.1$ and $\alpha =-1.4$, we obtain that the fraction
of reprocessed ionizing photons, $F_{MB}$, within a very thin
photoionized sheet\footnote{With these parameters, the total column
density is given by $ N_{H} = N_{H^+} \approx 2.3 \, 10^4 N_{H^0}$.}
is given by

$$F_{MB}\approx 1.8 \, 10^{-18} N_{H^0}= 1.1 \, 10^{-4}$$

\noindent The equivalent width of Ly$\alpha$ in emission as a result of
recombination is simply

$$EW^{em}_{Ly\alpha} \approx 390\, F_{MB} \, C_f \, {\rm \AA}=0.04\, C_f \,{\rm \AA}$$

\noindent where $C_f$ is the covering factor. If we integrate the
absorbed intensity over the same area as that given by $C_f$, the
ratio of scattered versus emitted Ly$\alpha$ is $\sim 4$ (the ratio
between the above two equivalent widths). Allowing for the fact that
the cloud will be exposed not only to the nuclear continuum but to the
BLR Ly$\alpha$ which typically peaks at twice the continuum level, the
ratio can be expected to increase up to 8 times the Ly$\alpha$ emitted
by recombination. To account for the observed FWHM $\le 1000$\,
km\,s$^{-1}$ in 3C294 (South), one most rely on a turbulent velocity
field for the the Ly$\alpha$ clouds.

We ought to consider seriously the possibility that Ly$\alpha$ on the
southern knot of 3C294 corresponds to large scale scattered light by
either HI or even ionized gas unless other bona fide emission lines
were detected. So far, no  lines other than Ly$\alpha$ are observed
and this is consistent with our suggestion. The only certain way of
discriminating between reflection and {\it in situ} emission would be
the detection at this radius of {\it any} non-resonance line in either
the optical ([OII], [OIII], H$\beta$ etc.) or the UV (CII], CIII]). To
conclude, we believe that HII regions or starbursts should not be
considered the default emission mechanism in cases where only
Ly$\alpha$ is present, especially along the axis where the AGN
supposedly collimates its intense nuclear continuum~+~Ly$\alpha$(BLR)
light.

\subsubsection{Ly$\alpha$ absorption in 0943-242}
	
The radio galaxy 0943-242 shows a Ly$\alpha$ profile with several
absorption features which cover the whole spatial extent of the
Ly$\alpha$ emission (R\"ottgering et~al. 1995). The strongest of the
absorbers is at least as extended spatially as the emitting region
($\sim 13$\,kpc) and its HI column density is $10^{19} \,{\rm cm}^{-2}$. The
absorption line is blue-shifted 250km~s$^{-1}$ with respect to the
emission peak. The `screen' is kinematically distinct from the
emitting gas and therefore clearly external to the emitting
clouds). This is a clear demonstration that HI screens --- or mirrors
depending on the perspective --- exist on galaxy scales at these early
epochs. From Fig.~5 of R\"ottgering et~al., we estimate that this
cloud would reflect $\sim 30$\% of the Ly$\alpha$. Thicker clouds
could exist around other objects but would be hard to detect when only
the faint wings at the extremities of the emission profile were
transmitted. In the case of 0943-242, even if the screen contained
dust, its effects would be negligible. The reason, as stated before,
is that Ly$\alpha$ photons (seen by the cloud) are incident from the
exterior and will be very effectively scattered away before being
absorbed by the dust.  Even the photons getting through (i.e., without
any scattering) to us because they are sufficiently far in the wings
of the profile would not see much dust in this particular cloud. With
such low column densities, $\tau_V \sim 5 \, 10^{-22} \mu N_H =
7.5\,10^{-5}$ for $\mu=0.015$ and $\tau_V \sim$ 0.005 for $\mu=1$. Much thicker
clouds may result in some non-negligible extinction but again this
does not arise because of the multiple scattering of Ly$\alpha$.

\section{Conclusions}

If the large-scale extended emission line regions (EELR) in radio galaxies are
photoionized predominantly by the collimated UV radiation emitted by a hidden
AGN, the line emission and transfer processes are characterized by what we call
an 'open' geometry with externally illuminated clouds. This produces an
emission line spectrum which is distinctly different from an internally ionized
HII region, especially in the ultraviolet. 

Our principal conclusion is that internal dust, even in proportions as large as
that found in the solar neighbourhood, does not satisfactorily explain the
large CIV/Ly$\alpha$ ratios commonly seen in groundbased observations of 
high redshift radio galaxies. We have demonstrated that geometrical effects
can, in most cases without invoking any dust, explain the observed trends in
the data. While internal dust can quite effectively kill the Ly$\alpha$
luminosity, it is essential to examine also its effects on the other UV lines, especially CIV which is also a resonant transition. Although some internal
dust within the ionized gas is allowed by our modeling and is required in a
few cases, we find that the effects of geometrical perspective dominate the
observed decrease of Ly$\alpha$ without affecting excessively the other lines.
This implies that the faintness of Ly$\alpha$ cannot necessarily be used as an
argument to demonstrate the existence of dust in these objects. 

The influence of these geometrical factors on the line spectrum gives us,
potentially, a natural explanation for some of the emission line asymmetries
seen in the 'alignment effect' radio galaxies.

We have also investigated the effects that material external to the ionized
cones will have on the appearance of the objects in the light of Ly$\alpha$. We
suggest that the existence of diffuse Ly$\alpha$ halos observed in some
high~$z$ radio galaxies could be due to the reflection of Ly$\alpha$ photons by
neutral clumps lying outside the ionization cones. In addition to reflecting
Ly$\alpha$, such neutral clouds could be responsible for the spatially extended
absorption seen within the Ly$\alpha$ emission profile. Pure Ly$\alpha$
emission clouds need not, therefore, necessarily be identified as star-forming
regions.

\vspace{0.5cm}

\addcontentsline{toc}{part}{Bibliography}
\parskip=1ex
{\Large \bf References}

\vspace{0.5cm}

\noindent Binette L., Wang J.C.L., Zuo L., Magris C.M., 1993a, \aj 105, 797

\noindent Binette L., Wang J.C.L., Villar-Mart\'\i n M., Martin P.G., Magris C.M., 1993b, \apj 414, 535
(BWVM3)

\noindent Chambers K.C., Miley G.K., van Breugel W.J.M.,
1987, {\it Nature}, 329, 604

\noindent Chini R., Kr\"ugel E., 1994, {\it A\&A}, 288, L33

\noindent Cimatti A., di Serego Alighieri S., Fosbury
R.A.E.,
Salvati M., Taylor D., 1993, \mnras 264, 421

\noindent  di Serego Alighieri, S., Binette, L., Courvoisier, T. J-
L., Fosbury, R.A.E. \& Tadhunter, C.N., 1988. {\it Nature}, 334, pp. 591-593

\noindent di Serego Alighieri S., Cimatti A., Fosbury
R.A.E.,
1993, \apj 404, 584

\noindent di Serego Alighieri S., Cimatti A., Fosbury
R.A.E., 1994, \apj 431, 123

\noindent  Dey A., Spinrad H., Dickinson M., 1995, \apj
440, 515

\noindent  Dunlop J.S., Hughes D.H., Rawlings S., Eales
S.A.,
Ward M.J., 1994, {\it Nature}, 370, 347 

\noindent  Eales S.A., Rawlings S., 1993, \apj 411, 67

\noindent Eisenhardt P., Soifer B.T., Armus L., Hogg D., Neugebauer G., Werner M., 1995, {\it BAAS}, 186, 5202

\noindent  Elston R., McCarthy P.J., Eisenhardt P.,
 Dickinson M., Spinrad H., Januzzi B.T., Maloney P., 1994, \aj 107, 910

\noindent Fosbury R.A.E., di
Serego Alighieri S., Courvoisier T., Snijders M.A.J., Tadhunter C.N.,
Walsh C.N., Wilson W., 1990, in Evolution in Astrophysics, Toulouse,
ESA SP-310 

\noindent  Hamman F., Ferland G., 1993, \apj 418, 11

\noindent  Heckman T.M., Chambers K.C., Postman M., 1992,
\apj
391, 39

\noindent  Hummer D.G., Kunasz P.B., 1980, \apj 236, 609

\noindent Isaak K.G., McMahon R.G., Hills R.E., Withington S., 1994, \mnras, 269, L28

\noindent  Ivison R.J., \mnras 275, L33, 1995

\noindent  Lilly S.J., 1988, \apj 333, 161

\noindent McCarthy P.J., van Breugel W.J.M., Spinrad H.,
Djorgovski S., 1987, \apj 321, L29

\noindent  McCarthy P.J., Spinrad H., van Breugel W.J.M., 
Liebert J., Dickinson M., Djorgovski S., Eisenhardt P., 1990a, \apj 365, 487

\noindent  McCarthy P.J., Kapahi V.K., van Breugel W.J.M., 
Subrahmanya C.R., 1990b, \aj 100, 1014

\noindent McCarthy P.J., van Breugel W.J.M., Kapahi V.K., 1991, \apj
371, 478

\noindent McCarthy P.J., Elston R., Eisenhardt P., 1992,
\apj 387, L29

\noindent McCarthy P.J., 1993, {\it ARA\&A} 31, 639

\noindent  van Ojik R., R\"ottgering H.J.A., Miley G.K.,
Bremer
M.N., Macchetto F., Chambers, K.C., 1994, {\it A\&A}, 289, 54

\noindent  van Ojik R., 1995, Ph.D.thesis, Rijksuniversiteit te
Leiden

\noindent  Robinson
A., Binette L., Fosbury R.A.E., Tadhunter C.N., 1987, \mnras 227, 97 (RBFT87)

\noindent  R\"ottgering H.J.A., Hunstead R.W., Miley G.K., 
 van Ojik R., Wieringa M.H., 1995, \mnras in press 

\noindent Serjeant S., Lacy M., Rawlings L.J., King L.J.,
Clements D.L., 1995, \mnras 276, L31

\noindent Spinrad H., Filippenko A.V., Wyckoff A., Stocke
J.T.,
Wagner R.M., Lawrie D.G., 1985, \apj 299, L7

\noindent  Sutherland R.S., Bicknell G.V., Dopita M.A.,
1993, \apj 414, 510

\noindent Tadhunter C.N., Fosbury R.A.E., Quinn P.J., 1989, \mnras 240, 255

\noindent  Tadhunter C.N., Robinson A., Morganti, R., 1989.
In: {\it ESO Workshop on Extranuclear Activity in Galaxies}, p. 293,
eds Meurs, E.J.A., Fosbury, R.A.E., ESO Conf. and Workshop Proc. No.
32, Garching


\chapter{Discussion}
\pagestyle{myheadings}

\markright{Discussion.}

        The projects developed in this thesis, although selfcontained,
provide results relevant, not only to the individual class of objects or
gaseous regions considered, but also to the nature of active galaxies in
general.         In this concluding chapter, I want to summarize what we have
learned about the  family of active galaxies and connect our results with the current ideas on this field.

        Active galaxies have been subdivided into several subgroups
defined according to
the strength and breadth of their emission lines and  their
luminosity over the entire electromagnetic spectrum.
 Sometimes, different physical conditions and emission mechanisms are
responsible for the differences between subgroups. In some cases, however, objects pertaining to distinct groups can be intrinsically similar, the differences being
due to pure geometrical effects.
This statement summarizes the basic ideas of the so called ``unified
scheme models''  (e.g., Lawrence \& Elvis 1982, Lawrence 1987, Barthel 1989),  mentioned several times in this work. Such model propose that
different classes of AGNs are the same kind of object but seen from different
directions. The directionality is taken to be due to an anisotropic
distribution of absorbing or scattering matter and a simple geometry which
would produce this situation  involves an absorbing torus around the
nucleus  (the proposed ``molecular torus''). In such
scenario, the continuum produced in the central, hidden  nucleus  is emitted
anisotropically into two wide, oppositely directed ionized cones, as a result
of the collimation by the torus.

    Where does the unified scheme enter our discussion? It is represented in the kind of geometry assumed in all our calculations,
in which a hard ionizing continuum illuminates  the gas {\it externally}:
a different situation from the one existing in HII regions, or planetary
nebulae, where the continuum source (the stars) is embedded within the gaseous
region.  In AGNs, numerous pieces of evidence indicate that the main ionization mechanism of the EELRs is 
photoionization by  a hard continuum, which is
produced in the central engine (shocks could also play a role, see section
6.2). Such situation is better described by an open geometry.

 The resulting models are able to explain the observed data which
suggests that the scene is basically correct. However, the relatively high number of input
parameters necessary to construct the models allows  an apparent high
degree of freedom to fit the observational results. One might argue that, using
another geometry, the same strategy would have driven us to the same
results.  The question is, how unique is the scheme that can reproduce the
observed data?

 Different observational pieces of evidence restrict  the n-dimensional space of
valid input parameters, to a much smaller space than that  of all the
models able to reproduce the observed emission line ratios. In this thesis,
each chapter where theoretical models are presented, justifies in detail
the assumed input parameters, based on our own or other authors' knowledge.
We select the input parameters in accordance with the
nature of the objects and regions under study. 

	This chapter is divided into several sections, each one referred 
to a specific aspect of the nature of active galaxies: dust, ionization mechanisms and geometry. Every section is completed with a brief overview of the current state of the field, as well as a comparison between other authors' results and ours.

\newpage

\section{ Dust in the ionized regions.}

        Dust has been the main  object of investigation in  this thesis.
We have provided a detailed
analysis of the interaction between the dust grains and the radiation impinging  on the 
gaseous component. We have also studied its effects  on the line emission from the ionized gas like changes on the ionization structure of the gas, electronic
temperature, metallicity, etc. Taking these effects into account is
necessary in order to interpret more correctly what is observed.

        This understanding is useful, for instance, to distinguish when
dust must be considered and when its effects can be neglected, allowing a simplification in the interpretation. Robinson et al. (1987, hereafter RBFT87) showed that in
several diagnostic diagrams involving optical lines,  the observed position of
the EELR of radio galaxies at low z were well represented by a sequence in
which one simply varies the ionization parameter, U (a measurement of the excitation level of the
emitting gas, see Chapter 2). Using the observed values of the line ratios, they inferred 
that the EELR spectra were not reddened and presented models which did not include dust effects.

        I used the same set of data employed by these authors in order to
investigate how dust mixed with the ionized gas would influence  the
predicted line ratios and, therefore, the positions within the diagnostic
diagrams (Villar-Mart\'\i n, 1993). The models showed  that  internal
dust does not affect severely the optical line ratios (unless species
susceptible of depletion are considered!). The influence is negligible for
low U values and increases with this parameter. For highly excited EELR,
like a few of the objects in RBFT87's sample, we should be more careful.
At such high U, the effects of dust are stronger (see Chapter 2) and it might
be necessary to  consider them when interpreting their observed ratios.
However, for typical EELRs at low redshift, where the ionization parameter
is in the range [-4,-3] in the logarithm, dust
effects on the optical line ratios are not relevant. The absence of reddening effects does 
not imply that dust is {\it not} present, instead is probably  indicative of  an open
geometry for the gaseous ionized regions, where the clouds are externally
illuminated. In a closed geometry, dust effects are stronger and the
high reddening of some NLR line ratios of Seyfert 2 suggests a much higher
covering factor than previously assumed (see Chapter 2).

        The situation described in the previous paragraph has advantages
and disadvantages. On the one hand, if dust effects are negligible, we can simplify
enormously our modeling when interpreting the optical line ratios or the
diagnostic diagrams that have been traditionally used to learn about the
nature of the ionizing source and the physical conditions  of the ionized
gas. We are therefore confident that our diagnostics are not wrong because of  dust
effects which we did not take into account.
On the other hand, these same optical line ratios, easy to measure,
do not tell us about the properties of the dust in the EELR, not even
whether dust exists or not. As explained in Chapter 3, we need to develop other strategies and tools which are
highly sensitive to dust, like  observations of emission lines emitted by
species {\it susceptible to depletion}, as shown in detail in Chapters 3 and
4. In these chapters we have studied (theoretically + observationally)  the emission of the [CaII]$\lambda\lambda$7291,7324 lines, which are appropriate because 1) calcium is an element very sensitive to depletion, and  2) as our models demonstrate, these lines should be strong and easy to detect under the physical conditions of  EELR whose the gas does not contain dust. Their weakness   demonstrates that {\it dust does exist in the EELR of
low z RGs}. This favours the idea that the origin of the EELR is
interaction or merging of two or more galaxies rich in gas, rather than
cooling from a hotter corona, a scenario which is  difficult to reconcile with the
existence of dust mixed with the emitting gas (see section 4.4.1).

\subsection{Dust at high $z$.}

        The subsequent question is: {\it  can we extrapolate our conclusion
to high z EELRs? Can we affirm that they also contain dust?} The answer is
{\it NO}. The techniques applied in this thesis, do not demonstrate the
existence of internal dust mixed with the ionized gas of high z radio
galaxies. However, as explained in Chapter 5 and below, there is strong evidence  that dust exists in objects at redshifts
as high as 2 and even more. Our work does not contradict this,
but demonstrates  that the line diagnosis based on UV (or optical) line ratios does not help us to infer the presence of dust because other factors, like geometry, dominate what we see. 
	We have studied the UV emission line spectrum of a sample of high $z$ radio galaxies analyzing the position in the diagnostic diagram CIV$\lambda$1549/Ly$\alpha$ {\it vs.} CIV$\lambda$1550/CIII]$\lambda$1909. 
Both line ratios are very sensitive to geometry and dust due to the resonance character of the CIV$\lambda$1549 and Ly$\alpha$ lines. We have demonstrated that dust, even in proportions as large as the one found in our ISM, can not explain the observed  data, unless the influence of geometry id considered. In fact, our study shows that geometrical effects can in most cases, without invoking any dust, explain the observed data. Our models allow some internal dust but it must be in very little amounts ($\sim$ 2\% of the dust/gas ratio in our
ISM)  in order to not contradict other
line ratios, like CIV/CIII, which is not reddened. Effects of geometrical perspective, and not dust, dominate the observed line ratios.

This can have important implications. As explained in Chapter 5, the
weakness of Ly$\alpha$ has been used as an argument to demonstrate the
existence of dust at high z but we have shown that there is an alternative
explanation based on purely
geometrical effects.
 The resonance character of the Ly$\alpha$ line makes neutral hydrogen a
very efficient mirror for the Ly$\alpha$ photons which, depending on the
{\it geometry}, prevent photons  from reaching the observer. In this case
the line becomes much fainter than expected while no photons have actually been 
destroyed but rather deflected from our line of sight. Conversely this same effect can be responsible for the apparent strength
of Ly$\alpha$ emitted from regions associated with  neutral hydrogen and which reflects 
Ly$\alpha$ photons back towards the observer.
This could explain the huge Ly$\alpha$ halos observed around some objects,
where the line originates from zones which seem to be located out of the ionizing cones (like PKS0943-242, see section 5.3). This
reflection effect
is very efficient, whether dust is present or not, which means that
Ly$\alpha$ can even be stronger than predicted in the absence of dust, if
the geometry is favourable.

        We have shown that the Ly$\alpha$ line (its intensity or ratio with
other emission lines) is {\it not} a reliable indicator of dust, but, on the
contrary, a good tool to learn about the gas distribution (ionized as well as 
neutral) and, thus, about the geometry of the object. I will discuss further
this
point in section $\S$6.4.

 	The existence of dust  at high $z$ is indicated by other observational  pieces of evidence like the extended polarized blue continuum mentioned several times in this work. Also, as explained in Chapter 4, there are several objects at high $z$ detected in the millimeter  spectral range, like 4C41.17 ($z$=3.8) and B2 0902+34 ($z$=3.5) (Chini \& Kr\"ugel 1994; Dunlop et al. 1994). Their fluxes imply the presence of dust masses of the order of 10$^8$ M$_{\odot}$. Also high $z$ quasars have been detected in the millimeter region by Andreani et al. (1993), McMahon et al. (1993) and Isaak et al. (1994).

	The  paragraphs above refer to objects where the UV line emission is dominated by the EELR. There can also be objects where the unresolved {\it nuclear emission
dominates} over the extended emission (EELR). In such cases, as 
demonstrated in Chapter 2, a rather closed geometry would be more appropriate.
A good example is the IRAS source
F10214+4724 ($z$=2.3) (Downes et al. 1992), a galaxy where the emission lines seem to reveal a very strong 
 reddening. IR measurements indicate the existence
of large quantities of dust, due in part to the
amplification effects of a gravitational lens (e.g. Eisenhardt et
al. 1995, Serjeant et al. 1995) which amplifies the IR luminosity. We
know that dust is present, also from polarization
measurements (Lawrence et al. 1993; Goodrich et al. 1996). But is its
emission line spectrum as reddened as some authors claimed
(e.g. Elston et al. 1994)?  Our results indicate rather little
reddening (or a very patchy distribution) while Elston et al. (1994)
reported strong line reddening by measuring H$\alpha$/H$\beta$
$\geq$20, which would imply A$_V \geq$5.4. However, according to these same authors, the
CIV/CIII ratio is quite high ($\sim$ 3.6) ($\sim$3.4 according to
Goodrich et al. 1996).  Also, the HeII/CIII$\sim$1.7 and CIV/HeII$\sim$2.0 ratios 
(Goodrich et al. 1996) are similar to the values observed for
other high z objects (see Chapter 5 and Villar-Mart\'\i n, Binette \&
Fosbury 1995).  Why is H$\alpha$/H$\beta$ so reddened 
and not the UV line ratios?.  Goodrich et al. (1996) for
instance detect broad components of several UV lines in the polarized
flux. For CIII, however, a broad component can be seen in direct
upolarized light. Maybe H$\alpha$ has similarly a broad component in
direct unpolarized light while the fainter H$\beta$ has none. This
would result in an {\it apparent} higher Balmer decrement.  By
themselves, the UV line ratios do not indicate high quantities of
dust and as explained in Chapter 5, if this object is a Seyfert 2
(Elston et al. 1994), a more closed geometry might be more appropriate
for describing the emitting gas. And, as shown in Chapter 2, only
small amounts of dust are required to strongly weaken the Ly$\alpha$
emission.

In the near future, I want to combine what we have learned from our
photoionization models,  with the knowledge that the powerful
polarization techniques have already provided.
 Much of what is known nowadays about the nature of the alignment effect observed at $z>$0.7 (see Chapter 4, Introduction)   is the result of
polarization studies, which distinguish between the scattered and the ``in
situ'' continuum emission and tell us about the nature and distribution of
the scattering material. In this way, for instance, di Serego Alighieri, Cimatti \& Fosbury (1994) were able to exclude the possibility of scattering in two radio
galaxies as due to  {\it hot electrons} in a gaseous halo.

        Comparisons of our models with the observations do not allow a precise  estimation  of 
the amounts of dust existing in the EELRs, both at high and low $z$. Studying the depletion of calcium, we have demonstrated that dust exists, but we can not say how much. Calcium is so sensitive to depletion 
that very tiny amounts of dust will deplet it completely: a dust /gas ratio
($\mu$) =0.1 decreases the abundance of calcium in the gaseous phase by a
factor of  $\sim$ 24 and if $\mu$=1, by a factor of 5000! (see chapter 3).
 On the other hand, although polarization measurements set strong constraints to the nature and distribution of scattering material, it is still difficult to distinguish between 
cool electrons and dust as the origin of the scattering (Dey, Spinrad \& Dickinson,  1995 and di Serego
Alighieri et al. 1996). It could be interesting to investigate if the 
low dust/gas ratios which we deduce, are able to explain the level of
polarization and the polarized flux observed in some objects. 

\section{The ionization mechanism.}

There is strong evidence which support photoionization by the hard
continuum emitted by the central AGN as the main ionization mechanism  of
the EELRs (and NLRs) of most radio galaxies and Seyferts, at least at low
redshift. At higher
$z$ the uncertainties due to the lower spatial resolution and sensitivity
of our instrumentation makes
such study much more difficult. Moreover, at
high $z$, most objects have been observed in the UV rest frame, a spectral
range that is not well known, and still not often used for diagnosis. The
optical line ratios that have provided so much information  about the nature
of low $z$ objects, are redshifted into the IR where the observations are still 
difficult. The existing IR instrumentation (MIRACLE MPE/ROE camera, UKIRT
CGS4 spectrometer,  the 3D Near Infrared Imaging Array Spectrometer (MPE), etc) has already allowed the study of the rest-frame
optical wavelengths of a few objects at high $z$ ($\geq$1) (e.g. Eales \&
Rawlings 1993, McCarthy, Elston \& Eisenhardt 1992), but much more effort needs to be
dedicated. 

        Our models support that photoionization by the hard continuum produced  in the central engine is the main ionization mechanism  both for the NLR and the EELR line ratios of radio galaxies and Seyferts.
  However, there is an 
important limitation of the optical line ratios as a diagnostic: with  
photoionization models it is possible to diagnose the mean ionizing photon 
energy (RBFT87), but not to distinguish between hard power laws and hot black 
bodies. Ionization by normal stars is ruled out, but a cluster of superhot 
stars, so called `Warmers' could also explain the optical line ratios.

        At this point, I want to center this discussion around shocks, which
have not been discussed yet in this thesis and deserve special attention.
 
\subsection{Shocks in active galaxies}

        In many active galaxies radio jets have been detected.  Very likely 
the energy which triggers them is originated in the central AGN. The jet contains relativistic particles accelerated to near the speed of  light
 which propagate through the gas of the host galaxy. In its way out the jet
will interact with the external gas and the shocks produced during such an 
interaction will disturb the kinematics and morphology of that gas. The hot gas behind a fast shock emits cooling radiation and this could also ionize the gas.
 However, as explained before, we expect that the central AGN
plays also an important  role in the ionization of the gas. But, which is the
dominant process?   Do shocks only alter
the morphology and density of the gas or do they also modify its ionization
state? Jets in AGNs are still surrounded by many mysteries and a clear
understanding of the jet/cloud interaction phenomena is needed.

	At least for most low $z$ objects, AGN photoionization has had better
agreement with the observations. Beautiful evidence  supporting this point came from the ionization {\it
cones} imaged in several nearby Seyfert galaxies, like NGC5252 (Tadhunter
\& Tsvetanov 1989; Acosta-Pulido et al. 1996; see also the clearly defined ionization cones in the WFPC images of NGC5728, Fig.~31). Wilson, Ward \& Haniff (1988) suggested a possible broad trend
for the radio axis to be aligned with the axis of the extended emission line
regions in Seyfert galaxies on a similar spatial scale. Wilson and
Tsvetanov (1994) confirmed this tendency, pointing
towards a common collimation mechanism of the radio plasma and  the ionizing
radiation.

\begin{figure}
\vspace{5in}
\caption[]{}{\small An HST WFPC (1992) image of the core of the barred spiral Seyfert galaxy NGC5728. The bi-conical structure of the ionized gas indicates that ionizing continuum emitted in the central AGN is beamed in two opposite cones (Wilson et al. 1993). }

\end{figure}

      In many objects, spectroscopic evidence favours AGN against shock 
photoionization. For instance, the already mentioned trend defined 
by the EELRs line
ratios in the diagnostic diagrams of RBFT87, which is well explained if they  are ionized by the
central AGN allowing for  a different ionization state  (i.e. U)  in each object. It is
difficult to understand this behaviour with the excitation parameter in the frame
of the shock theory.

	On the other hand,	
        there have been also problems with photoionization models that
shocks were able to explain, at least partially. Three were the main
discrepancies:
a) too weak high excitation lines, b) too low electronic temperatures, and c)
too small range in HeII/H$\beta$. A recent work by Binette, Wilson \& 
Storchi-Bergmann (1995)
demonstrates that a mixture of matter- and radiation-bounded clouds 
(proposed earlier by Viegas and Prieto, 1992; although not successfully) can solve these
discrepancies, without the need of additional heating sources, like shocks.
Moreover, from comparing U at different positions within the EELR in many RG (e.g., the radio galaxy 3C227, Prieto et al. 1993), the observations indicate that, in general, the ionization ($U$) does not change dramatically with radius, as we would expect if photoionization is due to the central AGN.
This is a difficult question with no answer yet within the frame of pure
AGN photoionization. \footnote{ If shocks are strongly influencing the 
spatial distribution of the excited gas, even if they do not ionize it, the variation of $U$ 
may be different than that expected from the pure AGN presence, as often assumed.}

        There are  some objects, where the presence
of shocks is clear: a clear example is the intermediate redshift
(z=0.310) radiogalaxy PKS 2250-41 which shows an emission line arc
circumscribing the western radio lobe (Tadhunter et al. 1994), which strongly indicates the
influence of the radio jet on the distribution and, possibly also on the
ionization of  the warm gas. Also, the highly disturbed kinematics and the
low ionization state near the radio structures is indicative of the
presence of shocks.  However, the source of ionizing energy is not clear:
are the shocks dominating over the AGN continuum and ionizing the gas or
are they  just compressing the clouds in the arc, which are then ionized by the
AGN?

        Although at  low redshifts, the line
emission spectra of most active galaxies
are well explained by photoionization by the central AGN it is important to
understand the jet/cloud interaction physics to disentangle on a case by case
 basis which effects are dominant.  

        In our models we have also assumed photoionization by a hard continuum
 to explain the
UV line ratios of high $z$ radio galaxies. The trend in U shown by the data
in the diagnostic diagram CIV/Ly$\alpha$ {\it vs.} CIV/CIII, is indicative
of this kind of excitation. But  the possibility that
shocks could also play an important role in the ionization of the
gas can not be excluded. 
At high redshifts, radio galaxies show a strong alignment between the
optical and the radio structures. One interpretation, well
supported by polarization studies, has already been mentioned here:
scattering of continuum emitted by the hidden AGN. Another
interpretation is that the propagation of the radio jet through the ISM of
the host galaxy compresses the gas and triggers intense star formation
({\it e.g.} Rees 1987, de Young 1989, Begelman \& Cioffe 1989). In fact the
optical structures revealed by ground based and HST images (Longair, Best \& R\"ottgering 1995; see Fig.~33)
point towards shocked gas, with much narrower spatial distribution than
the expected diffuse photoionized cone. High $z$ radio galaxies have been
selected on the basis of their powerful radio emission and line spectra. At
such high $z$ we are biased towards very powerful objects, where the
jet-cloud interaction might be very strong. Moreover, the shocked material,
more concentrated, emits stronger lines than the surrounding gas. This
could mean that due to the lack of sensitivity of our instrumentation, we
are only detecting the material that is interacting with the radio plasma.

\begin{figure}
\vspace{4in}
\caption[]{}{\small WFPC2 HST images of gaseous jets from three newly formed stars.}
\end{figure}

In order to reach a clear understanding of the jet/cloud
interaction physics, it is necessary to use  low and intermediate $z$
objects as laboratories, or even more local events, like the jets ejected from stars.
 HST WFPC2 images of jets originated by
newly forming stars (e.g. Stapelfeldt et al. 1995) show the complicated interactions that
take place when the ejected gas collides with the interstellar medium (see Fig.~32). The
general picture which could describe this scenario is a new star surrounded
by a  dusty disk, which feeds material onto the star. When the star is hot
enough, it will blow away much of the disk. The high speed material will
collide with the slower gas in the interstellar medium and a bow shaped
shock wave is created. In this way, a cavity is formed surrounding the
star.  If the shock wave encounters an obstacle (denser isolated clouds), a bow shaped shock wave is formed around the obstacle. Such a process will not only 
influence the ISM surrounding the star,
but also the future of the star itself. The jet can, for instance, modify
the mass accretion rate.
 
The geometry in this scene is quite similar to that which describes the
AGN phenomena, where the central source (presumably a black hole) is
surrounded by an accretion disk that feeds it. What we learn from stellar
jets can be very useful in understanding the role jets play  in active
galaxies.

\section{ Geometry}

        This thesis provides a tool for the understanding of the geometrical effects which
influence the emission line spectrum radiated by an ``open'' geometry
appropriate for the AGN phenomenon. This study demonstrates how important it
is to take perspective effects into account; the narrow line emission
depends on the orientation of the ionized cones with respect to the
observer. Such geometrical effects, studied in detail by us for the first
time, have been used to explain some observational effects, like the UV
spectrum of high $z$ radio galaxies and the discrepancies between the
predicted and the observed HI line ratios emitted by the Narrow Line Region
of Seyfert 2 galaxies (and, therefore, 
indirectly of Seyfert 1s according to the unification schemes). 
As explained in section 6.1, we have studied how neutral material external to the ionized 
cones can influence the
``appearance'' of an object in the Ly$\alpha$ light.

	The excited gas that we see does not necessarily map the distribution 
of material in the galaxy, but rather, the distribution of {\it excited} gas.
 Outside the radiation cones, there must be important reservoirs of neutral 
material. The work on NGC5252 by Prieto \& Freudling (1993) shows that the
 neutral hydrogen surrounds the nucleus but is located outside the radiation 
cones of ionized gas. Also, in many high $z$ radiogalaxies, the Ly$\alpha$ 
profile shows absorption features due to absorption of Ly$\alpha$ photons by 
intervening hydrogen gravitationally bound to the radio galaxy and spatially 
extended over a region of $\sim$50 kpc in some cases (van Ojik 1995). The HI 
absorption systems might be due to neutral gas located outside the ionized 
cones.
 It is therefore very important to distinguish between excitation and matter distribution.

  As explained in Chapter 5, the rest-frame UV spectral range
constitutes a powerful tool for exploring the 3-dimensional structure of
active galaxies in general: the existence of resonance lines in this
spectral range which are so sensitive to geometry has made this study
possible. In the ``open geometry'' that describes the EELRs, the two opposite
cones, with apex in the nucleus which are ionized by the central AGN will most  often
be forming a non negligible angle with the plane of the sky in such a way that one
of the cones will be closer to the observer. It is not a trivial task to
distinguish which is closer and which is farther from the observer: only
when there is some kind of observable asymmetry is this distinction possible. For
instance, in high luminosity FRII sources, one-sided jets are detected,
while two-sided jets are detected only in low luminosity (FR-I type) radio
galaxies. This jet asymmetry can be explained by an illusion caused by a
doppler-boosting effect due to the bulk relativistic motion of the jet.
This effect makes the nearest side (to the observer) brighter than the
other one.
Moreover, in  powerful radio galaxies with one-sided jets, the
depolarization is systematically stronger on the counter-jet side (Laing
1988).
Garrington \&  Conway  (1991) interpreted this effect as produced by an
external halo of ionized gas (identified by the authors with the extended
X-ray emission often observed in clusters). The depolarization asymmetry is
a geometrical effect, due to the different column densities of hot gas that
the radio emission from both sides  has to cross, the counter-jet 
side being seen through a longer path length of gas.

        When these asymmetries are present, we can readily distinguish which cone is
closer {\it to the observer} and which one is further. However, it is not
often easy to find these.  For radio galaxies with no visible jets (or which are symmetric) or
with the radio emission from both lobes too faint for the polarization to
be measured,
the distinction must be based on another strategy.

      We have proposed a new method to solve this problem using the strong
emission lines that are expected in the UV rest frame domain. Our models (Chapter
5) demonstrate that the line ratios involving resonance lines (like
Ly$\alpha$ and CIV$\lambda$1550) should be different on both sides of the
nucleus, with the differences increasing with the inclination angle with respect to the
plane of the sky. This suggested us a test to distinguish between
the cone closest to the observer and the furthest one: the cone which emits
the spectrum resembling our ``back'' clouds will be the nearest one, while the
furthest one should emit line ratios more similar to our ``front'' models.

\begin{figure}
\vspace{4in}
\caption[]{}{\small WFPC2 HST images of three high $z$ radiogalaxies 3C265 ($z=$0.81), 3C324 ($z=$1.21),  3C368 
($z=$1.13) (left to right). VLA data are superimposed on 
the images of 3C324 and 3C368. The alignment between the radio axis and the UV
rest-frame structures is clear in these two cases. The radio axis is marked by 
lines in the case of 3C265. The rest-frame UV continuum of this object has an 
elongated morphology misaligned by about 35$\deg$ from the radio axis. }
\end{figure}

        In this way, we expect to be able to describe the 3-dimensional
structure of the objects. Furthermore, this will also help to understand some of the
anisotropies that are observed in radio galaxies (some of them, in fact,
expected from the unified schemes). For instance,  the 3C256 radio galaxy
(Dey, Spinrad \& Dickinson 1995) has different levels of polarization in different
regions. The optical images obtained with the Keck 10m telescope, show two
clumps of emission in the south and a fainter more diffuse component
extending to the north west. Both regions have rather different
polarization levels. As the authors suggest, maybe there is a different
contribution of diluting radiation. However, we have to be careful with
this interpretation. It could also be a geometrical effect. The cross
sections for forward and backward scattering of dust grains are different.
This means that the dust scattering produced in both cones (when they are
not in the plane of the sky) is asymmetric, and, therefore, the relative
polarization measured will also be different. I plan to develop this
study in the near future.

        McCarthy, van Breugel \& Kapahi (1991) showed that in essentially all radio galaxies the
extended optical line emission is brightests at the cone closest  {\it to the nucleus}. The most plausible explanation that the authors propose
 is inhomogeneities in densities and distribution of dense clouds.
which means that one of the cones is intrinsically brighter than the other. Assuming this to be the situation, in some cases, the brightest cone must be nearer {\it the observer}
and in other cases, it will be further away. Therefore, (independently of which side is brighter) some objects can be expected to be 
dominated by a ``back'' type spectrum and others by a ``front'' type spectrum.
We find that this effect could explain the high dispersion observed in
the Ly$\alpha$/CIV ratio in very powerful radio galaxies at very high $z$,
while CIV/CIII remains nearly constant from one object to another.

        As we see, there is evidence to support our view. A very
strong test will be provided by spatially resolved UV spectroscopy of
active galaxies with extended emission on both sides of the nucleus
(objects devoid of any broad components will simplify enormously the study). The
data that we have used for our modeling were extracted from the literature
and correspond to high redshift radio galaxies where the spatial resolution
is generally insufficient to separate clearly the extended from the nuclear
emission. A project for the near future is to obtain long slit
spectroscopy (UV rest frame) of high $z$ radio galaxies with extended
Ly$\alpha$ already reported in the literature and where other strong UV
lines (like CIV, III and HeII) have also been detected. The aim is to analyze the
spatial distribution of the different lines and how the line
ratios vary on both sides of the nucleus. So far, the spectra have generally been 
presented in the literature with no spatial information, compressing the objects in the
spatial direction. Moreover, there has not been an attempt to obtain narrow
band image in other UV lines but Ly$\alpha$. Quotient maps, like
Ly$\alpha$/CIV will describe the geometry of the ionized gas and will
reveal regions of pure Ly$\alpha$ emission probably indicative of the
existence of neutral hydrogen.

        But, of course, the best tool for this research, available nowadays
is the HST, that, with the Faint Object Spectrograph, makes possible UV
spectroscopy at high spatial resolution for low $z$ objects, where the
distinction between the nuclear region and the extended ionized gas is much
easier. This is one of the main goals of our HST project (section 6.6).

\section{Our HST project}

I have mentioned the HST project in which I am involved and
which got observing time for the next observing cycle. I will describe it in more detail  in this section.

        We plan to obtain spatially resolved UV spectroscopy with the Faint
Object Spectrograph (FOS) covering the observed frame
Ly$\alpha$-CIII]$\lambda$1909, of four low $z$ radio galaxies with clear
extended emission line regions already observed from the ground in the
optical spectral range.        With the data to be obtained, we will apply our
newly developed ultraviolet line diagnostic and should be able to describe
the 3-dimensional geometry of the objects as well as study the dust effects on the
measured lines. We will be able to distinguish between the material on the
near and far side of the AGN and compare our results with other observed
asymmetries (see section 6.4). The main goal is, therefore, to elucidate
the nature of the geometrical effects which influence the line spectrum
radiated by the ``open geometry'' (externally ionized clouds) that describe
the EELRs.

In order to understand the nature of these objects, a clear separation
between the different components contributing to the spectral energy
distribution is needed: AGN + stars + scattered continuum + additional
sources (like nebular continuum). There are many phenomena that are a
direct, or indirect consequence of the AGN activity, like the extended scattered
continuum or the line emission
(both if the ionization is directly by the AGN or indirectly by shocks
induced by the radio jet crossing the ISM). The study of these phenomena will provide
information, not only on the AGN, but also on the stellar populations. Once
we have a clear understanding of how the AGN influences its environment, it
will be easier to separate the different components. What we learn will be
very useful to interpret phenomena observed at very high $z$, where
powerful radio galaxies are good laboratories to study the formation
and early evolution of galaxies. So far, it has been very difficult to
establish the state of evolution due to the difficulties in interpreting
the observations with a non clear understanding of the AGN influence. If
we knew the age of the oldest stars we could constrain the formation
redshift and test even the current cosmological models.

        When we were choosing our targets we searched for objects with
known anisotropies and for which we could distinguish which cone is on the near or far side. This will provide a
strong constraint to test the validity of the diagnostic method that we have proposed. A practical limitation appeared when we realized that not many objects exist in which the aforementioned distinction between cones has been possible. We looked for targets with a clear one-sided jet, indicative of the cone nearer to {\it the observer}.  However, objects containing jets are rather better  candidates of classes strongly
influenced by shocks produced by the interaction of the said jet with the cooler
gas of the galaxy, while our models are more appropriate and intended for objects where
photoionization by the central AGN is (more clearly) the main ionization mechanism of the
gas. Therefore, we decided to prepare a more complete proposal which would
allow us the study of both phenomena: the influence of geometry {\it and} the jet/cloud
interaction. We chose one target (PKS2152-69) which presents
clear evidence of interaction between the radio and optical plasmas. 
Moreover, we selected one source (PKS0349-27) which show EELR on both sides of the nucleus and with no evidences
of shocks. A new
UV line diagnosis to distinguish between shock and AGN  photoionization
will be developed. With our newly developed diagnostic method, we expect to infer important clues on the geometry of the ionized gas.

\vspace{0.5cm}

\addcontentsline{toc}{part}{Bibliography}
\parskip=1ex
{\Large \bf References}

\vspace{0.5cm}

\noindent Acosta-Pulido J.A., Vila-Vilar\'o B., P\'erez-Fourn\'on I., Wilson A.S., Tsvetanov Z.I., \apj 1996, {\it in press}

\noindent Andreani P., La Franca F., Cristiani S., 1993, \mnras, 261, L35

\noindent Barthel P.D., 1989, \apj 336, 606

\noindent Begelman M.C., Cioffe D., 1989, \apj 345, L21

\noindent Binette L., Wilson S.W., Storchi-Bergmann T., {\it A\&A}, in press

\noindent Cimatti A., di Serego Alighieri S., Fosbury
R.A.E.,
Salvati M., Taylor D., 1993, \mnras 264, 421

\noindent  Cimatti A., di Serego Alighieri S., Field G.B., Fosbury
R.A.E., 1994, \apj 422, 562

\noindent Chambers K.C., McCarthy P.J, 1990, \apj 354 L9

\noindent Chini R., Kr\"ugel E., 1994, {\it A\&A}, 288, L33

\noindent  Dey A., Spinrad H., Dickinson M., 1995, \apj
440, 515

\noindent de Young D.S., 1989, \apj 342, L59

\noindent di Serego Alighieri S., Cimatti A., Fosbury
R.A.E., 1994, \apj 431, 123

\noindent di Serego Alighieri S., Cimatti A., Fosbury
R.A.E., P\'erez-Fourn\'on, 1996, \mnras submitted

\noindent D'Odorico S., Cristiani S., Fontana A., Giallongo E., 1995, ESO PR 11/95; 15 Sept.

\noindent Downes D., Radford S.J.E., Greve A., Thum C., Solomon P.M., Wink J.E., 1992, \apj, 398, L25

\noindent  Dunlop J.S., Hughes D.H., Rawlings S., Eales
S.A., Ward M.J., 1994, {\it Nature}, 370, 347 

\noindent  Eales S.A., Rawlings S., 1993, \apj 411, 67

\noindent Eisenhardt P., Soifer B.T., Armus L., Hogg D., Neugebauer G., Werner M., 1995, {\it BAAS}, 186, 5202

\noindent  Elston R., McCarthy P.J., Eisenhardt P.,
 Dickinson M., Spinrad H., Januzzi B.T., Maloney P., 1994, \aj 107, 910

\noindent Garrington S.T., Conway R.G., 1991, \mnras 250, 644

\noindent Goodrich R.W., Miller J.S., Martel A., Cohen M.H., Tran H.D., Ogle P.M., Vermeulen R.C., 1996, \apj 456 L9

\noindent Isaak K.G., McMahon R.G., Hills R.E., Withington S., 1994, \mnras, 269, L28

\noindent Laing R.A., 1988, {\it Nature}, 331, 149

\noindent Lawrence A., Elvis M., 1982, \apj 256, 410

\noindent Lawrence A., 1987, \pasp 99, 309

\noindent Lawrence A. et al., 1993, \mnras 260, 28

\noindent Lilly S.J., 1988, \apj 333, 161

\noindent Lilly S.J., 1989, \apj 340, 77

\noindent Longair M.S., Best P.N., R\"ottgering H.J.A., 1995, \mnras 275 L47

\noindent McCarthy P.J., van Breugel W.J.M., Kapahi V.K., 1991, \apj
371, 478

\noindent McCarthy P.J., Elston R., Eisenhardt P., 1992,
\apj 387, L29

\noindent McCarthy P.J., 1993, {\it ARA\&A} 31, 639

\noindent Macchetto F.D., Lipari S., Giavalisco M., Turnshek D.A., Sparks W.B., 1993, \apj, 404, 511

\noindent McMahon R.G., Omont A., Bergeron J., Kreysa E., Haslam C.G.T., 1993, \mnras 267, L9

\noindent Prieto A., Walsh J., Fosbruy R.A.E., di Serego Alighieri S., 1993, \mnras 263, 10p

\noindent Prieto A., Freudling W., 1993, \apj 416 668

\noindent Rees M.J., 1987, \mnras 239, 1p

\noindent Robinson A., Binette L., Fosbury R.A.E., Tadhunter C.N., 1987, \mnras 227, 97 (RBFT87)

\noindent Serjeant S., Lacy M., Rawlings L.J., King L.J.,
Clements D.L., 1995, \mnras 276, L31

\noindent Stapelfeldt K.R. et al., 1995, {\it BAAS}, 185, 1802
 
\noindent Tadhunter C., Tsvetanov Z., 1989, {\it Nature}, 341, 422

\noindent Tadhunter C., Shaw M., Clark N., Morganti R., 1994,  {\it A\&A}, 288, L21                    

\noindent  van Ojik R., 1995, Ph.D.thesis, Rijksuniversiteit te
Leiden

\noindent  Viegas S.M., Prieto A., 1992, \mnras
258, 483 

\noindent Villar-Mart\'\i n, 1993, Proyecto de tesis: ``Influencia del polvo en modelos de fotoionizaci\'on aplicados a n\'ucleos activos de Galaxias''. Ito. de Astrof\'\i sica de Canarias y Universidad de La Laguna.

\noindent Villar-Mart\'\i n M., Binette L. \& Fosbury R.A.E, 1995, in Proc. for the Workshop "Cold Gas at high $z$", Hoogeveen, August 1995 (in press)

\noindent Wilson A.S., Ward M.J., Haniff C.A., 1988, \apj 334, 121

\noindent Wilson A.S., Braatz J.A., Heckman H.M., Krolik J.H., Miley G.K., 1993, \apj, 419, 61

\noindent Wilson A.S., Tsvetanov Z., 1994, \aj, 107, 1227

\chapter{Conclusions}

	We have considered for the first time the combined effects of geometrical perspective  and dust on the narrow emission line spectra of active galaxies. This is necessary to make a more correct interpretation of the observations. The kind of geometry determines strong differences on the emission line ratios
when dust is present, due to the much more destructive effects of dust when we approach  a close (spheric) geometry. Moreover,  when we study the line emission from the EELRs or objects dominated by such extended emission, the influence of perspective must be considered if resonant lines are involved, both if dust is or is not present.  In general the ionized cones have an inclination angle with respect the plane of the sky, which results on different viewing perspectives for both cones. This implies strong asymmetries on the UV line ratios of regions at both sides of the nucleus, due to the resonant character of Ly$\alpha$ and CIV and therefore, their strong dependence on perspective.

	We have demonstrated that the  NLR HI line ratios of Seyfert 2 galaxies  points towards a significantly self-covered distribution of gas mixed with dust which is very patchy.  The combination of line scattering and dust absorption effects can explain the observed ratios. Therefore, for those objects  where the line emission is dominated by the unresolved Narrow Line Region, a more close geometry than peviously assumed is appropriate.
 On the other hand, when the extended emission is dominant (EELR), an open geometry is more realistic. If such close geometry is valid for the NLR, we expect the central continuum source to be also  reddened.
	We claim that  the apparent
deficit of ionizing photons seen in many Seyfert~2's could in part be due to dust extinction of the ionizing continuum, and not only to the presence of a circumnuclear torus that collimates the ionization radiation along its axis, as proposed by other authors.

	Our studies on the UV emission line spectra of high $z$ radio galaxies demonstrate that the UV line ratios are strongly determined by geometrical perspective, rather than dust effects. Internal dust, even in proportions as large as that found in the solar neighbourhood, can not explain the UV line ratios measured for high $z$ radio galaxies, unless perspective effects are taken into account.

	 We claim that Ly$\alpha$ is not a good indicator of dust  but rather a good
tool to understand the geometry of the emitting gas.  Moreover, 
 the influence of geo\-metrical factors could give a natural explanation for some of the emission line asymmetries seen in the `alignment effect' radio galaxies.
 The reflection of Ly$\alpha$ photons by neutral material can 
determine the morphology of the object in the light of Ly$\alpha$.  We suggest that the diffuse Ly$\alpha$ halos observed in some high $z$ radio galaxies
could be due to the reflection of Ly$\alpha$ photons by clumps lying outside the ionization cones. 

	We have demonstrated that the gas in Extended Emission Line Regions in low $z$ radio galaxies is mixed with dust. The faintness of the [CaII]$\lambda\lambda$7291,7324 lines indicates that calcium is depleted onto dust grains. We have used this result to discriminate among the current theories about the origin of the extended ionized gas.
 If there is an universal origin, the existence of internal dust favours debris from mergers or tidal interactions as the most plausible scenario.

\newpage

\pagestyle{myheadings}
 
\markright{Apendix A}

\vspace{5cm}

\centerline{\huge \bf Apendix A}

\vspace{1cm}

\centerline{Binette, Wang, Villar-Mart\'\i n, Martin \& Magris 1993, ApJ, 414, 535}

\vspace{1cm}

	The term depletion refers to the underabundance of gas phase elements 
with respect to the solar standard abundances as a result of their assumed presence
 in dust. References to the studies of depletion in the local ISM can be found 
in Whittet (1992). We describe here a {\it theoretical} algorithm which attempts
 to deplete in a meaningful way the trace element abundances when computing line 
diagnostic for a plasma of arbitrary {\it gas} metallicity $Z_{gas}$ and dust content $\mu$. Both quantities $Z_{gas}$ and $\mu$ 
($\propto Z_{dust}$; see eq.[A5]) are known to depart significantly from solar 
neighborhood values in low-mass extragalactic systems or as a function of galactic 
radius in the Galaxy. For the purpose of modeling plasma conditions in a nonsolar environment, we defined an algorithm which scales in a self-consistent manner the depletion of the gas phase in proportion to the quotient $\mu/Z$, where $Z$ is the {\it total} metallicity of the plasma ($Z= Z_{dust} +Z_{gas}$) expressed {\it relative} to local ISM values:

$$ Z = \frac{\sum m_X (\{n_X\}_{dust} + \{n_X\}_{gas})_{plasma}}{\sum m_X\{n_X\}_{\odot}},  ~~[A1]$$ 

and $\mu$ is the dust content of the plasma expressed {\it relative} to that of the standard ISM dust-to-gas mass ratio. The reference solar abundances set which we adopt is taken from the compilation of Anders \& Greveese (1989) and is listed in Table 4 (see entry $\{n_X\}/\{n_H\}_{\odot})$. We define 
$D(X)$, the depletion index of element X, as

$$ D(X) =[\frac{X}{H}] = log \{\frac{n_X L}{n_H L}\}_{ISM} - log \{\frac{n_X}{n_H}\}_{\odot} ~~[A2]$$

where $\{n_X L / n_H L\}_{ISM}$ represents the ratio of the column densities of the gaseous element X (measured along various lines of sight of path length 
L[$\leq 6 \times 10^{20}$ cm]) over the corresponding measured column densities of H. In the local ISM, the number density of atoms $n_X$ of element X which are found either in gas form (or depleted into dust grains) is given relative to $n_H$ by $\{n_X/n_H\}_{\odot}$ 10$^{D(X)}$ (or $\{n_X/n_H\}_{\odot}$(1 - 10$^{D(X)}$). The adopted set of local ISM depletion indices $D(X)$ is listed in Table A1 and was derived from Table 2.2 of Whittet (1992) except for carbon for which the value -0.5 was adopted as suggested by some authors. This value is also consistent with our dust model which includes graphite grains.

	For the local ISM, for which $\mu\equiv 1$ (and $Z = Z_{dust} + Z_{gas}$ =1), the gas phase abundances are derived directly from the depletion indices of Table A1. As we are often interested in modeling line emission plasma with $Z \neq 1$, we could assume that the dust content $\mu$ scales linearly with Z. That this is approximately true is supported by the results of Issa, MacLaren, \& Wolfendale (1990) and Sauvage \& Vigroux (1991) who compared the metallicity and extinction in the Galaxy with those of the Magellanic Clouds or of nearby spirals. In our calculations, if we adjusted the dust content such that always $\mu/Z$ = 1, assuming that the depletion indices of Table A1 remain approximately valid, the procedure for depletion is straightforward and the gas phase abundance by number of element X is simply given by

	$$ \{n_X/n_H\}_{gas} = Z \{n_X/n_H\}_{\odot} 10^{D(X)}, ~~[A3] $$

	This is equivalent to saying that when $\mu/Z$ =1  both the dust grains as well as the gas phase are made up of the same relative proportion of trace elements.

\begin{table}
\centering
\small
\caption{Solar Abundances, ISM Depletion Indices and Correlation Coefficient Ratios.}
\vspace{0.5cm}
\begin{tabular}{lllll} \hline

Element &  & $\{n_X/n_H\}_\odot$ &  $D(X)_{ISM}$ &  $\{r_n/r_N\}^{ISM}_X$ \\ \hline

H   &	..........................	&  1.000 & ... & ... \\ 
He &	..........................	&  0.1  & 0.0 & ... \\ 
C &	..........................	&  3.6 $\times$ 10$^{-4}$ & -0.5 & 0.88 \\ 
N &	..........................	& 1.1 $\times$ 10$^{-4}$ &  -0.1 & 0.46\\ 
O &	..........................	& 8.5 $\times$ 10$^{-4}$  & -0.2 &  0.42\\ 
Ne &	..........................	& 1.2 $\times$ 10$^{-4}$  & 0.0 &  ...\\ 
Mg &	..........................	& 3.8 $\times$ 10$^{-5}$  & -0.7 & 1.11 \\ 
Si &	..........................	& 3.5 $\times$ 10$^{-5}$  & -1.6 & 1.19 \\ 
S &	..........................	& 1.9 $\times$ 10$^{-5}$  & -0.2 & 0.94 \\ 
Ar &	..........................	& 3.6 $\times$ 10$^{-6}$ &  0.0 & ... \\ 
Ca &	..........................	& 2.2 $\times$ 10$^{-6}$ & -3.7 & 1.42 \\ 
\hline

\end{tabular}
\end{table}

	For the more general case where $\mu \neq Z$, we  have no observational basis as to how the depletion indices might vary. If we use the local ISM indices as a guess of the depleted abundances, and then simply renormalize the whole abundance set by a constant factor until the depleted mass match the value implied by $\mu$ (see. eq. [A5] below), we obtain, in the case $\mu \ll Z$, rather implausible gas abundances for some elements. To alleviate this problem, we have chosen instead to vary the depletion indices in some proportion to the ratio $\mu/Z$ before linearly renormalizing the total mass of the depleted metals. An indication of the relative variation of the depletion indices $D(X)$ with the ratio $\mu/Z$ can be intuitively inferred from the measured quatities $\{r_n/r_N\}_X^{ISM}$ which represents the ratio for element X of the correlation coefficient of $D(X)$ with density, $r_n$, over the correlation coefficient of $D(X)$ with column density, $r_N$. The merits of the measured $\{r_n/r_N\}_X^{ISM}$ ratios is that they sample regions of somewhat different dust content $\mu$. The algorithm we devised for each case $\mu \neq Z$ is the following: we first define modified depletion indices $D(X,\mu /Z)$' which are dependent on $\mu/Z$ as follows

	$$ D(X, \mu/Z)' = D(X) (\frac{\mu}{Z})^{\gamma \{r_n/r_N\}X}, ~~[A4]$$

By trial and error, we established that $\gamma$ = 0.25 gave a reasonable behaviour of the depleted elements even when $\mu /Z \rightarrow$ 0. The values we adopted for $\{r_n/r_N\}_X^{ISM}$ (see Table 4) were derived from Figure 2.9 of Whittet (1992). Linear renormalization of the abundances implied by the modified indices $D(X,\mu /Z')$ is required in order that the mass of the elements depleted onto dust grain remains consistent with the value of $\mu$. This is done in an iterative fashion by determining the value of $\delta (\approx$ 1) until the following dust mass equation is satisfied

	$$ Z_{dust} \sum m_X \{n_X\}_\odot = 0.0089 \mu \frac{m_H\{n_H\}}{\sum m_X \{n_X\}} = $$

$$ Z \sum \{\frac{m_X n_X}{m_H n_H}\}_{\odot} [1 - \kappa ~min(1,\delta 10^{D(X,\mu/Z')})] , ~~[A5]$$

where the constant 0.0089 represents the dust-to-gas mass ratio relative to H (i.e., $\rho_{dust}/\rho_H$) of any of the extinction curves computed by P.G. Martin (this number is consistent with the ISM value if our dust composition within the grain model is valid). The constant $\kappa$ = 1.123 is a small correcting factor to allow for the fact that the summation is limited to the elements of Table 4 (considered by MAPPINGS) which does not include all the depleted elements (e.g., Fe). Once $\delta$ is determined, the gas phase abundance of any element X is

	$$\{n_X/n_H\}^{gas}_{Z,\mu} = Z \{n_X/n_H\}_{\odot} ~ min (1, \delta 10^{D(X,\mu/Z')}) , ~~[A6] $$

	The value of $\kappa$ is such that $\delta$ becomes unity if $\mu/Z$ = 1 in which case equation (A5) reduces to equation (A3). With the above depletion algorithm, as $\mu /Z \rightarrow $ 0, the uncertainty in the absolute abundances of the highly depleted elements (e.g., Ca) is probably quite high but since no observational depletion data for $\mu /Z \ll$ 1 is available, this situation cannot be helped. The main purpose of our algorithm was  that it gave a cummulative depletion of the gas phase elements always consistent with the dus content $\mu$. Furtermore, as $\mu/Z \rightarrow $ 0 it also has the property of giving for {\it all} the elements cocerned much more plausible values than any alternative linear scheme which we could devise. The need for a depletion algorithm even if imperfect is called for in any photoionized gas model since it is quite conceibable that the dust content is lower than in the ISM as a result of destruction of dust grains by sputtering (although the timescales are relatively long compared to the lifetime of an HII region; see Osterbrock 1989) or as a result of segregative acceleration away from the cloud of much of the smaller grains' population due to radiation pressure.

\newpage

{\Large \bf References}

\vspace{0.5cm}

\noindent Anders E., Grevesse N., 1989, Geochim. Cosmochim. Acta, 53, 197 

\noindent Issa M.R., MacLaren I., Wolfendale, A.W., 1990, {\it A\&A}, 236, 237

\noindent Osterbrock D.E., 1989, Astrophysics of Gaseous Nebulae and ACtive Galactic Nuclei
(Mill Valley: University Science Books)

\noindent Sauvage M., Vigroux L., 1991, in IAU Symp. 148, The Magellanic Clouds, ed. R. Haynes
\& D. Milne (Dordrecht: Kluwer), 407 

\noindent Whittet D.C.B., 1992, Dust in the Galactic Environment (Bristol:IOP)

\newpage

\pagestyle{myheadings}
 
\markright{Apendix B}

\vspace{5cm}

\centerline{\huge \bf Apendix B}

\vspace{1cm}

\centerline{Binette, Wang, Villar-Mart\'\i n, Martin \& Magris 1993, ApJ, 414, 535}

\vspace{1cm}

\centerline {1. CALCULATIONS OF $N_{HI}^{critic}$}

\vspace{0.5cm}

We assume a slab geometry for the screen. For an isothermal screen with HI column
 density $N_{HI}^{screen}$, the optical depth for Ly$\alpha$ resonant scattering is

$$\tau(x) = N_{HI}^{screen} \sigma_0(T) \phi(x) , ~~[B1]$$

where $x = (\omega - \omega_0)\omega_D,\sigma_0(T)\phi (x)$ is the thermally averaged
resonant scattering cross section,

\vspace{0.1cm}

\centerline {$\sigma_0(T) = 5.9 \times 10^{-14} (T/10^4 K)^{-1/2} $ cm$^{-2}$,~~[B2]}

$$ \phi(x) = \frac{a}{\pi} \int_{-\infty}^{+\infty} {\frac {e^{-y^2} dy} { a^2 + (x-y)^2}}, ~~[B3]$$

\vspace{0.1cm}

$a = \Gamma / 2\omega_D = 4.7 \times 10^{-4} (T/10^4 K)^{-1/2}$ is the damping constant
with $\Gamma^{-1}$ the radiative lifetime of the excited state, and $\omega_D = \omega_0
(2\kappa_B T/m_H c^2)^{-1/2}$ is the thermal Doppler width with $\omega_0/2\pi )$ = 2.47
$\times$ 10$^{15}$ Hz the line center (Ly$\alpha$ frequency (Bonilha et al. 1979). For
photons of frequency $\omega_i$ and direction cosine  ${\bf n_i}$ injected into a screen
moving at velocity ${\bf v}$, we evaluate the optical depth (a Lorentz scalar) in the
cloud (comoving) frame to obtain

	$$\tau(\omega_{cloud}) = N^{screen}_{HI} \sigma_0(T) \phi(\omega_{cloud},T), ~~[B4]$$

where $\omega_{cloud} = \gamma \omega_i (1-{\bf v . n})$ and $\gamma =
(1-v^2)^{-1/2}$. For line center injection, $\omega_i = \omega_0$, and the  column
density at which $\tau(\omega_{cloud}$ =1, $N_{HI}^{crit}$, is given by

$$ N_{HI}^{crit} = \frac{1}{\sigma_0(T) \phi(\omega_{cloud},T)} = N_{HI}^{crit}(v).~~[B5] $$

\vspace{0.5cm}

\centerline {2. CALCULATIONS OF $f$}

\vspace{0.5cm}

Let $N(x)$ be the flux of Ly$\alpha$ photons impinging on a screen which we model as a
plane parallel slab. The width of the impinging line feature is $\delta V_s$ (in
km s$^{-1}$ so that $x = (\omega - \omega_0)/(\omega \delta V_s)$. Let the width of the
Ly$\alpha$ resonant scattering line profile in the screen be given by $\delta V_a$ (in
km s$^{-1}$). The probability that a line photon will suffer an interaction when
traversing the screen is $1 - e^{-\tau(x')}$, where $\tau(x')$ is given by equation
(B1), and $x' = (\omega - \omega_0)/(\omega_0 \delta V_a) = \kappa x$ with $\kappa
\equiv V_s/\delta V_a$. The fraction of Ly$\alpha$ photons that suffer interactions
when traversing the screen is then

$$ f = \frac {\int_{-\infty}^{+\infty}{dx
N(x)[1-e^{-\tau(x')}]}}{\int_{-\infty}^{+\infty}{dx N(x)}} = 
f(N^{screen}_{HI},\kappa). ~~[B6]$$

Taking $N(x) \propto e^{-x^2}$ gives

$$ f = \pi^{-1/2} \int_{-\infty}^{+\infty}{dx e^{-x^2}}[ 1-e^{-\tau(x')}].~~[B7] $$

Equation (B6) also gives the fraction of the impinging Ly$\alpha$ energy flux that
interacts with the HI gas in the screen since the impinging energy flux of Ly$\alpha$
is approximately $\hbar \omega_0 N(x)$.

\vspace{0.5cm}

{\Large \bf References}

\vspace{0.5cm}

\noindent Bonilha J.R.M., Ferch R., Salpeter E.E., Slater G., Noerdlinger P.D., 1979, \apj 233,
649

\newpage

\pagestyle{myheadings}
 
\markright{Apendix C}

\vspace{5cm}

\centerline{\huge \bf Apendix C}

\vspace{1cm}

\centerline{Binette, Wang, Villar-Mart\'\i n, Martin \& Magris 1993, ApJ, 414, 535}

\vspace{1cm}

	The transfer of the UV continuum across an arbitrary thick slab of dust which is
discussed in $\delta$2.4.4 was carried out using the numerical solution described in
detail by Magris (1985; see also Bruzual et al. 1988). Let us consider an arbitrary
frequency with its corresponding absorption ($\sigma_{abs}$) and scattering
($\sigma_{sca}$) dust cross sections. We are interested in solving the following transfer
equation

$$ \mu \frac{dI(\tau ,\mu)}{dt} = I(\tau ,\mu) - S^S(\tau ,\mu), ~~[C1]$$

where $d\tau = (\sigma_{sca} + \sigma_{abs}) dx$ is the differential optical depth of the
medium. This equation which takes into account scattering as well as absorption
by homogeneously distributed dust particles is appropriate to the plane parallel
 geometry
case. The scattering source function due to the dust grains is given by

$$S^S =  \frac{a}{4\pi} \int {I(\tau ,\mu) - S^S(\tau ,\mu)}, ~~[C2]$$

where $a = \sigma_{sca}/(\sigma_{sca} + \sigma_{abs})$ is the albedo and $\Phi(\theta) =
(1-g^2)/(1+g^2-2g~cos\theta)^{3/2}$ is the scattering phase function. The phase function
is characterized by the asymmetry parameter $g = <cos \theta >$ with $\theta$ the
scattering angle and $g$ the Henyey-Greenstein asymmetry parameter characterizing our
dust grain models and tabulated by P.G.Martin as a function of frequency along with
$\sigma_{abs}$ and $\sigma_{sca}$ (see grain model in Martin \& Rouleau 1991). The
transfer equation was solved by using the finite difference solution method of Milkey,
Shine, \& Mihalas (1975) with the following boundary conditions:

\vspace{0.4cm}

\centerline {$I(\tau,\mu) = 0 $ ~~~ for ~~~ $\tau = 0, ~~~ \mu < 0$}

\vspace{0.2cm}

~~~~~~~~~~~~~~~~~~~~~~~~~~~~~~~~~~~~~~~~~~~~~~~~~~~~~~~~~~~~~~~~~~~~~~~~~~~~~~~~~~~~~~~~~~~~~[C3]

\vspace{0.35cm}

\centerline {$I(\tau,\mu) = I_c$ ~~~~~~ for ~~~  $\tau = T_{ext}, ~~~ \mu > 0,$}

\vspace{0.4cm}

where $T_{ext}$  is the optical depth of the slab and $I_c$ is the intensity of the
incident continuum radiation which illuminates {\it isotropically} one side of the dust
slab. To obtain the numerical solution, we use a logarithmically uniform mesh that is
symmetric about slab points.  In order to test the accuracy of our numerical solution,
we have compared our results with the analytical solution of Roberge (1983). In
particular, we could reproduce very well the behavior of the mean intensity versus
optical depth inside the slab when illuminated isotropically on both surfaces.

{\Large \bf References}

\vspace{0.5cm}

\noindent Bruzual A., Magris C.M., Calvet N., 1988, \apj 333, 673

\noindent Magris C.G, 1985, in senior physics thesis, Universidad Sim\'on Bol\'\i var and CIDA,
Venezuela

\noindent Martin P.G., Rouleau F., 1991, in Extreme Ultraviolet Astronomy, ed. R.F. Malina \& S.
Bowyer (Oxford: Pergamon), 341

\noindent Milkey R.W., Shine R.A., Mihalas D., 1975, \apj 202, 250

\noindent Roberge W.G., 1983, \apj 275, 292

\newpage

\pagestyle{myheadings}
 
\markright{Apendix D}

\vspace{5cm}

\centerline{\huge \bf Apendix D}

\vspace{1cm}

\centerline{Villar-Mart\'\i n \& Binette 1995, A\&A, in press}

\vspace{1cm}

By comparing the estimated photoionization rates from the excited
level 3d of Ca$^{+*}$, $\Pi_{3d}$, to that from the ground state 4s
of Ca$^+$, $\Pi_{4s}$, we show that photoionization of excited
Ca$^{+*}$ is a negligible process. 

 	A fundamental parameter which determines the importance of
the ionization rate from the metastable level is its relative population
with respect to the ground state: $\frac{n_{3d}}{n_{4s}}$, being 
$n_{4s}$ the density of Ca$^+$ ions in the ground level and $n_{3d}$
the density of Ca$^+$ ions in the metastable level, 3d. To evaluate this fraction,
we have solved analytically the statistical equilibrium equations.
To simplify the calculations  we have considered a three level atom, reducing the
two 4p sublevels (see Fig.~16, Chapter 3) to a single level and the same for the 3p sublevels. 
We have taken into account all the processes (collisional and radiative) which can
populate or depopulate  each of the levels. 
	
The resolution of the system of three equations gives us  the ratios
$\frac{n_L}{n_{Ca^+}}$, with L=3d,4s,4p, being $n_{Ca^+}$ the total
density of Ca$^+$ ions. The density and temperature we considered were
300$cm^{-3}$ and 10000K, respectively.  From this we deduced the
relative population with respect the ground state. The results turn
out to be:

$$\frac{n_{3d}}{n_{4s}} \sim 10^{-7}$$  and	$$\frac{n_{4p}}{n_{4s}} \sim 0$$

	The negligible population of the upper 4p level prevents any
contribution by cascade to the population of the 3d level, therefore
collisional excitation is the only important mechanism populating the
metastable 3d level. This is consistent with the fact that we do not
observe the triplet of Ca$^+$ (8498,8542,8662) (4p to 4s) in any EELR
although it is observed in the broad line region of AGN where
densities are higher by many order of magnitudes.

{\bf a) Ionization of Ca$^{+*}$ by soft continuum photons.}

The soft UV counterpart of the ionizing continuum provides a source of
ionization for both Ca$^{+}$ (IP: 11.9eV) and Ca$^{+*}$ (IP: 10.2eV). To
estimate the photoionization rates, we will make the following approximations:

\noindent 1) At the fairly large depth in the cloud where the specie
Ca$^+$ becomes abundant, we only need to consider  photons with
energies $<13.6eV$, the ionization potential (IP) of H$^0$, because photons
just above this energy have already been absorbed and also because of
the rapid decrease of the photoionization cross section with
increasing energies.

\noindent 2) The ionizing continuum which reaches the PIZ is considered to
be the soft UV counterpart of an ionizing PL of index $-1.4$ (but
unattenuated since the opacity due to dust or trace elements below
13.6eV is relatively small). The continuum impinging the cloud is
described by $F_\nu = F_s \nu^{-1.4}$. In number of photons this is
$F_{\nu}/h\nu = F_s/h \nu^{-2.4}$. The constants $F_s/h$ will cancel out
when taking the ratio $\Pi_{3d}/\Pi_{4s}$.

	The atomic data was taken from Osterbrock (1987) for H$^0$ and
from Shine \& Linsky (1975) for Ca$^+$ and Ca$^{+*}$. Tables 1, 2, 3
where we define $a'_{\nu}$ = $a_{\nu}*10^{18}$ and $\nu'=\nu/10^{16}$
show the relevant atomic data.  The threshold ionizing frequency of
H$^0$ is labeled $\nu_0$.

We  estimate the  quotient $\Pi_{3d}/\Pi_{4s}$ as follows:

$$\frac{\Pi_{3d}}{\Pi_{4s}}=
\frac{n_{3d}}{n_{4s}}
\frac{\int\limits_{\nu_{3d}}^{\nu_{0}}{\nu^{-2.4}a_{\nu}(Ca^{+*})
d\nu}}{\int\limits_{\nu_{4s}}^{\nu_0}{\nu^{-2.4}a_{\nu}(Ca^+) d\nu}}, ~~[D1]$$
If we define $a'_{\nu}$ = $a_{\nu}*10^{18}$ and $\nu'=\nu/10^{16}$, we have

$$\frac{\Pi_{3d}}{\Pi_{4s}}=
\frac{n_{3d}}{n_{4s}}
\frac{\int\limits_{\nu'_{3d}}^{\nu'_{0}}{\nu'^{-2.4}a'_{\nu}(Ca^{+*})
d\nu'}}{\int\limits_{\nu'_{4s}}^{\nu'_0}{\nu'^{-2.4}a'_{\nu}(Ca^+) d\nu'}} $$ 

$$= \frac{n_{3d}}{n_{4s}} \frac{I_1} {I_2}, ~~[D2]$$ 
The values of $a'_{\nu}(Ca^{+*})$ at $\nu'_0, \nu'_{4s}$ and
$a'_{\nu}(Ca^{+})$ at $\nu'_0$ and $\nu'_{4s}$ have been obtained by
interpolation from (see Tables 2 and 3).
Using the rule of the rectangle to approximate a given integral:

{\small $$\int\limits_{x_0}^{x_1}f(x) dx \sim (x_1-x_0)f(x_0) + 
(x_2-x_1)f(x_1) + ... $$
$$ + (x_n-x_{n-1}) f(x_{n-1})$$}

\begin{table}
\centering
\caption{$H^0$ photoionization cross section}
\begin{tabular}{lll} \hline
  $\nu'$ & $a'_{\nu}(H^0)$  & $\nu^{'-2.4}a'_{\nu}(H^0)$ \\ \hline
         0.3    &       6.4    &     115.1	 \\
         0.4    &       2.5     &    22.54	 \\
         0.6     &     1.15    &     3.919	 \\
         0.8     &      0.6    &     1.025	 \\
           1     &      0.3      &     0.3	 \\
         1.2     &      0.2    &    0.1291	 \\
         1.4     &     0.15   &    0.06689	 \\
         1.6    &       0.1  &     0.03237	 \\
         1.8    &      0.07  &     0.01708	 \\
           2     &     0.05   &   0.009473	 \\
         2.2    &      0.04  &    0.006029	 \\
         2.4    &      0.03   &     0.00367	 \\
         2.6    &      0.02   &    0.002019	 \\
         2.8    &     0.015  &    0.001267	 \\
           3    &     0.015   &   0.001074	 \\
         3.2    &      0.01  &   0.0006133	 \\
         3.4    &      0.01  &   0.0005302	 \\
         3.6    &      0.01  &   0.0004622	 \\

\end{tabular}
\end{table}

\begin{table}
\centering
\caption{Ca$^+$ 4s photoionization cross section}
\begin{tabular}{llll} \hline
$\lambda(\AA)$ &  $\nu'$ & $a'_{\nu}(Ca^{+})$  & $\nu^{'-2.4}a'_{\nu}(Ca^{+})$ \\ \hline
        1044 &     0.2873 ($\nu'_{4s}$) &     0.2036 &      4.063	\\
        1000 &        0.3 &     0.2097 &      3.772	\\
         950 &     0.3158 &     0.2145 &      3.412	\\
	  &    0.3288 ($\nu'_0$) &     0.2157 &      3.114     \\
         900 &     0.3333 &     0.2170 &      3.031	\\
         850 &     0.3529 &     0.2172 &      2.644	\\
         800 &      0.375 &     0.2149 &      2.262	\\
         750 &        0.4 &     0.2103 &      1.896	\\
         700 &     0.4286 &     0.2033 &      1.554	\\
         650 &     0.4615 &     0.1942 &      1.242	\\
         600 &        0.5 &     0.1830 &      0.966	\\
         550 &     0.5455 &     0.1700 &     0.7282	\\
         500 &        0.6 &     0.1554 &     0.5295	\\
         450 &     0.6667 &     0.1394 &      0.369	\\
         400 &       0.75 &     0.1225 &     0.2443	\\
         350 &     0.8571 &     0.1049 &     0.1518	\\
         300 &          1   &   0.0870  &   0.08697	\\		
         250 &       1.2  &     0.0691   &   0.0440	\\
         200 &        1.5  &    0.0519 &    0.0199	\\
         150 &          2  &    0.0352 &    0.00668	\\
         100 &          3 &      0.0199 &   0.001417	\\
          50 &          6  &    0.0198  & 0.0002685	\\

\end{tabular}
\end{table}

\begin{table}
\centering
\caption{Ca$^{+*}$ 3d photoionization cross section}
\begin{tabular}{llll} \hline
$\lambda(\AA)$ & $\nu'$ & $a'_{\nu}(Ca^{+*})$  & $\nu^{'-2.4}a'_{\nu}(Ca^{+*})$ \\ \hline
       1218 &     0.2462 ($\nu'_{3d}$) &      6.148 &      177.70  \\
        1200 &       0.25 &      6.086 &      169.50  \\
        1150 &     0.2609 &      5.907 &      148.60	\\
        1100 &     0.2727 &      5.716 &      129.20	\\
        1050 &     0.2857  &      5.511 &      111.4	\\
 	1044 & 	    0.2873 ($\nu'_{4s}$)  &    5.403  &    107.80  \\
        1000 &        0.3 &      5.295 &      95.22	\\
         950 &     0.3158 &      5.066 &      80.56	\\
	    &     0.3288 ($\nu'_0$) & 4.9465 &  71.39  \\
         900 &     0.3333 &      4.827 &      67.41	\\
         850 &     0.3529 &      4.576 &      55.72	\\
         800 &      0.375 &      4.315 &      45.43	\\
         750 &        0.4 &      4.044 &      36.46	\\
         700 &     0.4286 &      3.762 &      28.74	\\
         650 &     0.4615 &       3.47 &      22.19	\\
         600 &        0.5 &      3.168 &      16.72	\\
         550 &     0.5455 &      2.856 &      12.23	\\
         500 &        0.6 &      2.534 &      8.634	\\
         450 &     0.6667 &      2.202 &      5.827	\\
         400 &       0.75 &      1.862 &      3.714	\\
         350 &     0.8571 &      1.517 &      2.196	\\
         300 &          1 &      1.173 &      1.173	\\
         250 &        1.2 &     0.8424 &     0.5439	\\
         200 &        1.5 &     0.5401 &     0.2041	\\
         150 &          2 &     0.2865   &    0.05428 	\\        
     	 100 &  	3   &     0.1051   &   0.007524	\\
          50  &         6    &   0.01471  &   0.0001995 \\
   
\end{tabular}
\end{table}

[D2] reduces to 

$$\frac{\Pi_{3d}}{\Pi_{4s}} \sim \frac{n_{3d}}{n_{4s}} \frac{ 10.06 } {0.156}, ~~[D3]$$ 
If we take into account that $\frac{n_{3d}}{n_{4s}}\sim 10^{-7}$ we obtain

$$\frac{\Pi_{3d}}{\Pi_{4s}} \sim 6\times10^{-6}$$ which means that the
ionization of Ca$^+$ into Ca$^{++}$ originates overwhelmingly from the
ground level 4s.

{\bf b) Ionization of Ca$^{+*}$ by Ly$\alpha$ photons.}

Ly$\alpha$ photons have energies slightly higher than the treshold
energy of level 3d of Ca$^{+*}$ and constitute undoubtedly  an important source of
ionization for this level.  To compute the Ly$\alpha$ emissivity, we
assume the low density regime whereby $\simeq \frac{2}{3}$ of
recombinations of H$^+$ lead to the emission of a Ly$\alpha$ photon.
Since Ly$\alpha$ is a resonance line of large line scattering opacity,
we consider that the density of Ly$\alpha$ photons within the Ca$^+$
region result from the production of Ly$\alpha$ either locally or from
{\it deeper} regions.  The justification for this is that resonance line
photons generated from layers nearer the slab'surface would be
reflected outward as a result of the increasing fraction of H$^o$ with
depth (see Binette et~al 1993b).  Therefore the Ly$\alpha$ flux
potentially available to ionize level 3d is a small fraction $\eta$ of
the total Ly$\alpha$ flux emitted by the cloud. This fraction is of
order 0.2 corresponding to the fraction of ionizing photons of H$^0$
{\it not yet} absorbed at the typical depth where the Ca$^+$ specie is
abundant.

	We first compute the ${\Pi^{Ly\alpha}_{3d}/\Pi_{4s}}$ ratio
taking into account that resonance scattering will increase the
density of locally emitted line photons by a factor $\xi \sim 10^{7}$,
the mean number of scatterings before escape. Such a high number of
line scattering, however, characterizes only the locally produced
Ly$\alpha$ photons. This accumulation (or slowing down) effect which
we want to estimate is only effective within a zone of optical depth
of order a few. Let's take $\tau_{scat} \sim 10 = \sigma_{scat} n(H^0)
~\delta X$ where $\delta X$ is the geometrical depth and $n(H^0)$ the
local density of $H^{0}$.  Adopting $\sigma_{scat} \simeq 6 \times
10^{-14} cm^{2}$ (cf Appendix~B of Binette et~al 1993b), the column
density of recombining H$^+$ we should consider in generating
Ly$\alpha$ is within a thickness $N(H^+) \sim N(H^0)= n(H^0) ~\delta X
= 10/6\times10^{-14} = 1.7\times 10^{14}cm^{-2}$.

	Taking these considerations into account, the problem reduces
to estimating the quotient
	
$$ \frac{\Pi^{Ly\alpha}_{3d}}{\Pi_{4s}}=\frac{n_{3d}}{n_{4s}}
\frac{\frac{2}{3} \eta \xi 
a_{Ly\alpha}(Ca^{+*}) N(H^+)  \int\limits_{\nu'_0}^{\infty}{\nu'^{-2.4}a'_{\nu}(H^0)
d\nu'}}{\int\limits_{\nu'_{4s}}^{\nu'_0}{\nu'^{-2.4}a'_{\nu}(Ca^+) d\nu'}}$$
$$= \frac{n_{3d}}{n_{4s}} \frac{2}{3} \eta \xi a_{Ly\alpha}(Ca^{+*}) N(H^+),
\frac{I_3}{I_2} 
~~ [D4]$$
where $a_{Ly\alpha}($Ca$^{+*})$ is the photoionization cross
section of Ca$^{+*}$ at the Ly$\alpha$ energy ($\simeq 6\times10^{-18} cm^{2}$).
Taking again ${n_{3d}}/{n_{4s}} \sim 10^{-7}$, we obtain

$$ \frac{\Pi^{Ly\alpha}_{3d}}{\Pi_{4s}}\sim 1.36 \times 10^{-4} \frac{I_3}{I_2}, ~~
[D5]$$ 
Using again the rule of the rectangle we obtain

$$ \frac{\Pi^{Ly\alpha}_{3d}}{\Pi_{4s}}\sim 1.5\times10^{-2}.$$ 

	The effect of slowing down of resonance line photons is
interesting but appears insufficient as it involves too small a
fraction of the Ly$\alpha$ photons generated within the nebula. Let's
now estimate the importance of all the Ly$\alpha$ photons generated
within the deeper zones which, after having scattered far enough in
frequency to escape, must still cross the Ca$^+$ zone (without
appreciable scattering in that zone). The effect on the
photoionization of level 3d is given in this case by
	
$$ \frac{\Pi^{Ly\alpha}_{3d}}{\Pi_{4s}}=\frac{n_{3d}}{n_{4s}}
\frac{\frac{2}{3} \eta
a'_{Ly\alpha}(Ca^{+*}) \int\limits_{\nu'_0}^{\infty}{\nu'^{-2.4}
d\nu'}}{\int\limits_{\nu'_{4s}}^{\nu'_0}{\nu'^{-2.4}a'_{\nu}(Ca^+)
d\nu'}}, ~~ [D6]$$
$$= \frac{n_{3d}}{n_{4s}} \frac{2}{3} \eta  a'_{Ly\alpha}(Ca^{+*}) \frac{I_4}{I_2} 
~~[D7]$$
(Note that $a'$ is used here for Ca$^{+*}$). Using the rule of the rectangle  we obtain

$$ \frac{\Pi^{Ly\alpha}_{3d}}{\Pi_{4s}}\sim 2.43\times10^{-6}$$ 

In summary, Ly$\alpha$ emission and/or trapping are insufficient to
ionize Ca$^+$. As in the case of soft continuum photons, this is
basically the result of the extremely small population characterizing
the excited level 3d.

\vspace{0.5cm}

{\Large \bf References}

\vspace{0.5cm}

\noindent Roberge W.G., 1983, \apj 275, 292

\newpage

\pagestyle{myheadings}
 
\markright{}

~
\newpage

~

\vspace{2cm}

\centerline{\huge \bf Acknowledgements}

\vspace{1cm}

	I want to express my gratitude to Luc Binette, that, just a few monts ago, became co-director of this thesis, because during these years  he has oriented my thesis, sharing his knowledge and rigorous scientific method.  Thanks to Bob Fosbury, my local supervisor at the ST-ECF
(Space Telecope European Coordinating Facility, Garching-Germany) for his scientific support. I learned with him that the apparent complication of the Universe, is often the result of our too difficult questions.   To Ismael P\'erez, Reynier Pelletier y Jose Acosta 
who helped me with very constructive scientific discussions.

	Very specially, thanks to my parents and sisters, who have always helped me to advance in my work with their permanent support. Also to Jacco van Loon because he has contributed so much to my profesional and personal development, in the bad and good moments. And a special gratitude to Jes\'us A. Rodr\'\i guez (European Southern Observatory), who has supported me and followed step by step the evolution of my career towards the doctorate.

	I want to thank  the Deustche-Forschung Gemainschaft (DFG) for the grant I enjoyed during these years. Also, to the ST-ECF
group that supported me, not only with money and computer facilities, but also with helpful scientific advices. Thanks to the Itto. de Astrof\'\i sica de Canarias (IAC, Tenerife-Spain) which provided me with approprite tools to work in all my visits.

\newpage
~

\newpage

~

\vspace{2cm}

\centerline{\huge \bf Agradecimientos}

\vspace{1cm}

	Quiero expresar mi agradecimiento a Luc Binette, que se convirti\'o en co-director de esta tesis hace tan s\'olo unos meses, porque durante estos a\~nos ha orientado mi trabajo, compartiendo sus conocimientos y su m\'etodo ci\'entifico riguroso. Gracias tambi\'en a Bob Fosbury, mi supervisor en el ST-ECF (Space Telecope European Coordinating Facility, Garching-Germany), por su apoyo ci\'entifico. Con \'el he aprendido que la complicaci\'on aparente del Universo es a menudo el resultado de nuestras preguntas demasiado dif\'\i ciles. A Ismael P\'erez, Reynier Pelletier y Jose Acosta que me ayudaron con discusiones ci\'entificas muy constructivas.

	Muy especialmente, agradezco a mis padres y hermanas, que  han empujado mi trabajo con su apoyo permanente. Tambi\'en a Jacco  van Loon que
tanto me ha aportado profesional y personalmente, en los malos y los buenos momentos. Y un agradecimiento especial
a Jes\'us A. Rodr\'\i guez (European Southern Observatory) que me ha apoyado y ha seguido paso a paso la evoluci\'on de mi carrera hacia el doctorado.

	Agradezco a la Deustche-Forschung Gemainschaft (DFG) por la beca que he disfrutado estos a\~nos. Tambi\'en al grupo ST-ECF que me proporcion\'o facilidades econ\'omicas y de ordenadores, as{\'\i} como \'utiles consejos para avanzar en mi trabajo. 
Agradezco  al Itto. de Astrof\'\i sica de Canarias (IAC, Tenerife-Spain), que en mis visitas me proporcion\'o siempre medios adecuados para trabajar.

\end{document}